%% file: thesis.tex
\documentclass[12pt]{report}

\pdfoutput=1

\usepackage{xcolor}
\usepackage{appendix}
\usepackage{hyperref}

\usepackage{epsfig}
\usepackage{amssymb}
\usepackage{amsfonts}
\usepackage{amsmath}
\usepackage{euscript}
\usepackage{verbatim}
\usepackage{latexsym}
\usepackage{graphicx}
\usepackage{slashed}
\usepackage[utf8]{inputenc}
\usepackage{graphicx}	
\usepackage{subcaption}
\usepackage{wrapfig}
\usepackage[T1]{fontenc}
\usepackage{mathtools}
\usepackage{caption}
\usepackage{float}
\usepackage{subfloat}
\usepackage{calligra}
\usepackage{braket}
\usepackage{caption}
\usepackage{url}

\usepackage[spanish]{babel}

\def\pa{\partial}
\def\al{\alpha}
\def\be{\beta}

\def\de{\delta}
\def\et{\eta}

\def\si{\sigma}
\def\om{\omega}

\def\Si{\Sigma}
\def\Om{\Omega}
\def\non{\nonumber}
\def\ka{\kappa}

\def\m{\mathcal}

\newcommand{\br}{\biggr}
\newcommand{\bl}{\biggl}

\hypersetup{colorlinks=false, linkcolor=blue, citecolor=red}

\begin{document}
	
	\pagenumbering{gobble}

\input{./title.tex}

\pagebreak

\input{./dedicatoria.tex}

\pagebreak

\input{./contribuciones.tex}

\pagebreak

\input{./gracias.tex}

\pagebreak

\input{./notacion.tex}

\pagebreak

\pagebreak

\input{./Resumen.tex}

\pagebreak

\vspace*{7cm}

\begin{center}
	\textit{Dedico la presente trabajo a mis padres, Eleuterio Choque e Irene Quispe}\\[20pt]
\end{center}

\pagebreak

\pagenumbering{arabic}

\tableofcontents

\newpage

\leavevmode\thispagestyle{empty}\newpage

\newpage

\chapter*{\textbf{{Lista de publicaciones}}}
\addcontentsline{toc}{chapter}{{{ List of Publications}}}

\begin{flushleft}
	\textsf{\textbf{\large Esta tesis esta basada en las siguientes publicaciones}}
\end{flushleft}

\begin{enumerate}



	\item
A.~Anabalon, D.~Astefanesei, D.~Choque and C.~Martinez,
``Trace Anomaly and Counterterms in Designer Gravity,''
JHEP {\bf 1603}, 117 (2016)
doi:10.1007/JHEP03(2016)117
[arXiv:1511.08759 [hep-th]].

	\item
A.~Anabalon, D.~Astefanesei and D.~Choque,
``On the thermodynamics of hairy black holes,''
Phys.\ Lett.\ B {\bf 743}, 154 (2015)
doi:10.1016/j.physletb.2015.02.024
[arXiv:1501.04252 [hep-th]].

	\item
A.~Anabalon, D.~Astefanesei and D.~Choque,
``Hairy AdS Solitons,''
Phys.\ Lett.\ B {\bf 762}, 80 (2016)
doi:10.1016/j.physletb.2016.08.049
[arXiv:1606.07870 [hep-th]].

\end{enumerate}

\pagebreak

\leavevmode\thispagestyle{empty}\newpage


\input{./intro.tex}

\input{./capitulo2.tex}

\input{./capitulo3.tex}

\input{./capitulo4.tex}

\input{./capitulo5.tex}

\input{./capitulo6.tex}

\input{./conclusiones.tex}

\newpage

\appendix
\input{./apendice.tex}


\end{document}

%% file: title.tex
\begin{titlepage}
\pagenumbering{gobble}

\begin{center}

	\null 
	
	{\LARGE \emph{Agujeros Negros con Pelo y Dualidad AdS/CFT}}
	
	\vskip 1cm

	\large{Tesis presentada para el grado de Doctor en F\'isica\\ \textbf{Doctor of Philosophy}}
	
	\vspace{3cm}
	by\\
	{\Large\textbf{\textsf{{David Choque Quispe}}}}
	
	\vspace{3cm}
	Director de tesis\\
	{\Large\textbf{\textsf{{Dr. Dumitru Astefanesei}}}}
	
	\vspace{1.7cm}
	\begin{figure}[h]
		\centering
		\includegraphics[scale=0.35]{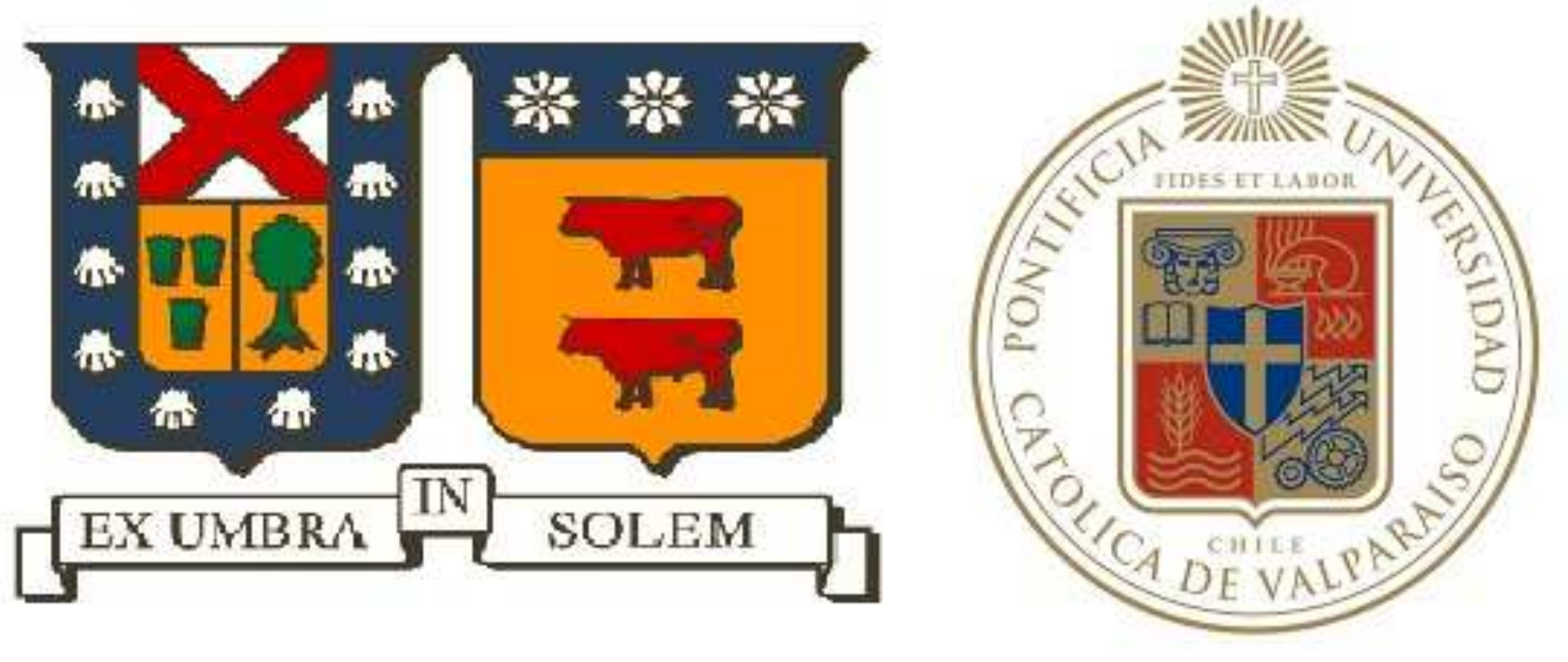}
	\end{figure}
	
	\vspace{.5cm}
	\textit{Universidad T\'ecnica Federico Santa Mar\'ia\\
		\ Av. Espa\~na - 1680. Chile-Valpara\'iso. \\}

\end{center}

\end{titlepage}

%% file: dedicatoria.tex

\section*{Dedicatoria}

Dedico este trabajo a mis padres
quienes sembraron en mi
grandes deseos de superaci\'on y
amor por la vida


%% file: contribuciones.tex

\section*{Contribuci\'on de los autores}

La tesis contiene materiales que fueron 
previamente publicados en las refs. \cite{Anabalon:2015ija, Anabalon:2015xvl, Anabalon:2016izw}. En el cap\'itulo $2$
presentamos una revisi\'on general de la literatura
necesarios para una adecuada compresi\'on del resto de 
cap\'itulos. No se hace ninguna afirmaci\'on en la originalidad del material presentado en este cap\'itulo.  En el cap\'itulo $3$ se presentan los
resultados de la refs. \cite{Anabalon:2012dw, Acena:2013jya, Anabalon:2013eaa} los que 
fueron publicados por Dumitru Astefanesei y sus 
colaboradoes. De la misma forma presentamos resultados
conocidos que permiten ganar intuici\'on. Cap\'itulo $4$ es una descripci\'on de los m\'etodos concretos que se usar\'on
en el desarrollo de la investigaci\'on y mostramos ejemplos 
intuitivos ampliamente conocidos en la literatura.
Los cap\'itulos $5$ y $6$ son la presentaci\'on de los art\'iculos de ref. \cite{Anabalon:2015ija,Anabalon:2015xvl,Anabalon:2016izw} los que fueron publicados por mi, Dumitru Astefanesei, Andres Anabalon y Cristian Martinez. Mi contribuci\'on a estos trabajos fue 
desarrollar los c\'alculos, construir las gr\'aficas mediante paquetes matem\'aticos, participar en las discusiones en todos los estados del desarrollo de los proyectos, dise\~nar y/o editar los art\'iculos.




%% file: gracias.tex

\section*{Agradecimientos}

Este trabajo no habr\'ia sido posible sin el apoyo y el est\'imulo de mi tutor, Prof. Dumitru Astefanesei,  bajo cuya supervisi\'on desarroll\'e los diferentes t\'opicos  
de la investigaci\'on. Al Prof. Andres Anabalon 
quien en diversas conversaciones y reuniones 
aprend\'i el uso de los paquetes matem\'aticos de Maple y Matem\'atica
para el desarollo de c\'alculos complejos, muy comunes en la F\'isica te\'orica.

Me gustar\'ia agradecer a la Direcci\'on
General de Investigaci\'on y Postgrado de la Universidad T\'ecnica Federico Santa Mar\'ia, quienes a mi llegada al programa de Doctorado me brindar\'on subvenci'on y dem\'as facilidades para incorporarme al programa de Doctorado. Especialmente al Prof. Issac Florez (Director de Postgrados Cient\'ifico-Tecnol\'ogicos), las secretarias Elizabeth Muga y Mar\'ia Loreto Vergara,  por la gran amabilidad y eficacia para con diversos tramites de becas y matr\'iculas. De igual forma agradezco al Instituto
de F\'isica de la PUCV y a sus investigadores quienes me acogieron y compartier\'on sus diversas experiencias en la investigaci\'on. 

Tambi\'en me gustar\'ia agradecer al programa de becas CONICYT, por su confianza y subvenci\'on a mis estudios de Doctorado. De la misma forma, mediante su apoyo realic\'e una pasant\'ia de investigaci\'on en Alemania, Berl\'in en el instituto Max Planck. Finalmente dicha subvenc\'on me permit\'io ampliar mi biblioteca personal, la que fue crucial para 
el desarrollo de la presente t\'esis.

No puedo terminar sin agradecer a mi familia, en cuyo est\'imulo constante y
amor he confiado a lo largo de mis a'nos en el postgrado. Estoy agradecido
tambi\'en con los profesores de mi alma mater, la Universidad Nacional San Ant\'onio Abad del Cusco, en especial con el Prof. Oswaldo Luizar que bajo su gu\'ia
postul\'e al postgrado en la UTFSM.


%% file: notacion.tex

\section*{Notaci\'on}

Se usan unidades naturales $\hbar=1=c$, donde $\hbar$ la 
constante de Planck reducida mientras que $c$
es la velocidad de la luz. \\

Usamos las siglas en Ingles CFT de \textit{Conformal Field Theory} en vez de su traducci\'on al espa\~nol que es \textit{Teor\'ia del  Campo Conforme}.\\

Usaremos las siglas QCD para \textit{Quantum Chromodynamics} en lugar de su versi\'on
en espa\~nol, la cual rara vez es usada en la literatura. \\

De la misma forma usamos las siglas en 
Ingles QFT de \textit{Quantum Field Theory}
en lugar de su traducci\'on al espa\~nol
 \textit{Teor\'ia Cu\'antica del Campo}.\\

Usamos la palabra $bulk$ en lugar de la traducci\'on al Espa\~nol $bolsa$. Esta se refiere al interior de una variedad, cuya
din\'amica esta determinada por las ecuaciones
de Einstein (espacio-tiempo).\\

Otras de las palabras que dejamos en Ingles son \textit{background} y \textit{ground}. Por ejemplo
usamos la expresi\'on \textit{m\'etrica del background} 
la cual se entiende como \textit{m\'etrica no din\'amica}. Y por supuesto
la frase \textit{ground state} que se refiere al \textit{estado basal, o estado no exitado}.\\

Los siglas, AdS y SAdS son para denominar las expresiones 
\textit{Anti de Sitter} y 
\textit{Schwarzschild Anti de Sitter}.

\begin{titlepage}
\end{titlepage}


%% file: Resumen.tex

\begin{center}
	\section*{Abstract}
\end{center}

The present thesis is based
in the material previously published in  \cite{Anabalon:2015ija,Anabalon:2015xvl,Anabalon:2016izw}, and it is organized in the following way.

The second chapter is shown as the isometric group of the space time AdS is exactly (isomorphic)
match to the conformal group of the dual CFT theory. We present the precise formulation of the duality AdS/CFT and we describe the behavior of a massive scalar field in AdS space-time.

In chapter 3 we present solutions of black holes widely known in the literature, which we will use in some later calculations to gain intuition, we explain the no-hair theorem and two examples of exact black hole solutions that can evade this theorem. These results were published in \cite{Anabalon:2012dw,Acena:2013jya}.

Chapter 4 we present details of the methods of
holographic renormalization and Hamiltonian formalism, which are applied in the
development of the later chapters. For the solution SAdS calculate the on-shell action and regularize, compare with the method of subtraction of the background. In the same way, we calculate the Brown-York stress tensor and its holographic dual.
The examples in this section are known results in the literature.

Chapter 5 is based on \cite{Anabalon:2016izw}. 
We investigated the boundary conditions of AdS in the presence of a scalar field
minimally coupled to gravity with conformal mass $m^{2}=-2/l^{2}$, for non-logarithmic and logarithmic branch. We calculate the on-shell action, we study the variational principle and propose new counter-terms that regularize the action. We build the Brown-York stress-tensor, calculate the stress tensor of the dual field theory and analyze the trace anomaly. With the Hamiltonian method we calculate the gravitational mass and study under what conditions the scalar field gives a contribution to the mass.  

Chapter 6, is based on \cite{Anabalon:2015ija,Anabalon:2016izw}, we describe the phase transitions of black holes with scalar hair and horizons flat and spherical. We build
the thermodynamic phase diagrams and discuss their implications in the dual quantum field theory. \\
Finally in chapter 7 the conclusions and future directions of the work are presented.


\newpage
\begin{center}
	\section*{Resumen}
\end{center}

La dualidad AdS/CFT es una realizaci\'on  concreta del principio hologr\'afico. Es una herramienta muy eficaz para extraer informaci\'on de teor\'ias gauge fuertemente acopladas (en d-dimensiones) de una teor\'ia gravitacional cl\'asica (en d+1 dimensiones). Los agujeros negros asint\'oticamente AdS juegan un rol importante en el entendimiento de la din\'amica y termodin\'amica de las teor\'ias del campo hologr\'aficas duales. En particular, estos agujeros negros son duales a los estados t\'ermicos de la teor\'ia del campo del borde.

Consideramos teor\'ias de campos escalares m\'inimamente acoplados a la gravedad con potenciales (escalares) no-triviales.
Los campos escalares desenpe\~nan un rol importante en cosmolog\'ia para modelar la materia oscura, en f\'isica de altas energ\'ias la part\'icula de Higgs es escalar. Calculamos la acci\'on on-shell regularizada con el m\'etodo de renormalizaci\'on hologr\'afica y de esta obtenemos las cantidades termodin\'amicas de soluciones exactas de agujeros negros. Estudiamos los diagramas de fase de estos agujeros negros y sus interpretaciones en la teor\'ia del campo dual. Constru\'imos el tensor de stress de Brown-York y calculamos la masa (hologr\'afica).
A continuaci\'on, para una soluci\'on general, calculamos la masa hologr\'afica, la masa Hamiltoniana y vemos que concuerdan para cualesquiera condiciones de borde. Estudiamos la anomal\'ia de traza y su contribuci\'on no-trivial a la masa. 

La presente tesis esta basada 
en el material previamente publicado en \cite{Anabalon:2015ija,Anabalon:2015xvl,Anabalon:2016izw},
y est\'a organizada de la siguiente manera.

El segundo capitulo se muestra
como el grupo de isometr\'ia del espacio tiempo AdS es ex\'actamente (isomorfa)
el grupo conforme de la teor\'ia cu\'antica dual CFT. Presentamos la formulaci\'on precisa de la dualidad AdS/CFT 
y describimos el comportamiento de un campo
escalar masivo en el espacio-tiempo AdS.

En el cap\'itulo 3 se presentan soluciones de agujeros negros ampliamente conocidos en la literatura, los cuales usaremos en algunos c\'alculos posteriores para ganar intuici\'on, explicamos el teorema de no pelo y dos ejemplos de soluciones \textit{ex\'actas} de agujeros negros que logran evadir dicho teorema. Estos resultados fueron publicados en \cite{Anabalon:2012dw,Acena:2013jya}.

Capitulo 4  presentamos detalles de los m\'etodos de 
Renormalizaci\'on hologr\'afica y el formalismo Hamiltoniano, los cuales se aplican en el
desarrollo delos cap\'itulos posteriores. Para la soluci\'on SAdS calculamos la acci\'on on-shell y regularizamos, comparamos con el m\'etodo de sustracci\'on del background. De la misma forma, calculamos el tensor de
de stress de Brown-York y de la teor\'ia cu\'antica dual.
Los ejemplos de esta secci\'on son resultados conocidos en la literatura.  

Capitulo 5 esta basada en \cite{Anabalon:2016izw}. 
Investigamos las condiciones de borde de AdS, en presencia de un campo escalar
m\'inimamente acoplado a la gravedad. 
Concr\'etamente para un campo escalar de 
masa conforme $m^{2}=-2/l^{2}$, para las ramas
logar\'itmica y no logar\'itmica. Calculamos
la acci\'on on-shell, estudiamos el principio variacional y planteamos nuevos contrat\'erminos que regularizan la acci\'on. Construimos el tensor de Brown-York, calculamos el tensor de stress de la teor\'ia cu\'antica dual y analizamos la anomal\'ia de traza. Con el m\'etodo Hammiltoniano calculamos la masa gravitacional y estudiamos en que condiciones el campo escalar da una contribuci\'on la masa.    

Capitulo 6, la que esta basada en \cite{Anabalon:2015ija,Anabalon:2016izw}, describimos las transiciones de fase
de agujeros negros con pelo escalar y de horizontes plano y esf\'erico. Construimos 
los diagramas de fase y discutimos sus implicaciones en la teor\'ia cu\'antica dual.\\
Finalmente en el capitulo 7 se presentan las conclusiones y futuras direcciones del trabajo.

%% file: intro.tex


\chapter{Introducci\'on}

Los agujeros negros han fascinado a las diversas
generaciones de f\'isicos, desde John Michell en 1783, nombrada por \'el como \textit{estrellas oscuras}. Pasando luego por el advenimiento de la Relatividad General en 1915 y un a\~no despu\'es Karl Schwarzschild obtuvo una de las primeras soluciones a la Relatividad General, \textit{el agujero negro de Schwarzschild}. 
En los a\~nos $60$, conocido como la \'epoca de oro de los agujeros negros, cuyos pricipales protagonistas fueron,  Wheeler, Bekenstein, Hawking entre otros, mostraron que los agujeros negros se comportan como un sistema termodin\'amico. Uno de los grandes problemas que surgi\'o fue describir el orig\'en microsc\'opico de la entrop\'ia de los agujeros negros, en este contexto, la teor\'ia de cuerdas fue capaz de explicar la entrop\'ia de algunos agujeros negros (extremos y cercanamente extremos) cargados y/o rotantes, donde $AdS$ forma parte de la geometr\'ia (altamente sim\'etrica) cerca del horizonte \cite{Sen:2005wa,Astefanesei:2006dd}.
Finalmente, es importante mencionar que observaciones astron\'omicas
demuestran la existencia de agujeros 
negros cerc\'anamente extremos, por ejemplo la fuente de rayos-X GRS 1915+105 es un agujero negro r\'apidamente rotante de Kerr \cite{McClintock:2006xd}. 

De forma paralela, las teor\'ias cu\'anticas 
fueron desarrollándose, explicando con mayor precisi\'on escalas de energ\'ia mayores (distancias muy peque\~nas). La teor\'ia que describe las interacciones 
fuertes de las part\'iculas (quarks) que constituyen 
a los mesones y bariones es QCD, se basa en el grupo gauge $SU(3)$. Esta teor\'ia tiene una car\'acteristica muy especial, ya que a bajas energ\'ias el acoplamiento es fuerte, por lo que los quarks est\'an confinados. En este ente r\'egimen fuerte de acoplamiento es dif\'icil realizar c\'alculos. En 1974 \textquoteright t Hooft sugiri\'o que cuando 
el n\'umero de colores $N$ es muy grande se pueden encontrar grandes simplificaciones \cite{hooft}.

Juan Maldacena  en 1997
propuso la dualidad AdS/CFT \cite{Maldacena:1997re},
mostrando un ejemplo concreto donde dos teor\'ias aparentemente distintas, como son, gravitaci\'on por un lado y una teor\'ia cu\'antica por el otro son duales. Esta dualidad nos permite 
obtener informaci\'on de una teor\'ia cu\'antica (en d-dimensiones) en su r\'egimen de acoplamiento fuerte a partir de una teor\'ia cl\'asica gravitacional en (d+1 dimensiones)  
cl\'asica. Esta es una realizaci\'on concreta del principio hologr\'afico. 
Interesantemente se mostr\'o que las simetr\'ias 
locales de AdS corresponden a las simetr\'ias
globales de la CFT. Es bien conocido que las simetr\'ias AdS
en el borde cambian en presencia de campos escalares, la cuesti\'on es investigar en que condiciones los 
campos escalares preservan la simetr\'ia conforme.
Los agujeros negros asint\'oticamente AdS, con temperatura de Hawking $T$, son los estados t\'ermicos de la teor\'ia cu\'antica dual, por lo tanto, es importante en el lado gravitacional definir adecuadamente la energ\'ia, temperatura, entrop\'ia, etc, del sistema gravitacional. En espacios-tiempo AdS, existen diversos
m\'etodos para calcular la energ\'ia gravitacional.  Mientras que el m\'etodo hologr\'afico esta basado en
la dualidad AdS/CFT de la que podemos calcular la masa y las dem\'as cantidades termodin\'amicas, el procedimiento Hamiltoniano nos da la energ\'ia debido a la simetr\'ia de las traslaciones temporales. Mostramos que ambos m\'etodos concuerdan en el resultado de la energ\'ia, a\'un cuando la simetr\'ia conforme se rompe.  

Finalmente es importante destacar la importancia
de los campos escalares en diversas \'areas de la f\'isica. En cosmolog\'ia el campo escalar conocido como \textit{inflat\'on} describe la etapa inflacionaria de nuestro universo temprano, cuyos efectos pueden ser  medidos del fondo c\'osmico de radiaci\'on. Adem\'as uno de los recientes candidatos a la materia oscura son las  part\'iculas escalares conocidas como \textit{Axiones}. 
En F\'isica de altas energ\'ias la part\'icula de Higgs, recientemente descubierta es una part\'icula escalar con una masa aproximada de 125 GeV.

%% file: capitulo2.tex
         \chapter{Dualidad gravedad/gauge}

La evidencia experimental es la base del modelo
est\'andar que describe la f\'isica a nivel fundamental.
Este modelo clasifica a toda la materia de acuerdo a 
su esp\'in, $Bosones$
y $Fermiones$, donde estas part\'iculas son excitaciones
de alg\'un campo. Las interacciones fuertes requieren
grandes cantidades de energ\'ia (peque\~nas distancias)
para explorar sus propiedades e implicaciones, que se muestran a distancias
del orden de la longitud de Planck. En esa escala los efectos de la gravedad
cu\'antica son significativos pero por el momento no se ha 
podido cuantizar la gravedad de forma consistente. Al 
rescate viene la teor\'ia de cuerdas, donde los elementos
fundamentales, las part\'iculas, son reemplazadas por cuerdas. 
          
Las cuerdas oscilantes dan un espectro 
de masas o energ\'ias, estas oscilaciones 
son part\'iculas a bajas energ\'ias, de hecho las oscilaciones de la cuerda pueden darnos distintos tipos de part\'iculas. Es interesante 
aclarar que hay varias teor\'ias de cuerdas y todas ellas incluyen part\'iculas de masa cero y esp\'in dos (gravit\'on). Actualmente, la teor\'ia de cuerdas 10-dimensional es descrita por cuerdas con excitaciones fermi\'onicas y da lugar a una teor\'ia supersim\'etrica. Y es posible ir a cuatro dimensiones considerando teor\'ia de cuerdas en $\mathbb{R}^{4}\times M_{6}$ donde $M_{6}$ es alguna variedad seis dimensional compacta.     

La relaci\'on entre teor\'ias gauge y las teor\'ias de cuerdas en espacios-tiempo AdS
fue motivado por el estudio de las D-branas 
y agujeros negros en teor\'ia de cuerdas.
Las D-branas son solitones en teor\'ia de 
cuerdas y vienen en distintas dimensiones. 
Por ejemplo la D-cero-brana es una part\'icula puntual tipo solit\'on. Cuando el acoplamiento de la cuerda es peque\~no, $g_{s}<<1$, 
las D-branas son m\'as fundamentales que las cuerdas. En la teor\'ia de cuerdas perturbativa, 
las Dp-branas se definen como superficies donde las cuerdas abiertas terminan y son fuentes de las cuerdas cerradas\footnote{ Estas Dp-branas, son hiperplanos (p+1)-dimensionales en el espacio-tiempo y
	pueden estar cargados dentro de potenciales gauge (p+1)-forma. \\
La letra "D" hace referencia a las condiciones de borde de 
Dirichlet de las cuerdas}.

En esta secci\'on discutiremos algunos aspectos
a cerca de la dualidad AdS/CFT (gravedad/gauge). En la primera parte mostramos las simetr\'ias (local/global) que comparten el espacio-tiempo AdS y la teor\'ia cu\'antica conforme. Presentamos la dualidad y su formulaci\'on precisa. Finalmente estudiamos el comportamiento de un campo escalar masivo en el espacio-tiempo AdS.

\section{Grupo conforme}
\label{seccion2}
Una extensi\'on interesante
de la invarianza de Poincar\'e es la 
adici\'on de la simetr\'ia de invarianza
de escala que relaciona la f\'isica 
a diferentes escalas\footnote{Es interesante mencionar que este es precisamente el comportamiento que muestran los materiales durante una transici\'on de fase cerca del punto cr\'itico.}. En relatividad especial, 
el \'algebra de Poincar\'e $ISO(d,1)$ define las simetr\'ias fundamentales 
del espacio-tiempo d+1 dimensional\footnote{En esta secci\'on las letras griegas corresponden a las coordenadas
del espacio-tiempo, $\mu,\nu=0,\cdots, d-1$}, con la signatura $\eta_{\mu\nu}=(-,+,+ .. +)$:
\begin{align}
\label{poincare}
i[M_{\mu\nu},M_{\rho\sigma}]&=\eta_{\nu\rho}M_{\mu\sigma}-\eta_{\mu\rho}M_{\nu\sigma}-\eta_{\sigma\mu}M_{\rho\nu}+\eta_{\sigma\nu}M_{\rho\mu}~ , \\ \non
i[P_{\mu},M_{\sigma\rho}]&=\eta_{\mu\rho}P_{\sigma}-\eta_{\mu\sigma}P_{\rho}~,\\ \non
[P_{\mu},P_{\nu}]&=0~, \\ \non
\end{align}
donde $P_{\mu}, M_{\mu\nu}$ son los generadores de traslaci\'on
y transformaci\'on de Lorentz respectivamente.
Cuando intentamos relacionar la f\'isica a diferentes escalas
(energ\'ias) agregamos la invarianza de escala al grupo de
Poincar\'e (\ref{poincare}):    
\begin{align}
\label{poincare5}
i[D,P_{\mu}]&=P_{\mu}~, \\ \notag
[M_{\mu\nu},D]&=0~. \\ \notag
\end{align} 
Hasta este punto, el grupo de Poincar\'e es extendida por $D$ el cual es el generador de dilataciones\footnote{En esta secci\'on usamos $d+1$ como la dimensi\'on del espacio-tiempo en lugar de $D$ para evitar confusiones con el operador de dilataci\'on $D$}. La transformaci\'on de escala $x^{\mu}\rightarrow\lambda x^{\mu}$ preserva la forma de la m\'etrica salvo un factor de escalar, esta 
s\'olo cambia la distancia entre los puntos de forma r\'igida,
como una fotograf\'ia que preserva la forma de la foto a diferentes tama\~nos. El grupo de generadores $P_{\mu}$, $M_{\mu\nu}$ y $D$, forman el grupo de Weyl y puede
extenderse mediante un generador vectorial, $K_{\mu}$,
que describe las transformaciones conformes especiales
\begin{equation}
K_{\mu}:\qquad x^{\mu}\rightarrow \frac{x^{\mu}+a^{\mu}x^{2}}{1+2x^{\nu}a_{\nu}+a^{2}x^{2}}~,
\end{equation}  
cuyas simetr\'ias est\'an generadas seg\'un el siguiente \'algebra 
en conjunto con (\ref{poincare}) y (\ref{poincare5})
\begin{align}
i[M_{\mu\nu},K_{\rho}]&=\eta_{\mu\rho}K_{\nu}-\eta_{\nu\rho}K_{\mu}~, \\ \notag
[D,K_{\mu}]&=iK_{\mu}~, \\ \notag
[P_{\mu},K_{\nu}]&=2i(M_{\mu\nu}-\eta_{\mu\nu}D)~, \\ \notag
[K_{\mu},K_{\nu}]&=0~. \\ \notag
\end{align}
Se tiene que, el grupo de Poincar\'e, m\'as transformaciones de escala y 
la transformaciones conformes especiales forman el grupo
conforme. Las teor\'ias cu\'anticas que son invariantes bajo este grupo
se conocen como Teor\'ias Cu\'anticas Conformes (CFT). El grupo conforme
es el grupo de transformaciones que preservan la forma de la m\'etrica 
salvo un factor conforme $g_{\mu\nu}(x)\rightarrow \Omega^{2}(x)g_{\mu\nu}(x)$.
Incluye adem\'as la simetr\'ia de inversi\'on, que es una simetr\'ia discreta actuando como $x^{\mu}\rightarrow x^{\mu}/x^{2}$.
%
%
\newpage
Esta \'algebra es isomorfa al \'algebra del grupo de isometr\'ia $SO(d,2)$, con generadores 
$J_{ab}=-J_{ba}$ $(a,b=0,...,d+1)$. Usando la descomposici\'on
de los \'indices $a=(\mu,d,d+1)$, se pueden realizar
las identificaciones de los generadores:
\begin{equation}
J_{\mu\nu}=M_{\mu\nu}~, \qquad J_{\mu d}=\frac{1}{2}(K_{\mu}-P_{\mu})~, \qquad J_{\mu(d+1)}=\frac{1}{2}(K_{\mu}+P_{\mu})~, \qquad J_{(d+1)d}=D~.
\end{equation}
Estos generadores forman una matriz antisim\'etrica $(d+1)\times(d+1)$ 
\begin{align}
J_{ab} = \left( \begin{array}{ccc}
J_{\mu\nu} & J_{\mu d} & J_{\mu (d+1)} \\
-J_{\mu d} & 0 & D \\
-J_{\mu (d+1)} & -D & 0 \end{array} \right)~.
\end{align}
Un caso especial es para un espacio-tiempo $d=2$ dimensional, donde el grupo es infinito dimensional, esto debido a que
tiene un conjunto infinito de generadores.
Para espacios-tiempo de dimesi\'on $d\geq 3$ el n\'umero 
de generadores es\footnote{En este conteo no se toma en cuenta la simetr\'ia de inversi\'on ya que esta es discreta.}: $\frac{(d+1)(d+2)}{2}$ 

\begin{itemize}
	\item $P_{\mu}: x_{\mu}\rightarrow x_{\mu}+a_{\mu}\Rightarrow d$~(generadores) ~, 
	\item $M_{\mu\nu}: x_{\mu}\rightarrow \Lambda^{\nu}_{\mu}x_{\nu}\Rightarrow \frac{d(d-1)}{2}$~(generadores)~, 
	\item $D: x^{\mu}\rightarrow\lambda x^{\mu} \Rightarrow 1$~(generador)~,
	\item $K_{\mu}: x_{\mu}\rightarrow \frac{x_{\mu}+a_{\mu}x^{2}}{1+2x_{\nu}a^{\nu}+a^{2}x^{2}}\Rightarrow d$~(generadores)~.
\end{itemize}
 
Los operadores (o campos) son
auto-funciones del operador de escala $D\phi=-i\Delta\phi$, 
donde $\Delta$ es la dimensi\'on escala ($scaling$ $dimension$) del campo
\begin{equation}
x\rightarrow \lambda x \Rightarrow \phi(x)\rightarrow \phi(x)^{'}=\lambda^{\Delta}\phi(\lambda x)~.
\end{equation}
Esquem\'aticamente hablando los generadores $K_{\mu}$ y $P_{\mu}$
cumplen la funci\'on de operadores destrucci\'on y creaci\'on respectivamente,
donde $K_{\mu}$ disminuye la dimensi\'on del campo y $P_{\mu}$ la aumenta, 
veamos, $P_{\mu}\phi=?$ Considerando 
el \'algebra conforme
\begin{equation}
[D,P_{\mu}]=-iP_{\mu}\Rightarrow D(P_{\mu}\phi)=-i(\Delta+1)(P_{\mu}\phi)~.
\end{equation}
Donde la acci\'on de $P_{\mu}$ sobre el campo $\phi(x)$ (de dimensi\'on $\Delta$)
da como resultado un campo de dimensi\'on $\Delta+1$.
Cada representaci\'on del grupo conforme tiene campos u operadores
de dimensi\'on m\'inima, estos se llaman Operadores Primarios ($primary$ $operators$),
y son tales que $K_{\mu}\phi=0$, entonces los operadores primarios
son como el estado de vac\'io $\vert 0\rangle$. \\
Considerando el \'algebra conforme la funci\'on de correlaci\'on
a dos puntos de dos campos de igual dimensi\'on es
\begin{equation}
\langle\phi(0)\phi(x)\rangle\equiv\frac{1}{(x^{2})^{\Delta}}~.
\end{equation}
Adem\'as la dimensi\'on del tensor
energ\'ia-momento de cualquier teor\'ia conforme es
un operador de dimensi\'on $\Delta=d$, y siempre que hayan
simetr\'ias globales las corrientes $J_{\mu}$ conservadas tienen
dimensi\'on $\Delta=d-1$.
\newpage 
\section{Espacio Anti-de Sitter}
El espacio anti-de Sitter (AdS), es un
espacio-tiempo (o variedad) de geometr\'ia Lorentziana, curvatura constante negativa y que posee
el n\'umero m\'aximo de isometr\'ias 
en todas las dimensiones. Es importante
diferenciar entre un espacio (variedad Riemanniana) y un espacio-tiempo (variedad Pseudo-Riemanniana o Lorentziana),
esta diferencia radica en la signatura de la m\'etrica,
la cual debe ser Lorentziana para espacios-tiempo.\\

Para entender la diferencia, revisaremos r\'apidamente 
la geometr\'ia de espacios con curvatura constante,
como la esfera y el espacio hiperb\'olico.\\

La esfera bidimensional $S^{2}$ puede definirse como
la compactificaci\'on del espacio Euclideo
plano $\mathbb{R}^{2}$  el que se consigue agregando 
un punto en el infinito, en donde podemos definir
naturalmente una teor\'ia conforme en $S^{2}$.
Otra manera de obtener $S^{2}$ es agregando una 
ligadura (v\'inculo) en el espacio Euclideo $\mathbb{R}^{3}$:

\begin{equation}
ds^{2}=dX^{2}+dY^{2}+dZ^{2}~,
\end{equation}
donde la esfera $S^{2}$ est\'a sumergida y se
define por la ligadura $L=cte$,
\begin{equation}
X^{2}+Y^{2}+Z^{2}=L^{2}~.
\end{equation}
Resolviendo esta ligadura la m\'etrica 
es
\begin{equation}
ds^{2}=L^{2}(d\theta^{2}+\sin^{2}{\theta} d \varphi^{2})~.
\end{equation}
Esta m\'etrica que define a la esfera $S^{2}$ tiene las 
isom\'etrias descritas por el grupo $SO(3)$, en otras palabras, la esfera de curvatura constante $R=2/L^{2}$ es invariante (homog\'eneo) bajo el grupo de rotaciones $SO(3)$. Similarmente el espacio de curvatura constante negativa es el hiperb\'olico. No debemos confundir el hiperbolide definido en el espacio Euclideo

\begin{equation}
ds^{2}=dZ^{2}+dX^{2}+dY^{2}~,
\end{equation}

cuya ligadura es

\begin{equation}
-Z^{2}+X^{2}+Y^{2}=-L^{2}~,
\end{equation} 
pero este hiperboloide no es homog\'eneo, es decir no hay un grupo de simetr\'ia $SO(3)$ que lo deje invariante\footnote{La signatura de la m\'etrica del hiperboloide en general tiene p-direcciones tipo tiempo y q-tipo espacio, $(-\cdots -, +\cdots +)$, la cual s\'i es invariante bajo el grupo $SO(p,q)$ pero no es invariante bajo el grupo $SO(p)$, donde $p+q=3$.}.\\    
El espacio hiperb\'olico $H^{2}$ no puede sumergirse en el 
espacio Euclideo pero si en el espacio 3-dimensional de Minkowski. 
Definimos el espacio hiperb\'olico como:

\begin{equation}
ds^{2}=-dZ^{2}+dX^{2}+dY^{2}~,
\end{equation}

\begin{equation}
-Z^{2}+X^{2}+Y^{2}=-L^{2}~,
\end{equation} 

donde el grupo de simetr\'ia es $SO(1,2)$
el cual es el grupo de "Lorentz". Parametrizando 
la ligadura como:

\begin{equation}
X=L\sinh{\rho}\cos{\varphi}~, \qquad Y=L\sinh{\rho}\sin{\varphi}~, \qquad 
Z=L\cosh{\rho}~,
\end{equation}

se obtiene la m\'etrica de curvatura constante negativa,
$R=-2/L^{2}$,

\begin{equation}
ds^{2}=L^{2}(d\rho^{2}+\sinh^{2}{\rho}~d\varphi^{2})~.
\end{equation}

Vemos que el espacio hiperb\'olico 
no es un espacio-tiempo (signatura Lorenziana). Este espacio 
es la versi\'on eucl\'idea del espacio $AdS_{2}$.
Esta m\'etrica puede ser conformemente mapeada 
a un disco $D_{2}$ cuyo borde $\partial D_{2}\equiv S^{1}$,
considerando que $S^{1}\equiv \{\infty\}U \{\mathbb{R}\}$.
Entonces, el borde del 
espacio hiperb\'olico es el espacio Euclideo compactificado (un c\'irculo).

 \subsection{Geometr\'ia del espacio-tiempo AdS}
 \label{geoAdS}
 
  El espacio-tiempo AdS$_{4}$ (en cuatro dimensiones) es una soluci\'on particular con simetr\'ia m\'axima a las 
  ecuaciones de Einstein con constante cosmol\'ogica negativa, $\Lambda<0$,
  
  \begin{equation}
  R_{\mu\nu}-\frac{1}{2}g_{\mu\nu}R=-\Lambda g_{\mu\nu}~,
  \end{equation}
  donde la constante cosmol\'ogica esta relacionada con
  el radio-AdS, $l$, mediante $\Lambda=-3/l^{2}$.
  Los tensores de Riemann, de Ricci y la curvatura escalar para el espacio-tiempo AdS$_{4}$, son respectivamente 
  
  \begin{equation}
    R_{\mu\nu\theta\sigma}=-\frac{1}{l^{2}}(g_{\mu\theta}g_{\nu\sigma}-g_{\mu\sigma}g_{\nu\theta})~, \qquad R_{\mu\nu}=-\frac{3}{l^{2}}g_{\mu\nu}~, \qquad R=-\frac{12}{l^2}~.
  \end{equation} 
  
   El espacio-tiempo $AdS_{4}$ puede entenderse como
   un hiperboloide Lorentziano sumergido en un espacio-tiempo $5$-dimensional con dos direcciones tipo tiempo $(X_{0},X_{4})$ 
\begin{equation}
ds^{2}=-dX_{0}^{2}-dX_{4}^{2}+dX_{1}^{2}+dX_{2}^{2}+dX_{3}^{2}~,
\label{AdS5}
\end{equation}
 \begin{equation}
 -X_{0}^{2}-X_{4}^{2}+X_{1}^{2}+X_{2}^{2}+X_{3}^{2}=-l^{2}~.
 \label{paraAdS5}
 \end{equation} 
 
 La ligadura (\ref{paraAdS5}) puede ser parametrizada en t\'erminos de las coordenadas $(\tau,\rho,\theta,\varphi)$ como
 
 \begin{align}
 X_{0}&=l\cosh{\rho}\cos{\tau}~, \\ \notag
 X_{4}&=l\cosh{\rho}\sin{\tau}~, \\ \notag
 X_{1}&=l\sinh{\rho}\sin{\theta}\sin{\varphi}~, \\ \notag
 X_{2}&=l\sinh{\rho}\sin{\theta}\cos{\varphi}~, \\ \notag
 X_{3}&=l\sinh{\rho}\cos{\theta}~. \\ \notag
 \end{align} 
Entonces la m\'etrica (\ref{AdS5}) toma la forma
\begin{equation}
\frac{ds^{2}}{l^{2}}=-\cosh^{2}{\rho}d\tau^{2}+d\rho^{2}+\sinh^{2}{\rho}(d\theta^{2}+\sin^{2}{\theta}d\varphi^{2})~.
\label{AdS1}
\end{equation} 
Estas coordenadas 
cubren todo el hiperboloide si $0\leqslant\tau<2\pi$ y $\rho\geqslant 0$, por consiguiente las coordenadas $(\tau,\rho,\theta,\varphi)$ son conocidas como coordenadas globales de AdS. La m\'etrica cerca $\rho=0$ se comporta como
\begin{equation}
\frac{ds^{2}}{l^{2}}\approx -d\tau^{2}+d\rho^{2}+\rho^{2}
(d\theta^{2}+\sin^{2}{\theta}d\varphi^{2})~,
\end{equation}
cuya topolog\'ia es $S\times\mathbb{R}^{2}$, donde $S^{1}$ son las curvas cerradas tipo tiempo en la direcci\'on $\tau$, las cuales violan causalidad, para evitarlo consideramos el cubrimiento universal, desenvolviendo (unwrap) el circulo $S^{1}$ (esto es, $-\infty <\tau<\infty$), entonces la topolog\'ia es $\mathbb{R}^{3}$. Las coordenadas can\'onicas $(t,r,\theta,\varphi)$ se obtienen con la siguiente transformaci\'on
 
 \begin{equation}
 r:=\sinh{\rho}~, \qquad t:=l\tau~,
 \end{equation}  
 entonces
 
 \begin{equation}
 ds^{2}=\biggl{(}1+\frac{r^{2}}{l^{2}}\biggr{)}dt^{2}+\biggl{(}1+\frac{r^{2}}{l^{2}}\biggr{)}^{-1}dr^{2}
 +r^{2}(d\theta^{2}+\sin^{2}{\theta}d\varphi^{2})~.
 \end{equation}
 La m\'etrica (\ref{AdS1}) en el sistema de coordenadas conformes $(\tau,\chi,\theta,\phi)$ 
 se obtiene considerando $\tan{\chi}:=\sinh{\rho}$ $(0\leq\chi<\pi/2)$
 
 \begin{equation}
 \frac{ds^{2}}{l^{2}}=\frac{1}{\cos^{2}{\chi}}[-d\tau^{2}+d\chi^{2}+\sin^{2}{\chi}(d\theta^{2}+\sin^{2}{\theta}d\varphi^{2})]
 \label{AdSconformal}~.
 \end{equation}
 
 Otro importante sistema de coordenadas es el de Poincar\'e, estas coordenadas cubren s\'olo la mitad del hiperboloide (\ref{paraAdS5}), la parametrizaci\'on es:
 
 \begin{equation}
 X_{0}=\frac{lr}{2}\bl{(}\vec{x}_{i}^{2}-t^{2}+\frac{1}{r^{2}}+1\br{)}~,  \qquad X_{i}=lrx_{i} \qquad (i=1,2)~,
 \end{equation}
 \begin{equation}
\hspace{-2.8cm} X_{3}=\frac{lr}{2}\bl{(}\vec{x}_{i}^{2}-t^{2}+\frac{1}{r^{2}}-1\br{)}~, \qquad X_{4}=lrt~,
 \end{equation}
 donde $(r,t,\vec{x})$, $(0<r, \vec{x}\in\mathbb{R}^{2})$
 
 \begin{equation}
 \frac{ds^{2}}{l^{2}}=r^{2}(-dt^{2}+d\vec{x}^{2})
 +\frac{dr^{2}}{r^{2}}
 \label{poincare1}~.
 \end{equation}
 
\begin{figure}[h!]
	\begin{center}
		\includegraphics[height=2.in]{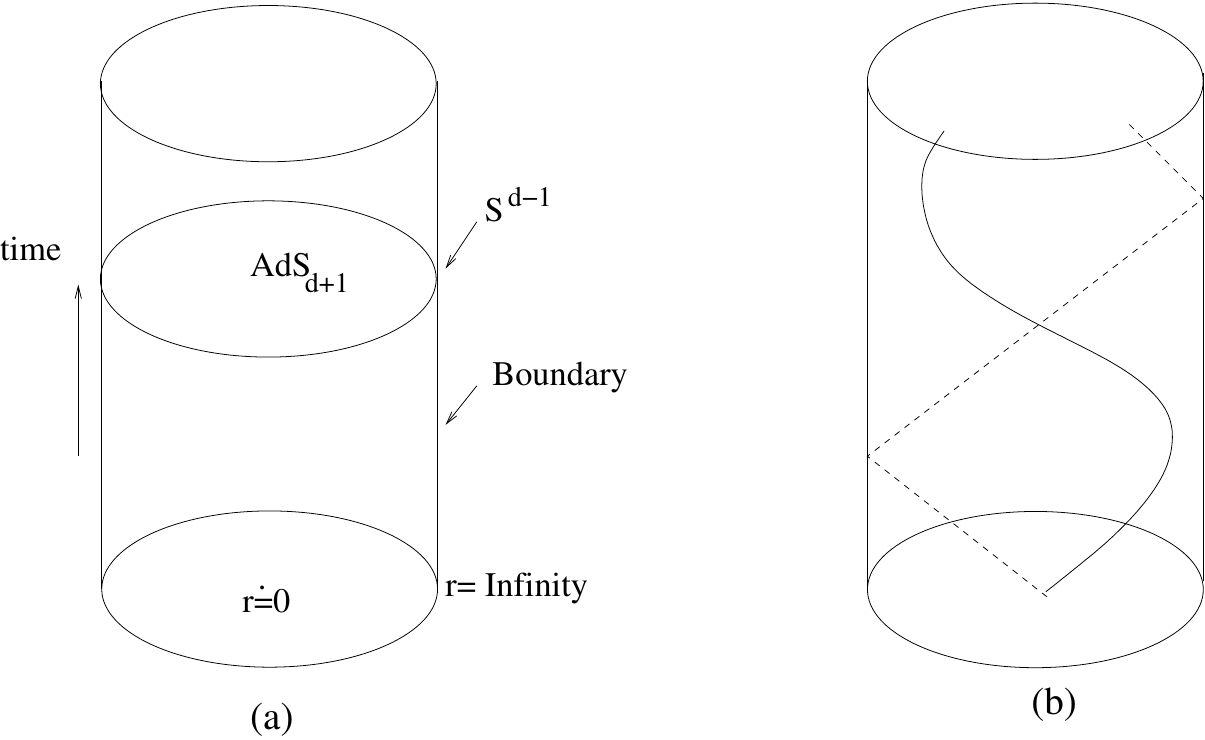}
	\end{center}
	\caption{(a) El diagrama de Penrose para el espacio-tiempo Anti-de-Sitter (\ref{AdSconformal}), es un cilindro s\'olido.
		El borde contiene la direcci\'on temporal y una esfera, $S^{d-1}$, representada como un c\'irculo. (b) Geod\'esica de una part\'icula masiva (linea s\'olida) y la geod\'esica de una part\'icula no masiva (linea punteada). }\label{Penrose}
\end{figure}

 Las coordenadas de Poincar\'e son usualmente escritas
 bajo el siguiente cambio de coordenadas $r=1/z$, de donde
 ahora el borde es $z=\infty$ y el interior es $z=0$,
 
 \begin{align}
 ds^{2}=\frac{l^{2}}{z^{2}}\bl{(}-dt^{2}+d\vec{x}^{2}+dz^{2}\br{)}
 \label{poincare2}~.
 \end{align}
 En este sistema de coordenadas es f\'acil ver que AdS$_{4}$
 es conformal-mente plano.
 Un espacio-tiempo es m\'aximamente sim\'etrico si admite 
 un m\'aximo n\'umero de generadores de simetr\'ia, este
 es el caso del espacio-tiempo AdS$_{4}$. El grupo de isometr\'ia para un espacio-tiempo $D$-dimensional AdS$_{D}$ es $SO(D-1,2)$, cuyo n\'umero de generadores es\footnote{Usamos en esta secci\'on "D" para referirnos a la dimensi\'on del espacio-tiempo AdS, no confundir con el generador de dilataciones $D$} $\frac{D(D+1)}{2}$. Es importante notar que, el borde conforme de AdS$_{D}$ 
es el espacio-tiempo de Minkowski con topolog\'ia
$\mathbb{R}^{1,D-2}$.\\
La conclusi\'on de estas dos secciones es: El n\'umero de generadores del grupo conforme en
un espacio-tiempo de $(D-1)$-dimensiones de Minkowski\footnote{Note que, $D=d+2$, donde $d$ es el n\'umero de dimensiones del espacio-tiempo de Minkowski del borde conforme de AdS.} es $\frac{D(D+1)}{2}$. Y el n\'umero de generadores
de isometr\'ias del espacio-tiempo  AdS$_{D}$
(en $D$-dimensiones) es $\frac{D(D+1)}{2}$.
Entonces el grupo de isometr\'ia de AdS$_{D}$ 
es el grupo conforme del espacio-tiempo de Minkowski 
en una dimensi\'on menor\footnote{En la secci\'on anterior, dejamos en claro que el grupo conforme y el grupo de isometr\'ia son isomorfos. Un enunciado que deja m\'as claro el esp\'iritu de la dualidad AdS/CFT es: El grupo de isometr\'ia (simetr\'ia local) del espacio-tiempo AdS$_{D}$, act\'ua como el grupo de simetr\'ia conforme (simetr\'ia global) del borde (conforme) de AdS$_{D}$ el cual, es el espacio-tiempo de Minkowski en $(D-1)$-dimensiones.}. \\
Otra de las caracter\'isticas del espacio-tiempo AdS es que la constante cosmolo\'gica act\'ua como 
un potencial atractivo, de manera m\'as
precisa, la constante cosmol\'ogica negativa 
da como resultado el diagrama conforme\footnote{Esta figura fue obtenida de \cite{Maldacena:2011ut}.} 
(\ref{Penrose}).

\newpage
Las part\'iculas masivas no llegan a alcanzar el borde en un tiempo finito, pero las part\'iculas sin masa (por ejemplo los fotones) pueden alcanzar el borde y retornar en un tiempo propio finito. Es claro que el espacio-tiempo AdS es como una caja, esto es \'util para explicar el porque un agujero negro en un espacio-tiempo asint\'oticamente AdS es estable del es punto de vista termodin\'amico, contrario a lo que sucede en en espacios-tiempo asint\'oticamente planos. 

\newpage
\section{Dualidad $AdS/CFT$}

En la teor\'ia de cuerdas, los objetos fundamentales son las cuerdas
las cuales pueden ser abiertas o cerradas. El \'unico par\'ametro
libre de la teor\'ia es la tensi\'on $T=\frac{1}{2\pi\alpha^{'}}=\frac{1}{2\pi l_{s}^{2}}$,
cuya dimensi\'on es $[T]=L^{-2}$, donde $l_{s}$ es la 
longitud de la cuerda y es del orden de $10^{-34}m$. En casos concretos, la 
consistencia de la teor\'ia requiere 10 dimensiones.
En estas dimensiones, las distintas oscilaciones de la cuerda 
describen distintas part\'iculas del modelo est\'andar. Es importante
recordar que la gravedad emerge cuando cuantizamos las cuerdas.
Hay dos tipos de cuerdas, abiertas y cerradas.
Las cuerdas abiertas en un espacio-tiempo de $4$ dimensiones 
tienen dos modos de oscilaci\'on que corresponden a dos grados de
libertad, los cuales son las dos polarizaciones de un campo gauge (fot\'on). Podemos concluir que las cuerdas abiertas representan a una teor\'ia gauge. Por otro lado, las oscilaciones de las cuerdas cerradas
representan al gravit\'on y dos part\'iculas escalares a\'un no descubiertas,
el dilat\'on y el axi\'on. Ya que los modos derecho e izquierdo tienen
dos grados de libertad, en total ser\'ian 4 grados de libertad (en 4 dimensiones), dos para el gravit\'on y el resto son los grados de libertad de las dos part\'iculas escalares restantes.

Dos cuerdas abiertas (teor\'ia gauge) pueden unirse y 
formar una nueva cuerda abierta, tambi\'en pueden unirse los
extremos (se requiere una ligadura no local) para formar una cuerda
cerrada (gravit\'on). La conclusi\'on es que la teor\'ia de cuerdas
puede describir de manera unificada las teor\'ias gauge (modelo est\'andar) y la gravitaci\'on. 

En 1997, Juan Maldacena present\'o la conjetura a cerca de 
la dualidad entre dos teor\'ias aparentemente disconexas.
B\'asicamente explica que existe una dualidad entre una teor\'ia
de gravitaci\'on (espacios-tiempo asint\'oticamente AdS) en D-dimensiones y una
teor\'ia cu\'antica conforme ($\mathcal{N}=4$ $SYM$ en 4-dimensiones) en $(D-1)$-dimensiones. Esta es una realizaci\'on concreta del pric\'ipio hologr\'afico, donde la coordenada radial en AdS$_{D}$ es la escala de energ\'ia para la teor\'ia cu\'antica dual en 
$(D-1)$-dimensiones. 

Esta dualidad (correspondencia) nos permite calcular ciertas cantidades en el lado de la gravitaci\'on (cl\'asica) las cuales tienen una correspondiente interpretaci\'on en el lado de la teor\'ia cu\'antica dual. Por ejemplo, la dualidad identifica a los agujeros negros en AdS con los estados t\'ermicos de la teor\'ia cu\'antica dual \cite{Witten:1998qj,Witten:1998zw}.
    
La formulaci\'on precisa de la dualidad AdS/CFT plantea que 
las funciones de partici\'on de la teor\'ia de cuerdas
tipo IIB en AdS$_{5}\times$S$^{5}$ coincide con la 
funci\'on de partici\'on de $\mathcal{N}=4$ super-Yang-Mills en el borde de AdS$_{5}$ \cite{Maldacena:1997re,Gubser:1998bc,Aharony:1999ti} 

\begin{equation}
Z_{gauge}=Z_{string}
\end{equation}

\subsection{L\'imite de \textquoteright t Hooft}

QCD es una teor\'ia gauge basada en el grupo gauge $SU(3)$. Es una teor\'ia asint\'oticamente libre, es decir, las 
constantes de acoplamiento efectivas decrecen a medida que la escala de energ\'ia se incrementa. A bajas energ\'ias
la teor\'ia es fuertemente acoplada y no es f\'acil realizar c\'alculos\footnote{En este r\'egimen de energ\'ia se estudia el fen\'omeno de confinamiento.}, de hecho una posible aproximaci\'on es usar simulaciones num\'ericas en el lattice. 
Sea la teor\'ia de Yang-Mills (puro) definida por la siguiente funci\'on de Lagrange, con el grupo gauge $SU(N)$, con $N$-colores
\begin{equation}
	\mathcal{L}=-\frac{N}{2g_{YM}^{2}}F_{\mu\nu}^{M}F_{M}^{\mu\nu}~.
\end{equation} 
Donde $F_{\mu\nu}$ es la fuerza del campo para el grupo $SU(N)$, cuyos generadores $T^{a}$. Para un valor finito de $N$ no hay un cambio esencial, pero en el l\'imite $N\rightarrow\infty$ manteniendo $g_{YM}^{2}$ fijo ocurren simplificaciones importantes. En el l\'imite de \textquoteright t Hooft el n\'umero de colores $N\rightarrow\infty$ es muy grande y el acoplamiento $g_{YM}^{2}\rightarrow 0$ es muy peque\~no
tal que el par\'ametro $\lambda=Ng_{YM}^{2}$ (conocido como acoplamiento de \textquoteright t Hooft) es fijo. Note que en las teor\'ias gauge $SU(N)$, el 
acoplamiento de \textquoteright t Hooft $\lambda=g_{YM}^{2}N$ es el que controla la expansi\'on perturbativa y no el
acoplamiento de Yang-Mills, $g_{YM}^{2}$.\\
\textquoteright t Hooft esperaba que se pudiese 
resolver ex\'actamente la teor\'ia con $N=\infty$,
y entonces un podr\'ia hacer una expansi\'on en $1/N=1/3$. La expansi\'on diagram\'atica de la teor\'ia del campo sugiere que la teor\'ia \textit{N-grande}
es una teor\'ia de cuerdas libre y que 
el acoplamiento de la cuerda es $g_{s}\sim 1/N$.
\newpage
\subsection{Prescripci\'on de Gubser-Klevanov-Polyakov y Witten}

En la presente tesis, nos concentramos 
en el r\'egimen de acoplamiento d\'ebil de 
la teor\'ia de cuerdas (gravedad cl\'asica), la que es dual al r\'egimen de acoplamiento fuerte de la teor\'ia gauge.\\

En el lado de la CFT, el r\'egimen de acoplamiento fuerte
de la teor\'ia $\mathcal{N}=4$ SYM,
esta controlada por el acoplamiento de \textquoteright t Hooft dada por $\lambda=g_{YM}^{2}N\sim g_{s}N>>1$ y $N\rightarrow\infty$, donde $g_{s}$
	es el acoplamiento de la cuerda y se relaciona con 
	el acoplamiento de Yang-Mills como
	$g_{YM}^{2}=2\pi g_{s}$. La funci\'on de partici\'on de la CFT
 esta dada por la funcional generatriz $W$ para la funci\'on de Green en la teor\'ia gauge\footnote{Trabajamos en la secci\'on Euclidea, donde $\tau=-it$}

\begin{equation}
Z_{CFT}=e^{-W}~.
\end{equation}

En la teor\'ia de cuerdas tipo IIB en 
$AdS_{5}\times S^{5}$, cuando el radio de curvatura de $AdS_{5}$ es m\'as grande en comparaci\'on a la longitud
de la cuerda $l>>l_{s}$ el r\'egimen de 
la supergravedad es v\'alida, concr\'etamente
\begin{equation}
\frac{l^{4}}{l_{s}^{4}}\sim g_{YM}^{2}N\sim g_{s}N>>1~,
\end{equation}  
de donde $g_{s}<1$ implica el l\'imite $N\rightarrow\infty$. Entonces la funci\'on
de partici\'on de la teor\'ia de cuerdas se aproxima
por la funci\'on de partici\'on 
\begin{equation}
Z_{cuerdas}\approx e^{-I^{E}_{SUGRA}}~,
\end{equation} 
donde, $I_{SUGRA}^{E}$ es la acci\'on cl\'asica de supergravedad
(en la aproximaci\'on de punto-silla y en la secci\'on Euclidea) evaluada en $AdS_{5}\times S^{5}$ (o en peque\~nas deformaciones de este espacio). Entonces,

\begin{equation}
Z_{cuerdas}\approx e^{-I_{SUGRA}^{E}}=e^{-W}=Z_{CFT}~.
\label{GKPW1}
\end{equation}

La ecuaci\'on (\ref{GKPW1}) es una f\'ormula concreta
para calcular observables en la teor\'ia gauge
a partir de c\'alculos geom\'etricos en el lado de 
supergravedad (gravedad cl\'asica). Esta fu\'e
propuesta en 1998 por Gubser-Klebanov-Polyakov e
independientemente por Witten (GKPW) \cite{Gubser:1998bc, Witten:1998qj}. A temperatura finita, 
$W$ esta relacionada con la energ\'ia libre de la 
teor\'ia gauge por $W=F/T$, donde $T$ es la temperatura. En el lado izquierdo de (\ref{GKPW1}), aplicando
por ejemplo al agujero negro de Schwarzschild-AdS, la 
acci\'on $I_{SUGRA}^{E}$ se calcula on-shell en las ecuaciones de movimiento. Entonces, dada la igualdad
(\ref{GKPW1}), se concluye que 
el agujero negro de Schwarzschild-AdS es dual a los estados t\'ermicos de la teor\'ia gauge (fuertemente acoplado) \cite{Witten:1998qj,Witten:1998zw}.\\ 

\subsection*{Correspondencia Campo $\leftrightarrow$ Operador}

Es bien sabido que el cambio 
de la constante de acoplamiento de
la teor\'ia gauge corresponde
a un cambio en el valor del borde
del dilat\'on \cite{Aharony:1999ti}. Entonces
(en un espacio-tiempo de $4$-dimensiones en el bulk) 
se tiene\footnote{La expresi\'on: $\phi\vert_{\partial AdS}$, se refiere al valor que toma el campo $\phi$ en el borde de AdS.}
%
\begin{equation}
\bl{\langle}e^{\int{d^{3}x}\phi_{0}(\vec{x})\mathcal{O}(\vec{x})}\br{\rangle}_{CFT}=e^{-I^{E}_{bulk}[\phi\vert_{\partial AdS}\rightarrow \phi_{0}]}~,
\end{equation}
donde $\phi(r,\vec{x})\vert_{r=\infty}=\phi_{0}(\vec{x})$, en el lado gravitacional son los valores del dilat\'on en 
el borde\footnote{$\phi(r,\vec{x})$ es el dilat\'on que se propaga en el bulk con la condici\'on de borde de que el campo $\phi$ en el borde tiene el valor de $\phi_{0}$. Aqu\'i consideramos las coordenadas en la secci\'on Euclidea $\vec{x}=(\tau,\theta,\phi)$ y $r$ la coordenada radial, donde el borde se encuentra en $r=\infty$.}. En el lado de la teor\'ia gauge, $\phi_{0}$ corresponde a la fuente del operador local $\mathcal{O}(\vec{x})$. Entonces 
las condiciones de borde (en el lado gravitacional), fijan los valores del dilat\'on $\phi_{0}$ (en el borde) y estas corresponden a operadores en la teor\'ia gauge.
De all\'i que, si consideramos perturbaciones en la m\'etrica (el gravit\'on) en el borde esta viene a ser la fuente del operador tensor de energ\'ia-momento
de la teor\'ia gauge\footnote{O, tambi\'en, que el gravit\'on se acopla al operador tensorial de energ\'ia-momento de la teor\'ia gauge. Los \'indices se 
	descomponen como $\mu=(a,r)$.}.	
\begin{equation}
\bl{\langle}e^{\int{d^{3}x}~h_{ab}^{0}T^{ab}}\br{\rangle}_{CFT}=e^{-I^{E}_{bulk}[h_{\mu\nu}\vert_{\partial AdS}\rightarrow h_{ab}^{0}]}~.
\end{equation}

\subsection{Part\'iculas y campos en el espacio-tiempo AdS}
\label{defor233}
De acuerdo a la prescripci\'on de GKPW, el campo
escalar en el borde es la fuente en la teor\'ia cu\'antica dual de un operador local $\mathcal{O}$, por lo que es importante 
estudiar la din\'amica de campos escalares y sus condiciones de borde en el espacio-tiempo AdS. En especial, nos concentramos en un campo escalar masivo en un espacio-tiempo AdS de $4$-dimensiones acoplado minimamente a la gravedad y sin potencial.\\
Sea la ecuaci\'on de Klein-Gordon en las coordenadas
de Poincar\'e (\ref{poincare2}), donde consideramos la
descomposici\'on de los \'indices $(t,\vec{x},z)=(x^{n},z)$,
donde el borde es $z=0$
\begin{equation}
(\nabla^{\mu}\nabla_{\mu}-m^{2})\Phi(z,x^{n})=0~,
\end{equation}
descomponiendo $\Phi(x^{\mu},z)$ en sus modos de Fourier
\begin{equation}
\Phi(x^{n},z)=\int{\frac{ d^{2}\vec{k}}{(2\pi)^{2}}}d\om f_{k}(z)e^{ik_{\mu}x^{\mu}}~,
\end{equation}
la ecuaci\'on de Klein-Gordon queda de la siguiente forma
\begin{equation}
\frac{d^{2}f_{k}}{dz^{2}}-\frac{2}{z}\frac{df_{k}}{dz}-(k^{2}+\frac{m^{2}l^{2}}{z^{2}})f_{k}=0~.
\label{AdSklein}
\end{equation}
La soluci\'on general es
\begin{equation}
f_{k}(z)=a_{1}z^{3/2}K_{\nu}(kz)+a_{2}z^{3/2}I_{\nu}(kz)~,
\end{equation}
donde $I_{\nu}(kz)$ y $K_{\nu}(kz)$ son
las funciones de Bessel de primer y
segundo tipo. El comportamiento de la ecuaci\'on (\ref{AdSklein}) en el interior
del espacio AdS ($z\rightarrow\infty$)
\begin{equation}
\frac{d^{2}f_{k}}{dz^{2}}-k^{2}f_{k}=0~,
\end{equation}
mientras las soluciones se comportan como:  $K_{\nu}(kz)\sim e^{-kz}$ 
and $I_{\nu}(kz)\sim e^{kz}$. El campo escalar
debe ser regular en el interior de AdS, por lo que
 $a_{2}=0$. Entonces la soluci\'on es: 
\begin{equation}
f_{k}(z)=a_{1}z^{3/2}K_{\nu}(kz)~, \qquad K_{\nu}(kz)=\frac{\pi}{2}\frac{I_{-\nu}(kz)-
	I_{\nu}(kz)}{\sin(\nu\pi)}~.
\end{equation}
La soluci\'on $f_{k}(z)$ cerca del borde $z\rightarrow 0$ es
\begin{align}
f_{k}(z)&=a_{1}z^{3/2}\frac{\pi}{2\sin{\pi\nu}}\bl{[}\frac{1}{\Gamma(1-\nu)}\bl{(}\frac{kz}{2}\br{)}^{-\nu}-\frac{1}{\Gamma(1+\nu)}\bl{(}\frac{kz}{2}\br{)}^{\nu}\br{]}~, \\ \non
f_{k}(z)&=\phi_{0}z^{\Delta _{-}}+\phi_{1}z^{\Delta_{+}}~, \\ \non
\end{align}
donde $\phi_{0}=a_{1}2^{-1+\nu}k^{-\nu}\Gamma(\nu)$, 
$\phi_{1}=a_{1}2^{-1-\nu}k^{\nu}\Gamma(-\nu)$ y $\Delta_{\pm}=3/2\pm\nu$. Sustituyendo en (\ref{AdSklein})
obtenemos informaci\'on adicional para el \'indice
$\nu$ 
\begin{equation}
\Delta_{\pm}(\Delta_{\pm}-3)-ml^{2}=z^{2}k^2\xrightarrow{z=0}
\Delta_{\pm}(\Delta_{\pm}-3)-ml^{2}=0~,
\end{equation}
cuyas soluciones son
\begin{equation}
\Delta_{\pm}=\frac{3}{2}\pm\nu~, \qquad \nu=\sqrt{\frac{9}{4}+m^{2}l^{2}}~.
\end{equation}
Entonces, la dimensi\'on conforme $\Delta$ del campo
escalar depende de su masa $m$. 
La condici\'on de Breitenlohner-Freedman (BF) $m^{2}\geq -\frac{9}{4l^{2}}$, asegura que los exponentes sean reales ya que exponentes complejos son signos
de inestabilidad lineal. Sea la masa BF 
 $m_{BF}^{2}=-\frac{9}{4l^{2}}$ que define una cota inferior,
el l\'imite superior para la masa se le conoce como l\'imite unitario
\begin{equation}
m_{BF}^{2}\leq m^{2}<m_{BF}^{2}+\frac{1}{l^{2}} \Rightarrow 0\leq\nu<1~.
\end{equation} 
En adelante trabajamos en las coordenas can\'onicas $(t,r,\Sigma_{k})=(r,x^{n})$, donde, $\Sigma_{k}$ es la secci\'on 
transversal, la que puede ser planar $k=0$, esf\'erica $k=1$
e hiperb\'olica $k=-1$. Entonces el fall-off del campo escalar
es 

\begin{equation}
\phi(r,x^{n})=\frac{\alpha(x^{n})}{r^{3-\Delta}}+\frac{\beta(x^{n})}{r^{\Delta}}+... ~,
\end{equation}  

\begin{equation}
\Delta=\frac{3}{2}+\nu~, \qquad \nu=\sqrt{\frac{9}{4}+m^{2}l^{2}}~,
\end{equation}

cuando el cuadrado de la masa del campo escalar se encuentra
dentro de las cotas unitaria y la de BF (conocida como \textit{la ventana de BF}), ambos 
modos $\alpha/r^{3-\Delta}$ y $\beta/r^{\Delta}$ son normalizables\footnote{Definiendo un producto
	interior $(\phi_{1},\phi_{2})$, en una foliaci\'on tipo espacio, la norma $(\phi,\phi)$ es convergente cuando $0\leq\nu<1$.}.
Entonces las posibles condiciones de borde son:
\begin{itemize}
	\item Condici\'on de borde de Dirichlet: $\alpha=0$, $\beta\neq 0$~.\\
	\vspace{-5ex}
	\item Condici\'on de borde de Neumann: $\alpha\neq 0$, $\beta=0$~.\\
	\vspace{-5ex}
	\item Condici\'on de borde mixta: $\alpha\neq0$, $\beta\neq0$~.\\
\end{itemize}

Veamos un ejemplo concreto para un
campo escalar de masa $m^{2}=-2/l^{2}$
en cuatro dimensiones, cuyas potencias relevantes
del fall-off son: $\nu=1/2$, $\Delta=2$, $3-\Delta=1$ 

\begin{equation}
\phi=\frac{\alpha}{r}+\frac{\beta}{r^{2}}+...
\end{equation}

{\bf Dirichlet:} El modo $\beta/r^{2}$ cae m\'as 
r\'apidamente que $\alpha/r$ y es en general normalizable
y puede ser cuantizado, entonces $\alpha=0$ corresponde
a $\phi$ fijo en el borde, es decir $\phi(x^{m},r=\epsilon)=\epsilon^{-2}\phi_{0}$.\\

{\bf Neumann:} En es este caso $\pa_{r}\phi$ es fijo 
en el borde, veamos:

\begin{equation}
\pa_{r}\phi=-\frac{\alpha}{r^{2}}-\frac{2\beta}{r^{3}}+...=\frac{\alpha^{'}}{r^{2}}+...
\end{equation}

note que $r^{-3}$ es un modo irrelevante y esto equivale a 
fijar $\beta=0$.

{\bf Mixto:} Finalmente, este caso es cuando $a\phi+b\pa_{r}\phi$ fijo en el borde, veamos:

\begin{equation}
a\phi+b\pa_{r}\phi=a\bl{(}\frac{\alpha}{r}+\frac{\beta}{r^{2}}\br{)}+b\bl{(}-\frac{\alpha}{r^{2}}-\frac{2\beta}{r^{3}}\bl{)}=
\frac{\alpha^{'}}{r}+\frac{\beta^{'}}{r^{2}}+...
\end{equation}

lo que equivale a considerar $\alpha\neq 0$ y $\beta\neq 0$.\\

A partir de la prescripci\'on de GKPW, se verifica que la 
dimensi\'on conforme del operador $\mathcal{O}$ es $\Delta$.  

\subsection*{Deformaciones en AdS/CFT}
Uno de los motivos m\'as importantes para estudiar teor\'ias
cu\'anticas conformes es comprender algunos de los aspectos
poco entendidos de teor\'ias del campo no conformes como QCD. La pregunta es, cuanto podemos aprender de las
teor\'ias no conformes (QCD) de las teor\'ia conformes. 
Una de las formas de romper la invarianza conforme es estudiar
la teor\'ia a temperatura finita\footnote{De hecho, CFT a temperatura finita es muy pr\'oxima a QCD a temperatura finita (a altas temperaturas).}. Pero tambi\'en es posible 
romper la invarianza conforme y al mismo tiempo preservar la simetr\'ia 
de Lorentz, deformando la acci\'on mediante operadores locales,

\begin{equation}
I_{CFT}\rightarrow I_{CFT}+p\int{d^{3}x\mathcal{O}(x)}~,
\end{equation}
para alg\'un operador escalar de Lorentz y alg\'un coeficiente $p$.
Las diferentes deformaciones dependen de la dimensi\'on de escala 
$\Delta$ del operador $\mathcal{O}$.\\
\vspace{-2ex}
\begin{itemize}
	\item Si, $\Delta-3<0$, la deformaci\'on es conocida como $relevante$, en la que
	la deformaci\'on es fuerte en el r\'egimen $IR$ ($r\rightarrow 0$) y 
	d\'ebil en el r\'egimen $UV$.
	\item Si, $\Delta-3>0$, la deformaci\'on es llamada $irrelevante$, y su 
	efecto es fuerte en el r\'egimen $UV$. 
	No discutiremos sobre las deformaciones 
	irrelevantes, ya que divergen cerca del borde $r=\infty$.
	\item El \'ultimo caso son las $deformaciones$ 
	$marginales$, donde $\Delta=3$ el cual no rompe
	la invarianza conforme en ordenes relevantes a la deformaci\'on.
\end{itemize}


%% file: capitulo3.tex
                    \chapter{Agujeros negros}

La velocidad de escape de un objeto celeste (estrellas, 
planetas, etc) puede calcularse usando mec\'anica Newtoniana y esta dada por $v=\sqrt{2GM/r}$. Esta es la m\'inima velocidad necesaria para escapar
al campo gravitacional de objeto celeste (con simetr\'ia esf\'erica) de masa $M$ y de radio $r$~\footnote{Por ejemplo las velocidades de escape de la tierra y del sol son: $v_{tierra}=11,2$ km/s y $v_{sol}=617.5$ km/s, donde los radios de la tierra y del sol son: $r_{tierra}=6~371$ km y $r_{sol}=695~700$ km.}. Un agujero negro es un objeto celeste cuya velocidad de escape es la velocidad de la luz $c$ y ya que no es posible alcanzar velocidades mayores a esta velocidad entonces nada puede escapar de su atracci\'on gravitacional\footnote{Jhon Michell, 1783.}. La teor\'ia Newtoniana es muy limitada para describir correctamente campos gravitacionales muy intensos, en esos casos la Relatividad General es la m\'as adecuada.

En esta secci\'on describimos varias de las soluciones de agujeros negros y la manera de evadir el teorema de no pelo, para obtener soluciones de agujeros negros con pelo escalar. Mostramos detalles de como obtener una familia de soluciones de agujeros negros con pelo escalar y de secci\'on transversal plana. Los detalles del c\'alculo del potencial se encuentran en el ap\'endice \ref{calculopotencial}. 
Es interesante, que esta familia de soluciones, para algunos de los valores del par\'ametro pelo $\nu$, se reducen a soluciones particulares encontradas en la literatura.
Finalmente, en el l\'imite de la constante cosmol\'ogica, $1/l^{2}=0$ en la teor\'ia y en las soluciones, se obtiene de forma consistente soluciones asint\'oticamente planas, y nuevamente este potencial escalar evade el teorema de no pelo. 
%
\section{Principio variacional}
Estudiamos el principio
variacional de la acci\'on de Hilbert-Einstein,
la que describe gravedad pura (sin materia) y sin
constante cosmol\'ogica.
Como ya vimos en el cap\'itulo anterior, el c\'alculo de la acci\'on on-shell en la secci\'on Euclidea (en la aproximaci\'on del punto silla), nos da la energ\'ia libre $F=\beta I_{on-shell}^{E}$\footnote{Donde $F$ puede ser la energ\'ia libre de Helmholtz, Gibbs o el Gran potencial, dependiendo del ensamble termodin\'amico.} con la que podemos calcular las cantidades termodin\'amicas y estudiar las posibles fases t\'ermicas de los agujeros negros. Las ecuaciones de movimiento son obtenidas variando la acci\'on, donde los t\'erminos de borde no tienen contribuci\'on a las ecuaciones, pero son importantes para obtener las cantidades conservadas y la energ\'ia libre.\\ 
El principio variacional esta bien definido
si $\delta I=0$ en todo el espacio-tiempo.
La acci\'on de Hilbert-Einstein con un t\'ermino de borde es
\begin{equation}
I=\frac{1}{2\kappa}\int{d^{4}x\sqrt{-g}R}+I_{B}~.
\end{equation}
Variando la acci\'on respecto a la m\'etrica, 
se obtiene
\begin{equation}
\delta I=\frac{1}{2\kappa}\int{d^{4}x\sqrt{-g}G_{\alpha\beta}\delta g^{\alpha\beta}}+\int{d^{4}x\sqrt{-g} g^{\alpha\beta}\delta R_{\alpha\beta}}+\delta I_{B}~,
\end{equation}
donde $G_{\alpha\beta}$ es el tensor de Einstein
\begin{equation}
G_{\alpha\beta}=R_{\alpha\beta}-\frac{1}{2}g_{\alpha\beta}R
\end{equation}
y el t\'ermino $\delta I_{B}$ debe ser elegido de tal forma que $\delta I=0$, esto es
\begin{equation}
\delta I_{B}=-\int_{\mathcal{M}}{d^{4}x\sqrt{-g} g^{\alpha\beta}\delta R_{\alpha\beta}}=-\oint_{\pa\mathcal{M}}{\epsilon v^{\mu}n_{\nu}\sqrt{-h}d^{3}x}~,
\end{equation}
donde
\begin{equation}
g^{\alpha\beta}\delta R_{\alpha\beta}=v^{\mu}_{;\mu}~, \qquad v^{\mu}=g^{\alpha\beta}\delta\Gamma^{\mu}_{\alpha\beta}-g^{\alpha\mu}\delta\Gamma_{\alpha\beta}^{\beta}~, \qquad \epsilon=n^{\mu}n_{\mu}=\pm 1~.
\end{equation}
El vector unitario $n_{\mu}$ es la normal 
en el borde y es un vector tipo tiempo ($\epsilon=-1$)
o tipo espacio ($\epsilon=1$), dependiendo si el borde
es una variedad de tipo espacio o tiempo.
Se puede mostrar que 
\begin{equation}
I_{B}=\int_{\pa\mathcal{M}}{d^{3}x\sqrt{-h}} K~.
\end{equation}
Donde la curvatura extr\'inseca $K_{\alpha\beta}=\nabla_{(\alpha}n_{\beta)}$, la m\'etrica inducida $h^{\alpha\beta}=g^{\alpha\beta}+\epsilon n^{\alpha}n^{\beta}$ y la traza  $K=h^{\alpha\beta}K_{\alpha\beta}$.
Este es el t\'ermino de borde de Gibbons-Hawking 
que da un principio variacional bien definido $\delta I=0$~\footnote{En las condiciones de borde de Dirichlet, en la que $\delta h=0$.}. Aqu\'i $\kappa=8\pi G$ con $G$ la constante gravitacional de Newton, $h$ es el determinante de la m\'etrica del borde. Entonces la acci\'on con la contribuci\'on de Gibbons-Hawking es
\begin{equation}
I=\frac{1}{2\kappa}\int_{\mathcal{M}}{d^{4}x\sqrt{-g}R}+\frac{1}{\kappa}\int_{\pa\mathcal{M}}{d^{3}x\sqrt{-h}}K~.
\end{equation} 
Este t\'ermino de borde tiene contribuci\'on 
a la acci\'on on-shell. En especial, el c\'alculo
de la energ\'ia libre para el agujero negro Schwarzschild
asint\'oticamente plano en el que el t\'ermino
de $bulk$ (referida a la integral en el volumen del espacio-tiempo) desaparece ya que $R=0$~\footnote{La ecuaciones para la m\'etrica son $R_{\alpha\beta}=0$, de donde $R=0$.}. La \'unica contribuci\'on a la acci\'on (energ\'ia libre) esta dada por el t\'ermino de borde Gibbons-Hawking. En general
esta es IR divergente. En las secciones posteriores mostramos
m\'etodos concretos para eliminar estas divergencias y obtener la acci\'on finita. Para m\'as detalles, ver la el ap\'endice \ref{apendice1}.  

\section{Agujeros negros asint\'oticamente planos, $\Lambda=0$}
La soluci\'on de las ecuaciones de Einstein (sin constante consmol\'ogica) de un
objeto est\'atico de masa $M$ y con simetr\'ia esf\'erica 
es el \textit{agujero negro de Schwarzschild} 
descubierto en 1916 por Karl Schwazschild. Esta soluci\'on
esta caracterizada por su masa y tiene un horizonte de sucesos que esconde la singularidad.
Si consideramos un campo electromagn\'etico m\'inimamente acoplado a la
gravedad, obtenemos la soluci\'on de Reissner-Nordstr\"om, este es una agujero negro est\'atico y con carga el\'ectrica.
El ansatz m\'as general con simetr\'ia esf\'erica y est\'atico, para ambas soluciones, es:
\begin{equation}
ds^{2}=-N(r)dt^{2}+H(r)dr^{2}+r^{2}(d\theta^{2}+\sin^{2}{\theta}d\varphi^{2})~.
\label{anzats1}
\end{equation} 
En estas coordenadas el sector asint\'otico corresponde a 
$r=cte\rightarrow\infty$.
Las soluciones asint\'oticamente planas tienen las siguientes propiedades :
\begin{itemize}
  \item La m\'etrica se aproxima a la m\'etrica de Minkowski:\\
  \vspace{1.9ex}
         $ds^{2}\sim -dt^{2}+r^{2}(d\theta^{2}+\sin^{2}{\theta}d\varphi^{2})$ en la regi\'on $r=cte\rightarrow\infty$~.
  \vspace{-2ex}
  \item El horizonte se obtiene de $N(r_{h})=0$~ y esta localizado en $r=r_{h}$~\footnote{Es importante recalcar que para el caso de Reissner-Nordstr\"om hay dos soluciones para $N(r_{h})=0$, por lo que hay dos horizontes, $r_{-}$ y $r_{+}$, tal que $r_{-}<r_{+}$.}. \\ 
  \vspace{-3.5ex}
  \item La \'unica topolog\'ia permitida por las ecuaciones de Einstein, para la secci\'on transversal, es la esf\'erica: 
        $d\Sigma_{k=1}^{2}=d\theta^{2}+\sin^{2}{\theta}d\varphi^{2}$~.\\
\vspace{-4.5ex}
  \item El invariante de Kretschmann:
\end{itemize}
\vspace{-2ex}
\begin{equation}
Krets=R^{\alpha\beta\gamma\sigma}R_{\alpha\beta\gamma\sigma}~.
\end{equation}   
Evaluada para la soluci\'on de Schwarzschild es $Kret\sim 1/r^{6}$ y para
Reissner-Nordstr\"om es $Krets\sim 1/r^{8}$. Eso nos dice que la singularidad
del espacio-tiempo se encuentra en $r=0$, regi\'on donde la gravedad es infinitamente fuerte.
\subsection*{Agujero negro de Schwarzschild}
Sea la acci\'on 
\begin{equation}
I[g_{\mu\nu}] = \frac{1}{2\kappa}\int_{\mathcal{M}}{d^{4}x\sqrt{-g}R} + \frac{1}%
{\kappa}\int_{\partial\mathcal{M}}{d^{3}xK\sqrt{-h}}~, \label{action3}%
\end{equation}
donde $\kappa=8\pi G$, $G$ es la constante de Newton y el t\'ermino de borde asegura que el  
principio variacional se satisface $\delta I=0$. 
Ya que no hay materia $T_{\mu\nu}=0$, entonces 
las ecuaciones para la m\'etrica son
\begin{equation}
R_{\mu\nu}=0~.
\end{equation}
Dado el ansatz (\ref{anzats1}), la soluci\'on es
%
\begin{equation}
ds^{2}=-\biggl{(}1-\frac{\mu}{r}\biggr{)}dt^{2}+\biggl{(}1-\frac{\mu}{r}\biggr{)}^{-1}dr^{2}+r^{2}(d\theta^{2}+\sin^{2}{\theta}d\phi^{2})~,
\label{schwarzshildflat}
\end{equation} 
donde el horizonte es tal que: $-g_{tt}=0\Rightarrow r_{h}=\mu$. La masa AMD o energ\'ia del agujero negro es~\footnote{Mas adelante calcularemos esta masa mediante tres diferentes formas para un caso mucho m\'as general y veremos que en el caso especial de Schwarzschild la masa es j\'ustamente $M=\mu/2G$.}
\begin{equation}
M=\frac{4\pi\mu}{\kappa}=\frac{\mu}{2G}~.
\end{equation}
Es f\'acil ver que en la regi\'on cerca del borde, $r\rightarrow\infty$, esta soluci\'on se aproxima a la m\'etrica de Minkowski~.

\subsection*{Agujero negro de Reissner-Nordstr\"om}
La acci\'on del campo gravitacional acoplado 
m\'inimamente a un campo gauge $U(1)$, es\footnote{En esta secci\'on usaremos algunas expresiones usando las formas diferenciales. Para una breve revisi\'on de las formas diferenciales, ver el ap\'endice \ref{Formas11}}
\begin{equation}
I[g_{\mu\nu},A_{\mu}] = \frac{1}{2 \kappa} \int_{\mathcal{M}}{d^4x \sqrt{-g} \left(R-\frac{1}{4}F_{\mu\nu}F^{\mu\nu}\right)}+ \frac{1}%
{\kappa}\int_{\partial\mathcal{M}}{d^{3}xK\sqrt{-h}}~, \label{action33}%
\end{equation}
donde $F_{\mu\nu}\equiv\partial_\mu A_\nu-\partial_\nu A_\mu$ 
es el tensor de Maxwell\footnote{El tensor de Maxwell satisface la identidad de Bianchi $F_{\mu\nu;\alpha}=0$}, 
$A_\mu$ es el potencial gauge. 
Las ecuaciones de movimiento para la m\'etrica 
y el campo gauge son, respectivamente,
\begin{equation}
R_{\mu\nu}-\frac{1}{2} R g_{\mu\nu}=\frac{1}{2}T^{EM}_{\mu\nu}~,
\label{RNeqmotion}
\end{equation}
\begin{equation}
\nabla_{\mu}F^{\mu\nu}=0~,
\label{RNgauge}
\end{equation}
donde el tensor de stress para el campo gauge es
\begin{equation}
T^{EM}_{\mu\nu}=F_{\mu\alpha}F^{\cdot\alpha}_{\nu}-\frac{1}{4}g_{\mu\nu}F^{2}~.
\end{equation}
Las ecuaciones para el campo gauge (\ref{RNgauge}) se satisfacen considerando que
\begin{equation}
A\equiv A_{\mu}dx^{\mu}=\left(\frac{q}{r}-\frac{q}{r_+}\right) dt~, \qquad F=-\frac{q}{r^2}~dr\wedge dt~.
\label{RNgaugepot}
\end{equation}
Las condiciones de borde para el campo gauge son tales que
es cero en el horizonte de sucesos del agujero negro, $A(r_{+})=0$ y constante en el borde $A(\infty)=-q/r_{+}$.
El horizonte $r_{+}$ es la ra\'iz mayor de la ecuaci\'on $-g_{tt}=0$~\footnote{Las soluciones para $-g_{tt}(r_{h})=0$, son $r_{h}=r_{\pm}$, es decir $r_{-}$ y $r_{+}$, donde $r_{-}\leq r_{+}=r_{h}$, se tiene dos horizontes.}.\\
Considerando el anzats (\ref{anzats1}), la distribuci\'on de materia es tal que $N(r)=H(r)^{-1}=f(r)$ y
la soluci\'on para la m\'etrica es
\begin{equation}
ds^{2}=-f(r)dt^{2}+f(r)^{-1}dr^{2}+r^{2}(d\theta^{2}+\sin^{2}{\theta}d\varphi^{2})~,
\end{equation}
\begin{equation}
f(r)=1-\frac{\mu}{r}+\frac{q^2}{4r^2}=\frac{(r-r_{-})(r-r_{+})}{r^{2}}~.
\label{RNf}
\end{equation}
donde $r_{\pm}=\frac{1}{2}(\mu\pm\sqrt{\mu^{2}-q^{2}})$
con $\mu^{2}\geq q^{2}$.
Este agujero negro esta caracterizado por su masa $M$ y carga $Q$. La masa AMD puede leerse f\'acilmente de $-g_{tt}$
\begin{equation}
M=\frac{4\pi\mu}{\kappa}=\frac{\mu}{2G}~,
\end{equation}
y la carga el\'ectrica de la ley Gauss es el
flujo del campo electromagn\'etico en la
secci\'on transversal en el borde, con elemento
de superficie $d^{2}\Sigma_{\mu\nu}$:
\begin{equation}
Q\equiv\frac{1}{\kappa}\oint{d^{2}\star F}=-\frac{q}{4 G}~.
\label{RNcharge}
\end{equation}
La ecuaci\'on para el horizonte ($-g_{tt}=0$),
expresada en t\'erminos de $(M,Q)$ tiene
dos ra\'ices $(r_{-},r_{+})$, $r_{-}\leq r_{+}=r_{h}$
\begin{equation}
r_{\pm}=G (M\pm\sqrt{M^{2}-4Q^{2}})~.
\end{equation}
El horizonte de sucesos existe si $M\geq 2\vert Q\vert$, por lo que hay una cantidad m\'axima de carga que el 
agujero negro puede tener. A diferencia de los agujeros negros Schwarzschild, estos agujeros
negros cargados pueden llegar a tener temperatura cero cuando $r_{-}=r_{+}$ y al mismo tiempo tener un horizonte finito\footnote{O tambi\'en $M=2\vert Q\vert$}. \\  
El potencial qu\'imico $\Phi$ se define como la energ\'ia $\Delta E$
necesaria para traer una carga $\Delta Q$ de la regi\'on asint\'otica
$r=\infty$ al horizonte del agujero negro
\begin{equation}
\Phi=A_{t}\vert_{r=\infty}-A_{t}\vert_{r=r_{+}}=\frac{4GQ}{r^{+}}~.
\end{equation} 
Es interesante mencionar que los agujeros negros de RN se 
generalizan a p-branas en mayores dimensiones \cite{Aharony:1999ti,charged50,charged55}. 
%

\section{Agujeros negros asint\'oticamente anti-de Sitter}
Los agujeros negros asint\'oticamente anti-de Sitter son aquellas
soluciones a las ecuaciones de Einstein con constante
cosmol\'ogica negativa $\Lambda=-3/l^{2}$. Nuevamente consideramos un ansatz radialmente 
sim\'etrico y est\'atico.
Estas soluciones son tales que el borde es
\begin{equation}
ds^{2}=\frac{r^{2}}{l^{2}}\biggl{(}-dt^{2}+l^{2}d\Sigma_{k}^{2}\biggr{)}~.
\label{bordeAdS}
\end{equation}
A diferencia de las 
soluciones asimt\'oticamente planas, la constante cosmol\'ogica
permite soluciones con secciones transversales planos, esf\'ericas e hiperb\'olicas. 
El ansatz de un agujero negro est\'atico y esf\'ericamente 
sim\'etrico es
\begin{equation}
ds^{2}=-N(r)dt^{2}+H(r)dr^{2}+S(r)d\Sigma_{k}^{2}~.
\label{anzats2}
\end{equation}  
donde las posibles geometr\'ias para la secci\'on transversal son
\begin{equation}
\label{sectrans}
d\Sigma^2_{k} =\left\{ \begin{array}{ll}
\vphantom{\sum_{i=1}^{n-1}}
d\theta^{2}+\sin^{2}{\theta}d\varphi^{2}& {\rm for}\; k = +1~,\\
\frac{1}{l^2}\sum_{i=1}^{2} dx_i^2&{\rm for}\; k = 0 ~,\\
\vphantom{\sum_{i=1}^{n-1}}
d\theta^{2}+\sinh^{2}{\theta}d\varphi^{2} &{\rm for}\; k = -1~.\ 
\end{array} \right.
\end{equation}
Distintas elecciones de la geometr\'ia
de la secci\'on transversal $k=-1,0,1$ nos dan distintas
topolog\'ias del borde \footnote{La topolog\'ia del borde nos dice como debemos foliar radialmente el espacio-tiempo}
\begin{equation}
\mathbb{R}\times H^{2}~, \qquad \mathbb{R}\times \mathbb{R}^{2}~, \qquad \mathbb{R}\times S^{2}~,
\end{equation} 
respectivamente. Este tipo de soluciones se conocen como agujeros negros topol\'ogicos y donde $\mathbb{R}$ corresponde al tiempo y $H^{2}$, $\mathbb{R}^{2}$, $S^{2}$ a $k=-1,0,1$. Estas geometr\'ias son el background 
donde la CFT se define. Es importante aclarar que el borde
de AdS~\footnote{Este borde (conforme), salvo el factor conforme $r^{2}/l^{2}$, es Minkowski, es decir, el borde conforme del espacio-tiempo $AdS_{4}$ es Minkowski en tres dimensiones.} es no din\'amico ya que no hay grados de libertad
 gravitacionales en la teor\'ia cu\'antica dual.\\
Una expresi\'on muy \'util
para la secci\'on transversal a la hora de usar 
paquetes matem\'aticos de c\'alculo es~\footnote{Esta es una forma compacta de escribir la expresi\'on (\ref{sectrans}), y adem\'as puede generalizarse para dimensiones mayores, ver ap\'endice \ref{apendice1}.}
\begin{align}
d\Sigma_{k}^{2}=\frac{dy^{2}}{1-ky^2}+(1-ky^{2})dz^{2}~.
\end{align}
En la que para $k=0$ cuyo dominio es $-\infty<(y,z)\leq\infty$. Para 
$k=1$ se tiene, $y=\cos{\theta}$, $z=\varphi$ por lo que
el dominio es $0\leq\theta\leq\pi$,~ $0\leq\varphi\leq 2\pi$. Finalmente para $k=-1$ se tiene, $y=\sinh{\theta}$ y $z=\varphi$ por lo que
el dominio es $-\infty<\theta<\infty$,~ $0\leq\varphi\leq 2\pi$.
\subsection*{Agujero negro de Schwarzschild-AdS}
La acci\'on de Hilbert-Einstein con constante consmol\'ogica $\Lambda=-3/l^{2}$ es
\begin{equation}
I[g_{\mu\nu}] = \frac{1}{2\kappa}\int_{\mathcal{M}}{d^{4}x\sqrt{-g}(R-2\Lambda)} + \frac{1}{\kappa}\int_{\partial\mathcal{M}}{d^{3}xK\sqrt{-h}}~. \label{action1}%
\end{equation}
Las ecuaciones de Einstein son
\begin{equation}
R_{\mu\nu}-\frac{1}{2}g_{\mu\nu}R=-\Lambda g_{\mu\nu}~.
\end{equation}
La soluci\'on es
%
\begin{equation}
ds^{2}=-\biggl{(}k-\frac{\mu}{r}+\frac{r^{2}}{l^{2}}\biggr{)}dt^{2}+\biggl{(}k-\frac{\mu}{r}+\frac{r^{2}}{l^{2}}\biggr{)}^{-1}dr^{2}+r^{2}d\Sigma_{k}^{2}~.
\label{schwarzshildAdS}
\end{equation} 
El horizonte es la soluci\'on mayor del 
polinomio dado por $-g_{tt}(r_{h})=0$
\begin{equation}
k-\frac{\mu}{r_{h}}+\frac{r_{h}^{2}}{l^{2}}=0~,
\end{equation}
donde la masa AMD es
\begin{equation}
M=\frac{\sigma_{k}\mu}{\kappa}~.
\end{equation} 
Fijando $\mu=0$, recuperamos el espacio-tiempo
AdS (global). Si el radio del horizonte del agujero negro esf\'erico AdS,
es peque\~no en comparaci\'on al radio AdS, $r_{+}<<l$, este se parece a la soluci\'on asint\'oticamente plana. De hecho, si fijamos $1/l=0$,
se obtiene la teor\'ia (la acci\'on) y la soluci\'on 
de Schwarzschild de manera consistente. 
%
\section{Agujeros negros con pelo escalar} 
 J. A. Wheeler (1971) propuso su famosa conjetura, de que los agujeros negros no tienen pelo de campos de materia mas generales
que el Maxwell en 4 dimensiones \cite{Wheeler:1971}~\footnote{El pelo se refiere a cantidades conservadas, las mismas que est\'an relacionadas a las constantes de integraci\'on. En general los agujeros negros tienen tres pelos, masa, momento angular y carga el\'ectrica.}. Sin embargo se mostr\'o que la hip\'otesis de Wheeler no se cumple, por ejemplo para teor\'ias Einstein-Yang-Mills, Einstein-Skyrme y en otras varias
combinaciones con campos dilat\'onicos \cite{Hertog:2004dr}.
\subsection{Teorema de no pelo}
En esta secci\'on presentamos los teoremas de no pelo y como pueden evadirse para obtener soluciones de agujeros negros con pelo. En la literatura existen varias soluciones exactas pero nos concentraremos en las soluciones estudiadas en \cite{Anabalon:2012dw}. El primer teorema de no-pelo-escalar fue planteado por Bekenstein (1974)     
%
 \begin{itemize}
 	\item No existen soluciones regulares de agujeros negros esf\'ericos asint\'oticamente planos con pelo escalar, para
 	campos escalares acoplados m\'inimamente con potenciales convexos (m\'inimo):  $\partial^{2} V/\partial\phi^{2}>0$
 \end{itemize}
Este teorema fue generalizado para campos escalares no-m\'inimamente
acoplados y para potenciales arbitrariamente positivos \cite{Heusler:1992ss}. Mas tarde estos teoremas fueron 
obtenidos para agujeros negros asint\'oticamente AdS \cite{Sudarsky:2002mk, Hertog:2006rr}.
 \begin{itemize} 	
 	\item No hay soluciones de agujeros negros con pelo escalar asint\'oticamente AdS cuyo potencial tiende asint\'oticamente a un m\'inimo (negativo) del potencial,  $\partial^{2} V/\partial\phi^{2}>0$ 
 \end{itemize}
 Sea la acci\'on de la gravedad acoplada m\'inimante
 a un campo escalar
 \begin{equation}
 I[g_{\mu\nu},\phi]=\int d^{4}x\sqrt{-g}\left[
 \frac{R}{2\kappa}-\frac{1}{2}\partial_{\mu}\phi\partial^{\mu
 }\phi-V(\phi)\right]~,
 \label{hairy2}
 \end{equation}
donde usamos la convenci\'on $\kappa=8\pi G$. Consideramos
unidades naturales $c=1=\hbar$, $\left[
\kappa\right]  =M_{P}^{-2}$ donde $M_{P}$ es la masa reducida de Planck.\\
La ecuaci\'on de movimiento para el campo escalar es 
\begin{equation}
\frac{1}{\sqrt{-g}}\partial_{\mu}\left(  \sqrt{-g}g^{\mu\nu}\partial_{\nu}%
\phi\right)  -\frac{\partial V}{\partial\phi}=0~. 
\label{dil}%
\end{equation}%
Para evadir el teorema de no-pelo, el potencial
debe cumplir con las siguientes condiciones, las mismas que aseguran la existencia del vac\'io estable
AdS en la regi\'on asint\'otica\footnote{Consideramos que el campo escalar en el borde es cero.}:
\begin{equation}
 \frac{dV}{d\phi}\br{\vert}_{\phi=0}=0~, \qquad
 V(0)=-\frac{3}{\kappa l^{2}}~, \qquad
  \frac{d^{2}V}{d\phi^{2}}\br{\vert}_{\phi=0}<0~.
\end{equation}
Esta nos dice que el potencial tiene un extremo, que es negativo y es m\'aximo.\\
Por ejemplo los potenciales escalares construidos en \cite{Acena:2012mr, Anabalon:2013qua}, evaden
el teorema de no pelo. El potencial $V(\phi)_{AdS}$ 
tiene dos par\'ametros $\alpha$ y $\Lambda=-3/l^{2}$,
\begin{equation}
V(\phi)_{AdS}=\bl{(}-\frac{1}{l^{2}}+\alpha \phi\br{)} (4+2\cosh{\phi})-6\alpha \sinh{\phi}~
\end{equation}
Para graficar,
fijamos $\alpha=(constante)l^{-2}$.
Donde, para cualquier sistema de coordenadas
en el borde el campo escalar tiende a cero $\phi=0$. 
Es interesante que este potencial tambi\'en permite obtener soluciones si, adem\'as de haber un campo escalar, hay un campo gauge que se acopla al campo escalar en la forma no minimal, $e^{\phi}F^{2}$, en la acci\'on.\\
Un segundo potencial se obtiene del anterior, fijando $1/l^{2}=0$, con este potencial se pueden obtener soluciones
asint\'oticamente planas.
\begin{equation}
V(\phi)_{flat}=2\alpha\phi(2+\cosh{\phi})-6\alpha\sinh{\phi}~.
\end{equation}
\begin{figure}[h!]
	\centering
	\begin{subfigure}[b]{0.4\textwidth}
		\includegraphics[width=1.2\textwidth]{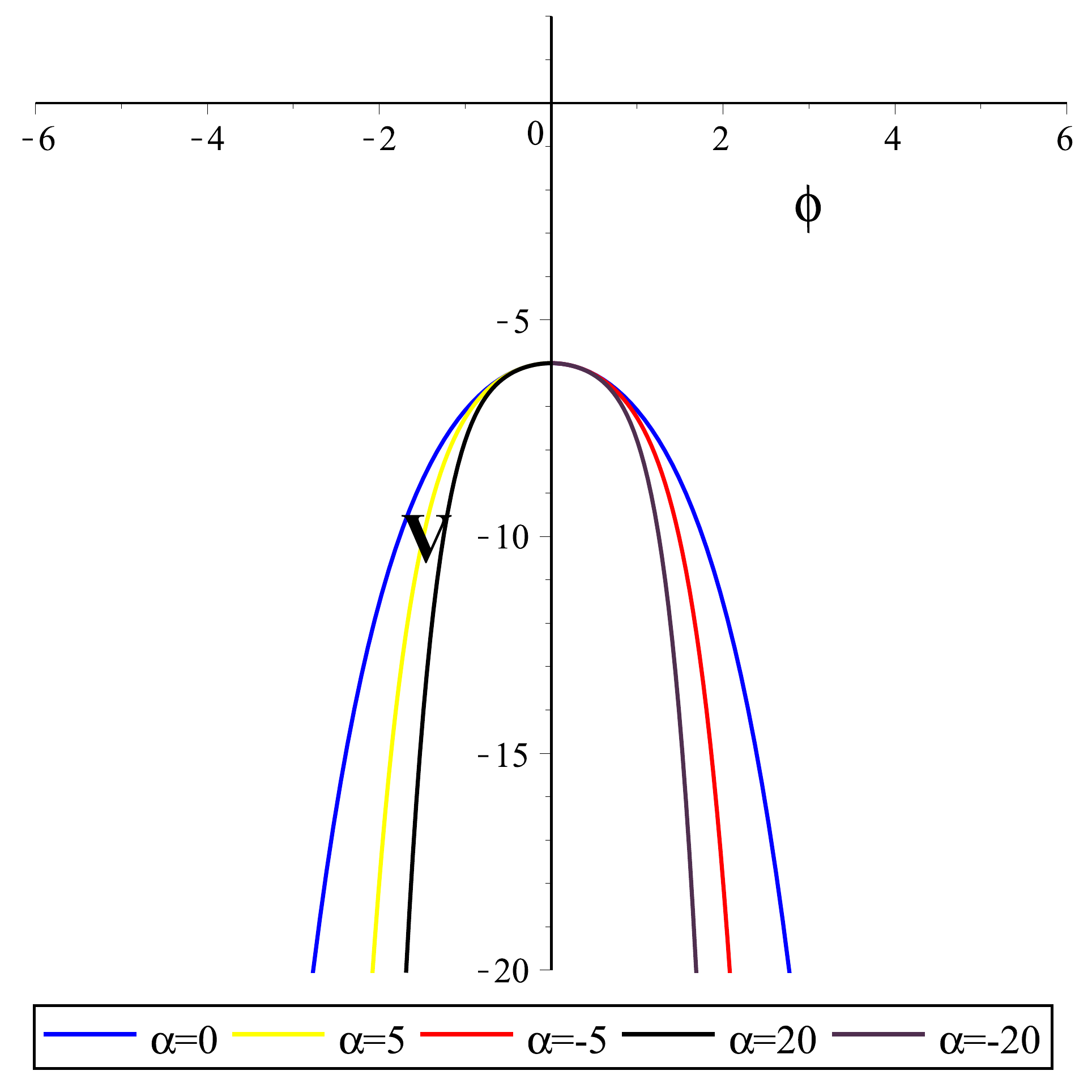}
	\end{subfigure}
	\begin{subfigure}[b]{0.4\textwidth}
		\includegraphics[width=1.2\textwidth]{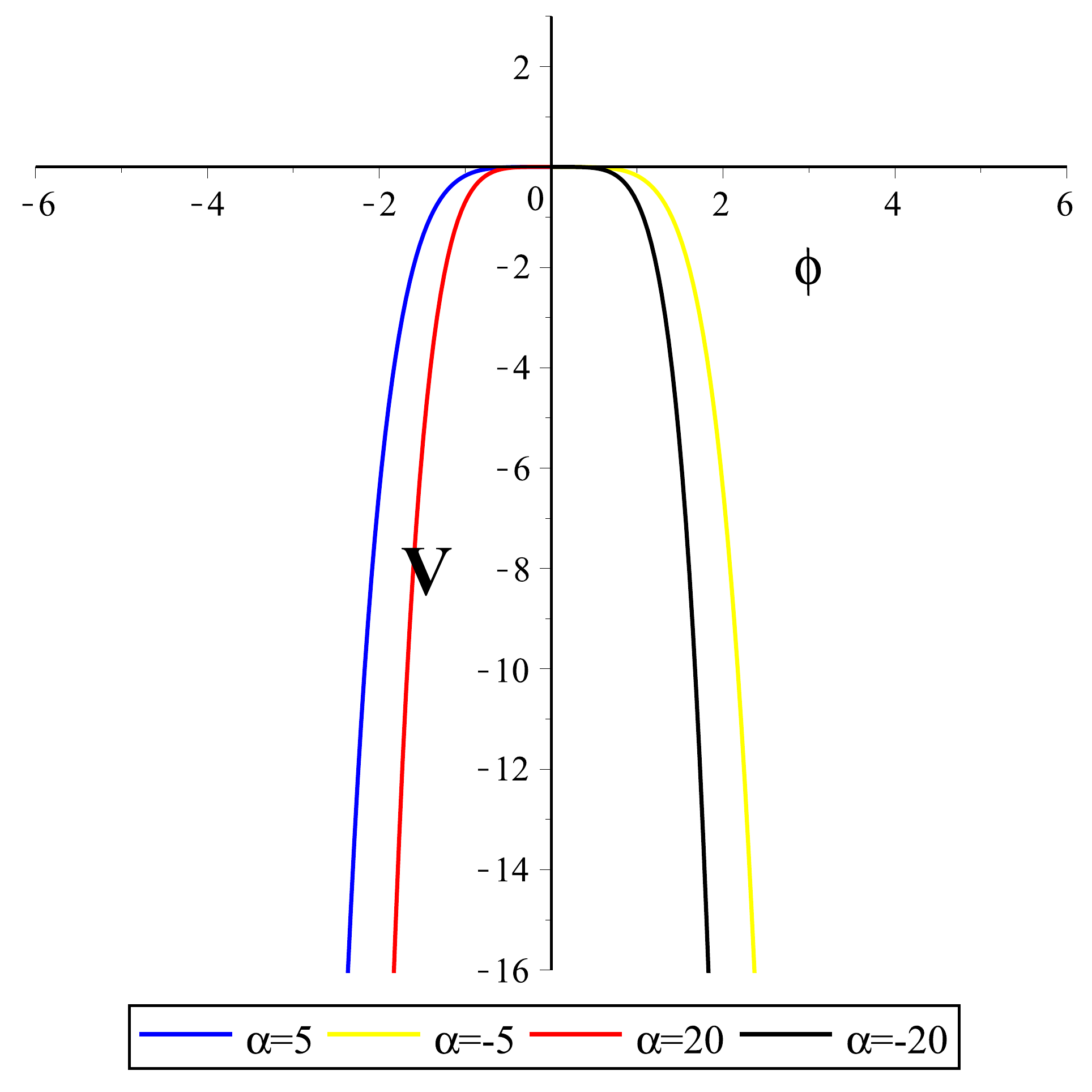}
	\end{subfigure}
	\caption{En nuestra notaci\'on: $V=V(\phi)$, $\phi(x)=\ln{x}$. La figura de la izquierda es el potencial para teor\'ias asint\'oticamente AdS. La figura de la derecha es el potencial para teor\'ias asint\'oticamente planas.}
	\label{vvsx}
\end{figure}
De acuerdo a la figura \ref{vvsx}, para ambos casos hay dos ramas (familias)
$\phi<0$ y $\phi>0$, en la que $\phi=0$ corresponde al valor 
del campo escalar en el borde y donde el potencial es un m\'aximo global (en cada rama). Note adem\'as que $V(\phi,\alpha)=V(-\phi,-\alpha)$.
\subsection{Agujero negro neutro con pelo escalar}
\label{soluhair}
Sean las ecuaciones de movimiento para el campo
escalar dada en (\ref{dil}) y la m\'etrica obtenidas a partir de la acci\'on (\ref{hairy2})
\begin{equation}
E_{\mu\nu}=R_{\mu\nu}-\frac{1}{2}g_{\mu\nu}R-\kappa T_{\mu\nu}^{\phi}~,%
\label{eqmotion}
\end{equation}
donde el tensor de stress del campo de materia es
\begin{equation}
T_{\mu\nu}^{\phi}=\partial_{\mu}\phi\partial_{\nu}\phi-g_{\mu\nu}\left[
\frac{1}{2}\left(  \partial\phi\right)  ^{2}+V(\phi)\right]~.
\label{stressmatter}
\end{equation}
Usaremos el siguiente ansatz para la m\'etrica de secci\'on transversal plana
\begin{equation}
ds^{2}=\Omega(x)\left[  -f(x)dt^{2}+\frac{\eta^{2}dx^{2}}{f(x)}+\frac{dy^{2}}{l^{2}}%
+\frac{dz^{2}}{l^{2}}\right]~.  \label{Ansatz}%
\end{equation}
Las ecuaciones para m\'etrica son:
\begin{align}
&E_{t}^{t}-E_{x}^{x}=0 \rightarrow \phi^{'2}=\frac{3\Om^{'2}-2\Om^{''}\Om}{\Om^{2}}~, \\ \notag
&E_{t}^{t}-E_{y}^{y}=0 \rightarrow f^{''}+\frac{\Om^{'}f^{'}}{\Om}=0~, \\ \notag
&E_{t}^{t}+E_{y}^{y}=0 \rightarrow V(\phi)=-\frac{1}{\Om^{2}\eta^{2}}\bl{(}f\Om^{''}+f^{'}\Om^{'}\br{)} ~.\\ \notag
\end{align} 
Las ecuaciones de movimiento 
pueden ser integradas si escogemos el siguiente
factor conforme
\cite{Acena:2013jya,Acena:2012mr,Anabalon:2013qua}:
\begin{equation}
\Omega(x)=\frac{\nu^{2}x^{\nu-1}}{\eta^{2}(x^{\nu}-1)^{2}}~,
\label{omegaandres}
\end{equation}
donde $\eta^{2}$ es s\'olo una constante de integraci\'on que esta relacionada con la masa y $\nu$
es el par\'ametro de pelo. El borde se encuentra en $x_{b}=1$ ya que $\Omega(x)$ diverge en ese punto.\\ 
Con esta elecci\'on de $\Omega(x)$, la ecuaci\'on para el campo escalar puede ser inmediatamente integrada 
\begin{equation}
\phi^{'2}=\frac{(\nu-1)^{2}}{x^{2}}-\frac{4\nu(\nu-1)x^{\nu-2}}{x^{\nu}-1}+\frac{4\nu^{2}x^{\nu-1}}{(x^{\nu}-1)^{2}}+\frac{2(\nu-1)}{x^{2}}+\frac{4\nu(1-\nu-x^{\nu})x^{\nu-2}x^{\nu-2}}{(x^{\nu}-1)^{2}}~,
\end{equation}
simplificando e integrando tal que en el borde $x_{b}=1$
el campo escalar es $\phi(x_{b})=0$, se tiene que  
\begin{equation}
\phi^{'2}=\frac{\nu^{2}-1}{2\kappa x^{2}} \rightarrow \int_{\phi}^{\phi=0}{d\phi}=\sqrt{\frac{\nu^{2}-1}{2\kappa}}\int_{x}^{1}{\frac{dx}{x}}~,
\end{equation}
\begin{equation}
\phi(x)=l_{\nu}^{-1}\ln{x}~, \qquad l_{\nu}^{-1}=\sqrt{\frac{\nu^{2}-1}{2\kappa}}~.
\label{escalhairy}
\end{equation}
La t\'ermino $l_{\nu}^{-1}$ le llamamos longitud del dilat\'on. 
El campo escalar es cero, $\phi=0$, cuando $\nu=1$, $l_{1}^{-1}=0$, es decir el caso $\nu=1$ el es de no-pelo. De la misma forma, con la elecci\'on de este factor conforme, podemos proceder a integrar f\'acilmente la segunda ecuaci\'on para m\'etrica 
\begin{align}
(f^{'}\Om)^{'} &=0~, \\ \nonumber
f(x)&=\frac{c_{2}\eta^{2}}{\nu^{2}}\int{\frac{(x^{\nu}-1)^{2}}{x^{\nu-1}}dx}+c_{1}~, \\ \nonumber
f(x)&=c_{1}+\frac{c_{2}\eta^{2}}{\nu^{2}}\bl{(}\frac{x^{2+\nu}}{2+\nu}+\frac{x^{2-\nu}}{2-\nu}-x^{2}\br{)}~. \nonumber
\end{align}
Las constantes\footnote{Aclaro que s\'olo hay una constante de integraci\'on, ya que una de ellas es en realidad la constante cosmol\'ogica y esta es un par\'ametro de la teor\'ia.} se fijan de tal forma que en el borde $x_{b}=1$, $f(x_{b})=1/l^{2}$ y $c_{2}=\alpha/\eta^{2}$, entonces
se obtiene\footnote{Esta elecci\'on es importante porque cuando calculamos el potencial on-shell, $\eta^{2}$ no debe aparecer en el potencial ya que es una constante de integraci\'on y no un par\'ametro de la teor\'ia.}
\begin{equation}
f(x)=\frac{1}{l^{2}}+\alpha\bl{[}\frac{1}{\nu^{2}-4}-\frac{x^{2}}{\nu^{2}}\bl{(}1+\frac{x^{-\nu}}{\nu-2}-\frac{x^{\nu}}{\nu+2}\br{)}\br{]}~.
\end{equation}
Esta soluci\'on es invariante bajo la transformaci\'on $\nu\rightarrow -\nu$, por lo que es lo mismo
estudiar en el rango de valores $\nu\geq 1$ y $\nu\leq -1$, mostramos que el potencial tambi\'en es invariante bajo el cambio de signo de $\nu$. Trabajaremos en el rango 
de valores $\nu\geq 1$.
Es f\'acil verificar que esta soluci\'on es bien comportada para cualquier
valor del par\'ametro pelo (hairy) $\nu\geq 1$ incluso en $\nu=2$, ya que en el l\'imite de $\nu=2$, $f(x)$ es convergente.
Es interesante que con $\Omega(x)$ y $f(x)$ podemos obtener una expresi\'on para el potencial $V(\phi)$,
despu\'es de un largo y cuidadoso c\'alculo, considerando que $x=e^{l_{\nu}\phi}$ se obtiene
\begin{align}
V(\phi)  &  =\frac{\Lambda(\nu^{2}-4)}{6\kappa\nu^{2}}\biggl{[}\frac{\nu
-1}{\nu+2}e^{-\phi l_{\nu}(\nu+1)}+\frac{\nu+1}{\nu-2}e^{\phi l_{\nu}(\nu
-1)}+4\frac{\nu^{2}-1}{\nu^{2}-4}e^{-\phi l_{\nu}}\biggr{]}\\
&  +\frac{\alpha}{\kappa\nu^{2}}\biggl{[}\frac{\nu-1}{\nu+2}\sinh{\phi l_{\nu
}(\nu+1)}-\frac{\nu+1}{\nu-2}\sinh{\phi l_{\nu}(\nu-1)}+4\frac{\nu^{2}-1}%
{\nu^{2}-4}\sinh{\phi l_{\nu}}\biggr{]}~.\nonumber
\end{align}
Este potencial tiene dos partes, una que depende de la constante cosmol\'ogica $\Lambda$ y otra que es controlada por $\alpha$. Los par\'ametros que definen la teor\'ia son: $\Lambda,G,\alpha, \nu$. Este par\'ametro $\alpha$ es importante, ya que si $\alpha=0$ se obtiene una singularidad desnuda, pero cuando $\alpha\neq 0$ se tiene un horizonte regular.
Por lo que $\alpha$ tiene una relevancia importante en el interior
del espacio-tiempo.\\
El comportamiento asint\'otico del potencial se obtiene expandiendo alrededor de $\phi=0$
\begin{equation}
V(\phi)=\frac{\Lambda}{\kappa}-\frac{\phi^{2}}{l^{2}}+\frac{\kappa\Lambda}{18}\frac{(\nu^{2}-3)}{\nu^{2}-1}\phi^{4}-\frac{l_{\nu}^{3}}{90}(\Lambda\nu^{2}-4\Lambda-6\alpha)\phi^{5}+O(\phi^{6})~,
\end{equation}  
de donde vemos que esta soluci\'on exacta evade el teorema de no-pelo y acepta el vac\'io est'andar $AdS$:
\begin{equation}
V(0)=\frac{\Lambda}{\kappa}~, \qquad \frac{dV}{d\phi}\br{\vert}_{\phi=0}=0~, \qquad \frac{d^{2}V}{d\phi^{2}}\br{\vert}_{\phi=0}=-\frac{2}{l^{2}}~.
\end{equation} 
El campo escalar tiene masa conforme $m^{2}=-2/l^{2}$ cuyo valor pertenece a la ventana de BF. El campo escalar vive en la parte exterior del horizonte de sucesos y se mantiene relativamente estable por la autointeracci\'on del campo escalar (back-reaction). Este pelo escalar r\'apidamente disminuye en las proximidades del borde.\\ 
\newpage
Finalmente, es importante destacar que hay dos familias
de soluciones bien comportadas, tales que\footnote{M\'as adelante mostramos que el comportamiento termodin\'amico es muy similar para ambas familias de soluciones} $\eta>0$.
\begin{itemize}
\item En el int\'ervalo $x\in (0,1)$ el campo escalar 
es negativo $\phi<0$, en ese caso, la singularidad se encuentra en $x=0$ y el borde es $x_{b}=1$. \\
\item En el int\'ervalo $x\in (1,\infty)$ el campo escalar es positivo $\phi>0$, en donde, $x_{b}=1$ sigue siendo el borde y la singularidad se encuentra en $x=\infty$. \\
\end{itemize}
Siguiendo un procedimiento similar se puede obtener la soluci\'on
de secci\'on transversal esf\'erica ($k=1$).
El potencial que obtuvimos para el caso $k=0$ es 
el mismo para este caso, y el campo
escalar es (\ref{escalhairy}). El ansatz y la soluci\'on es
\begin{equation}
ds^{2}=\Omega(x)\left[  -f(x)dt^{2}+\frac{\eta^{2}dx^{2}}{f(x)}+d\theta^{2}+\sin^{2}{\theta}d\varphi^{2}\right]~,  \label{Ansatz2}%
\end{equation}
\begin{equation}
f(x)=\frac{1}{l^{2}}+\alpha\bl{[}\frac{1}{\nu^{2}-4}-\frac{x^{2}}{\nu^{2}}\bl{(}1+\frac{x^{-\nu}}{\nu-2}-\frac{x^{\nu}}{\nu+2}\br{)}\br{]}+\frac{x}{\Omega (x)}~.
\label{Ansatz3}
\end{equation}
Cuando fijamos el par\'ametro $\nu=1$
el campo escalar es $\phi=0$ y el potencial escalar
se reduce a $V=\Lambda/\kappa$, entonces $\alpha$ deja 
de ser un par\'ametro de la teor\'ia y junto con $\eta$
forman una s\'ola constante de integraci\'on. 
Para ver esto realizamos la siguiente transformaci\'on de coordenadas
\begin{equation}
\Omega(x)\vert_{\nu=1}=r^{2}\Rightarrow x=1\pm\frac{1}{\eta r}~,
\end{equation} 
donde $\eta>0$ y $\pm$ es para la rama positiva y negativa respectivamente, se obtiene la soluci\'on de Schwarzschild-AdS 
\begin{equation}
-g_{tt}=\Omega(x)f(x)=k-\frac{\mu}{r}+\frac{r^{2}}{l^{2}}~, \qquad \mu=\mp\frac{\alpha+3\eta^{2}}{3\eta^{3}}~.
\end{equation} 
Aqu\'i $\mu$ es la nueva constante de integraci\'on relacionada con la masa del agujero negro en las coordendas $(t,r,\Sigma_{k})$. 
\subsection{Agujeros negros el\'ectricamente cargados con pelo}
Veamos un ejemplo simple de una soluci\'on
de agujero negro cargado con pelo
escalar 
\begin{equation}
I[{g_{\mu\nu},A_{\mu},\phi}]=\frac{1}{16\pi G_{N}}\int{d^{4}x\sqrt{-g}}\bl{[}R-\frac{1}{4}e^{\gamma\phi}F^{2}-\frac{1}{2}\pa_{\mu}\phi\pa^{\mu}\phi-V(\phi)\br{]}~,
\label{action2}
\end{equation}
donde el acoplamiento del campo gauge y el potencial 
son funciones del dilat\'on, y usamos la convenci\'on
$\kappa=8\pi G$. Las ecuaciones de movimiento
para el campo gauge, el dilat\'on y la m\'etrica, son
\[
\nabla_{\mu}\left(  e^{\gamma\phi}F^{\mu\nu}\right)  =0~,
\]
\[
\frac{1}{\sqrt{-g}}\partial_{\mu}\left(  \sqrt{-g}g^{\mu\nu}\partial_{\nu}%
\phi\right)  -\frac{\partial V}{\partial\phi}-\frac{1}{4}\gamma e^{\gamma\phi
}F^{2}=0~,
\]

\[
R_{\mu\nu}-\frac{1}{2}g_{\mu\nu}R=\frac{1}{2}\left[  T_{\mu\nu}^{\phi}%
+T_{\mu\nu}^{EM}\right]~,
\]
donde el tensor de stress de los campos de materia es
\[
T_{\mu\nu}^{\phi}=\partial_{\mu}\phi\partial_{\nu}\phi-g_{\mu\nu}\left[
\frac{1}{2}\left(  \partial\phi\right)  ^{2}+V(\phi)\right]~,
\,\,\,\,\,\,\,\,\,T_{\mu\nu}^{EM}=e^{\gamma\phi}\left(  F_{\mu\alpha}F_{\nu
}^{\cdot\alpha}-\frac{1}{4}g_{\mu\nu}F^{2}\right)~.
\]
\subsubsection*{Soluci\'on, $\gamma=1$}
El caso $\gamma=1$,  asint\'oticamente AdS, el caso 
asint\'oticamente plano fue explicado en \cite{Anabalon:2013qua}.
El potencial de la teor\'ia en este caso es
\begin{equation}
V(\phi)=\left(\frac{\Lambda}{3}+\alpha\phi\right)  \left(4+2\cosh
(\phi)\right)  -6\alpha\sinh(\phi)~.
\end{equation}
Considerando el ansatz
\begin{equation}
ds^{2}=\Omega(x)\left[  -f(x)dt^{2}+\frac{\eta^{2}dx^{2}}{x^{2}f(x)}%
+d\theta^{2}+\sin^{2}{\theta}d\varphi^{2}\right]~,
\end{equation}
se obtienen las soluciones para la m\'etrica y
el campo gauge
\begin{equation}
f(x)=\frac{1}{l^2}+\alpha\left[  \frac{(x^{2}-1)}{2x}-\ln{x}\right]
+\eta^{2}\frac{(x-1)^{2}}{x}-\frac{q^{2}\eta^{2}}{2x^{2}}\left(  x-1\right)
^{3}~,
\end{equation}
\begin{equation}
\Omega(x)=\frac{x}{\eta^{2}\left(  x-1\right)  ^{2}} \,\,\,\,\,\,\,\,\, ,
\,\,\,\,\,\, \qquad\phi(x)=\ln{x}~,
\end{equation}%
\begin{equation}
A=q\bl{(}\frac{1}{x}-\frac{1}{x_{+}}\br{)}dt ,\qquad F=-\frac{q}{x^{2}}dx\wedge dt~.
\end{equation}
Es interesante que para $\alpha=0$ (a diferencia del caso neutro descrito en la secci\'on anterior) a\'un existen soluciones regulares, en el sentido en el que existen horizontes (los ceros de la ecuaci\'on $f(x_{+})=0$) que esconden la singularidad. Actualmente estamos investigando sobre la transici\'on
de fase de estos agujeros negros cargados con pelo en diversos escenarios. Por ejemplo si $1/l^{2}=0$ las soluciones asint\'oticamente planas son regulares (por lo tanto permitidas). Otro caso interesante se da cuando estudiamos soluciones sin potencial escalar $V(\phi)=0$. En esos casos la responsabilidad para evadir el teorema de no pelo recae en el t\'ermino $e^{\phi\gamma}F^{2}$. Varias de estas soluciones pueden ser encontradas en \cite{Acena:2012mr, Anabalon:2013qua}.

%% file: capitulo4.tex
       \chapter{Renormalizaci\'on hologr\'afica y formalismo 
       Hamiltoniano}

\section{Sustracci\'on background vs contrat\'erminos}
\label{backsubtrac1}
El \textit{background} es el campo
de fondo, usualmente el estado base respecto al cual se comparan los estados o campos.
En general, la acci\'on on-shell
para soluciones de agujeros negros tienen divergencias\footnote{Incluso para el espacio-tiempo AdS (global) la acci\'on contiene divergencias.}. Estas divergencias  pueden eliminarse si calculamos la acci\'on on-shell respecto de su estado base.
El m\'etodo de sustracci\'on background 
se basa en hecho de que se conoce a prior\'i
el estado base de la soluci\'on del agujero
negro a estudiar. Pero hay soluciones en la literatura donde
el estado ground es patol\'ogico y casos en los
que es imposible obtener el estado ground. Por esta raz\'on ser\'ia mucho mejor tener un m\'etodo independiente del background. 
Es bien sabido que la acci\'on on-shell
contiene divergencias debido a la integraci\'on
en un volumen infinito. Estas divergencias IR (infrarrojas) en el lado de la teor\'ia gravitacional (invocando la dualidad AdS/CFT) se interpretan como las
divergencias UV (ultravioleta) de la teor\'ia cu\'antica dual. 
La t\'ecnica usual en teor\'ias cu\'anticas del campo consiste en agregar contrate\'rminos que eliminen estas divergencias. Entonces, en el lado gravitacional podemos agregar contrat\'erminos locales que dependen de la geometr\'ia 
intr\'inseca del borde \cite{Balasubramanian:1999re}\footnote{Este m\'etodo no solo 
se puede aplicar a soluciones de agujeros negros, tambien pueden usarse para regularizar
el tensor de stress hologr\'afico y obtener la energ\'ia de otras soluciones asintoticamente AdS, por ejemplo \cite{Astefanesei:2005yj,Astefanesei:2005eq,Astefanesei:2004kn,Astefanesei:2004ji,Balasubramanian:2002am,Balasubramanian:2005bg,He:2007ji}.}
\begin{equation}
I_{g}=-\frac{1}{8\pi G_{N}}\int_{\pa\mathcal{M}}{d^{3}x\sqrt{-h}~\Xi(l,\mathcal{R,\nabla\mathcal{R}})}~.
\end{equation}
Donde $\Xi(l,\mathcal{R,\nabla\mathcal{R}})$ depende del radio AdS $l$ y de los 
invariantes constru\'idos a partir 
de la curvatura de Riemann del borde
$\mathcal{R}^{a}_{~bcd}$\footnote{El hecho de que dependa s\'olo de la geometr\'ia intr\'inseca del borde garantiza que las condiciones de borde para la m\'etrica del borde $\delta h=0$ (Condici\'on de borde de Dirichlet) no se violen.}. La m\'etrica inducida es $h_{ab}$ y usamos la descomposici\'on de los \'indices del espacio-tiempo $\mu=(r,a)$.
Sea el espacio-tiempo AdS en coordenadas est\'aticas
 \begin{equation}
 ds^{2}=-\bl{(}k+\frac{r^{2}}{l^{2}}\br{)}dt^{2}+\bl{(}k+\frac{r^{2}}{l^{2}}\br{)}^{-1}dr^{2}+r^{2}d\Sigma_{k}^{2}~,
 \end{equation}
de la cual, la secci\'on transversal $d\Sigma_{k}^{2}$ 
admite diferentes topolog\'ias. La foliaci\'on $r=R=constante$ nos da la m\'etrica inducida 
\begin{equation}
h_{ab}dx^{a}dx^{b}=-\bl{(}k+\frac{R^{2}}{l^{2}}\br{)}dt^{2}+R^{2}d\Sigma_{k}^{2}~.
\end{equation}
Es claro que los contrat\'erminos gravitacionales\footnote{De ahora en adelante nombraremos $I_{g}$ como contrat\'erminos gravitacionales ya que dependen fundamentalmente de la geometr\'ia intr\'inseca del borde} $I_{g}$ que dependen de $\mathcal{R}$ y sus derivadas son distintas
segun la foliaci\'on que escogamos (en este caso, seg\'un la topolog\'ia de la secci\'on transversal). 
Concretamente, el proceso de regularizaci\'on depende
de la elecci\'on del sistema de coordenadas en la regi\'on asint\'otica de AdS. Entonces diferentes foliaciones dan como resultado diferentes teor\'ias 
cu\'anticas del campo duales. 
El borde conforme de AdS como vimos anteriormente es
el espacio-tiempo de Minkowski
\begin{equation}
ds^{2}=\frac{r_{b}^{2}}{l^{2}}(-dt^{2}+l^{2}d\Sigma_{k}^{2})~,
\end{equation}
donde $r_{b}$ es el borde $r=\infty$
\section*{Sustracci\'on background}
%
Sea la acci\'on con constante cosmol\'ogica $\Lambda=-3/l^{2}$
dada en (\ref{action1}) y dada la soluci\'on de Schwarzschild-AdS
de secci\'on transversal esf\'erica $k=1$, dada en (\ref{schwarzshildAdS})
\begin{equation}
ds^{2}=-\bl{(}1-\frac{\mu}{r}+\frac{r^{2}}{l^{2}}\br{)}dt^{2}+\bl{(}1-\frac{\mu}{r}+\frac{r^{2}}{l^{2}}\br{)}^{-1}dr^{2}+r^{2}(d\theta^{2}+\sin^{2}{\theta}d\varphi^{2})~
\end{equation}
El horizonte es tal que $-g_{tt}=f(\mu,r_{+})=0$
 y la temperatura es
\begin{equation}
T=\frac{f'}{4\pi}\br{\vert}_{r_{+}}=\beta^{-1}=
\frac{1}{4\pi}\bl{(}\frac{3r_{+}^{2}+l^{2}}{l^{2}r_{+}}\br{)}~. 
\end{equation}
El t\'ermino on-shell del bulk es:
\begin{equation}
I^{E}_{bulk}=\frac{12\pi\beta}{\kappa l^{2}}\int_{r_{+}}^{R}{r^{2}dr}=\frac{4\pi\beta}{\kappa l^{2}}(R^{3}-r_{+}^{3})~.
\end{equation} 
Donde $R$ es un cut-off tal que el 
borde es $R\rightarrow r_{b}=\infty$.
El t\'ermino de Gibbons-Hawking
se obtiene considerando la foliaci\'on
$r=R=constante$\footnote{No confundir $R$ con 
la curvatura escalar de Ricci}
\begin{equation}
h_{ab}dx^{a}dx^{b}=-\bl{(}1-\frac{\mu}{R}+\frac{R^{2}}{l^{2}}\br{)}dt^{2}+R^{2}(d\theta^{2}+\sin^{2}{\theta}d\varphi^{2})~,
\end{equation}
donde el proyector $P^{ab}=g^{ab}-n^{a}n^{a}$, la normal, la curvatura extr\'inseca y la traza $K=P^{ab}K_{ab}$ de la foliaci\'on son, respectivamente
\begin{equation}
n_{a}=\frac{\de_{a}^{r}}{\sqrt{g^{rr}}}~,\qquad  K_{ab}=\frac{\sqrt{g^{rr}}}{2}\pa_{r}h_{ab}~, \qquad
K=\frac{1}{l^{2}R^{2}}\bl{(}1-\frac{\mu}{R}+\frac{R^{2}}{l^{2}}\br{)}^{-1/2}\bl{(}-\frac{3l^{2}\mu}{2}+3R^{3}+2Rl^{2}\br{)}
\end{equation}
y obtenemos
\begin{equation}
I_{GH}^{E}=-\frac{4\pi\beta}{\kappa l^{2}} \bl{(}-\frac{3l^{2}\mu}{2}+3R^{3}+2Rl^{2}\br{)}~.
\end{equation}
Entonces la acci\'on Euclidea para el agujero negro  expl\'icitamente
divergente es\footnote{La acci\'on Euclidea se relaciona
	con la energ\'ia libre de la siguiente forma $F=\beta^{-1}I^{E}_{bh}$, consideramos unidades tales que la constante de Boltzman es igual a $1$.}:
\begin{equation}
I^{E}_{bh}=I^{E}_{bulk}+I_{GH}^{E}=\frac{4\pi\beta}{\kappa l^{2}} \bl{(}\frac{3l^{2}\mu}{2}-2R^{3}-2Rl^{2}-r_{+}^{3}\br{)}~.
\end{equation}
El estado ground 
es el espacio-tiempo
AdS (global) a temperatura finita $\beta_{0}=1/T_{0}$
(AdS t\'ermico), el cual 
se obtiene a partir de la
soluci\'on Schawarzschild-AdS
considerando $\mu=0$
\begin{equation}
ds^{2}=-\bl{(}1+\frac{r^{2}}{l^{2}}\br{)}dt^{2}+\bl{(}1+\frac{r^{2}}{l^{2}}\br{)}^{-1}dr^{2}+r^{2}(d\theta^{2}+\sin^{2}{\theta}d\varphi^{2})~.
\end{equation}
Necesitamos calcular la acci\'on on-shell
para esta m\'etrica, en la que $0\leq r\leq r_{b}$.
La coordenada temporal en la secci\'on Euclidea $t\rightarrow -i\tau^{E}$ tiene periodicidad (temperatura) $\beta_{0}=T_{0}^{-1}$ 
\begin{equation}
I^{E}_{AdS}=I^{E}_{bulk}+I_{GH}^{E}=\frac{4\pi\beta_{0}}{\kappa l^{2}} \bl{(}-2R^{3}-2l^{2}R\br{)}~.
\end{equation}
Para comparar
las dos acciones necesitamos que est\'en
a la misma temperatura (ambas soluciones
con la misma periodicidad en el borde)
\begin{equation}
\beta_{0}\sqrt{1+\frac{R^{2}}{l^{2}}}=\beta\sqrt{1+\frac{R^{2}}{l^{2}}-\frac{\mu}{R}}~.
\end{equation}
Con esta relaci\'on podemos 
comparar las acciones on-shell (energ\'ias libres)
en el borde $R\rightarrow r_{b}$
\begin{equation}
I^{E}=I^{E}_{bh}-I^{E}_{AdS}=\frac{4\pi\beta}{\kappa l^{2}}\bl{[} \bl{(}\frac{3l^{2}\mu}{2}-2R^{3}-2Rl^{2}-r_{+}^{3}\br{)}-\frac{\beta_{0}}{\beta}\bl{(}-2R^{3}-2l^{2}R\br{)}\br{]}~,
\end{equation}
de donde obtenemos la energ\'ia libre
\begin{equation}
F=\beta^{-1}I^{E}=\frac{4\pi}{\kappa l^{2}} \bl{(}\frac{l^{2}\mu}{2}-r_{+}^{3}\br{)}~.
\label{Free11}
\end{equation}
\section*{Contrat\'erminos}
\begin{figure}[h]
	\centering
	\begin{subfigure}[b]{8cm}
		\includegraphics[width=1.1\textwidth]{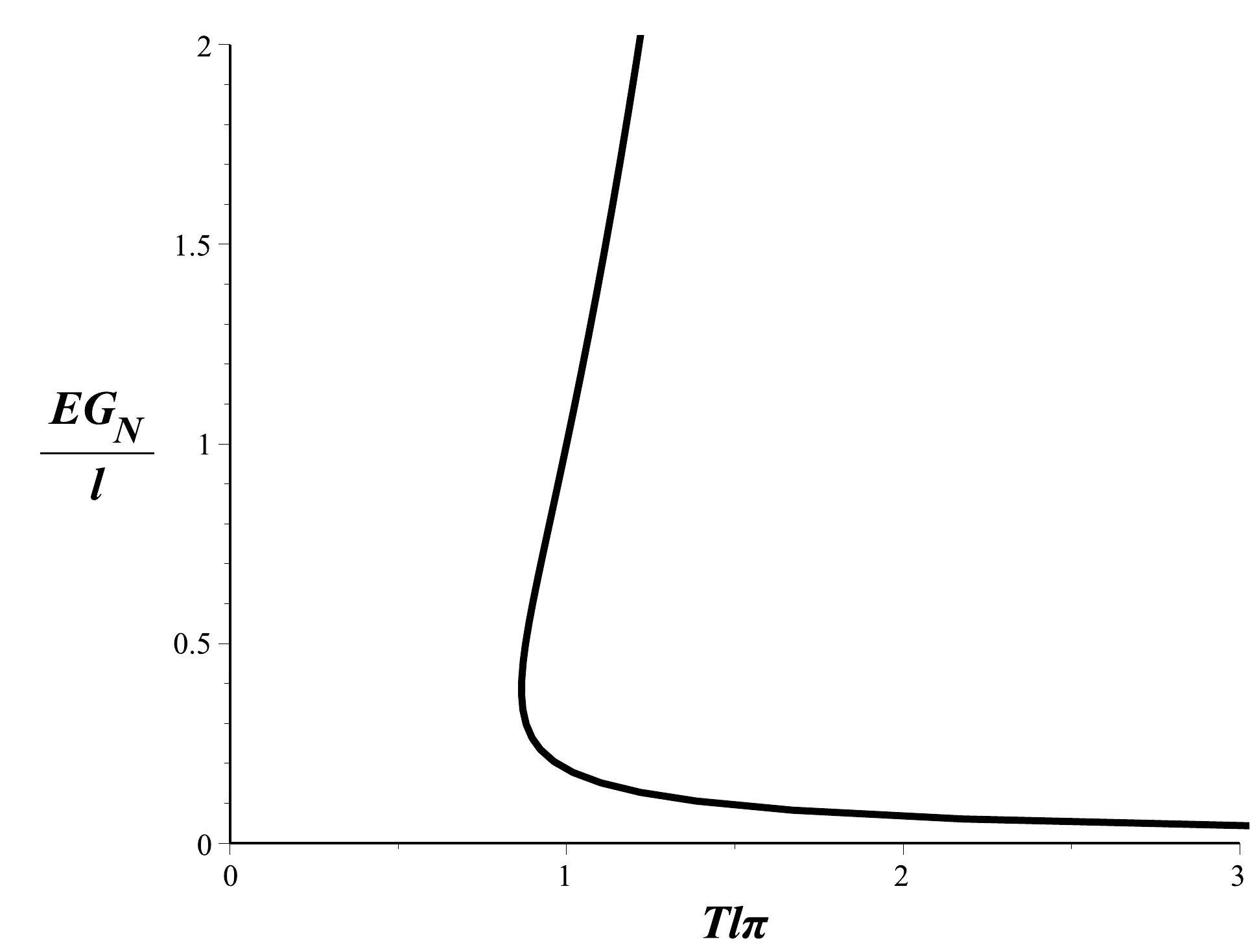}
	\end{subfigure}
	\begin{subfigure}[b]{8cm}
		\includegraphics[width=1.1\textwidth]{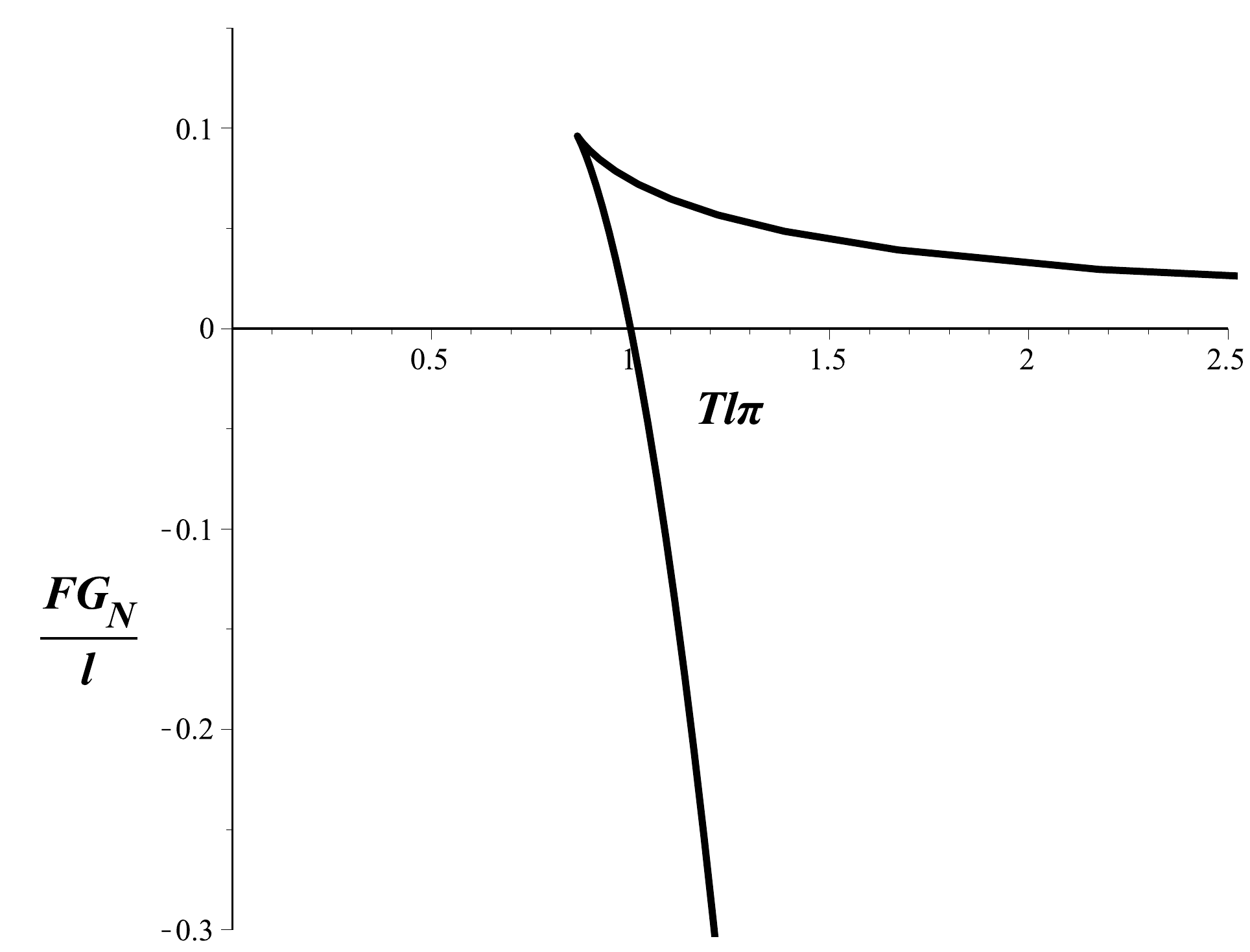}
	\end{subfigure}
\caption{Ambas figuras son para los agujeros negros de Schwarzschild-AdS. Izquierda, E vs T y Derecha F vs T. En la figura $G_{N}$ es la constante de Newton, pero a lo largo de la tesis usaremos $G$.}
	\label{trans}
\end{figure}

En este caso consideramos que desconocemos
la m\'etrica del estado ground. Entonces
consideramos contrat\'erminos que est\'an en funci\'on de de cantidades geom\'etricas 
intr\'insecas al borde, estas eliminan 
las divergencias infrarrojas que aparecen en el lado de la gravedad. Estos contrat\'erminos en el espacio-tiempo se interpretan como los contrat\'erminos que eliminan las divergencias ultravioletas de la teor\'ia cu\'antica dual y tienen la forma\footnote{Para espacios-tiempo de $4$-dimensiones, estos dos contrat\'erminos son suficientes. Para mayores detalles ver el ap\'endice \ref{apendice1}.}
\begin{equation}
I_{g}=-\frac{1}{\kappa}\int_{\partial M}{d^{3}x\sqrt{-h}\biggl{(}\frac{2}{l}+\frac{l\mathcal{R}}{2}\biggr{)}}~.
\end{equation}
En $4$-dimensiones, estos dos
primeros contrat\'erminos son suficientes
para eliminar las divergencias 
c\'ubicas de la contribuci\'on del $I_{bulk}^{E}$
y las divergencias c\'ubicas y lineales de $I^{E}_{GH}$.
La acci\'on Euclidea on-shell del $bulk$ y de Giboons-Gawking (GH) es 
\begin{equation}
I^{E}_{bulk}+I_{GH}^{E}=\frac{4\pi\beta}{\kappa l^{2}} \bl{(}\frac{3l^{2}\mu}{2}-2r_{b}^{3}-2r_{b}l^{2}-r_{+}^{3}\br{)}~.
\end{equation}
Los contrat\'erminos evaluados on-shell son
\begin{equation}
I_{g}^{E}=\frac{4\pi\beta}{\kappa l^{2}}\bl{(}1+\frac{l^{2}}{R^{2}}-\frac{\mu l^{2}}{R^{3}}\br{)}^{\frac{1}{2}}(2R^{3}+kl^{2}R)\br{\vert}_{R=r_{b}}=
\frac{4\pi\beta}{\kappa l^{2}}(2r_{b}^{3}+2l^{2}r_{b}-\mu l^{2})~,
\end{equation}
entonces la acci\'on Euclidea on-shell con esta nueva contribuci\'on es
\begin{equation}
I^{E}=I^{E}_{bulk}+I_{GH}^{E}+I_{g}^{E}=\frac{4\pi\beta}{\kappa l^{2}} \bl{(}\frac{l^{2}\mu}{2}-r_{+}^{3}\br{)}~.
\end{equation}
Conclu\'imos que ambos m\'etodos nos dan el mismo resultado comparando con (\ref{Free11}). 
Un procedimiento similar muestra que para los agujeros negros asint\'oticamente planos, mostrados en (\ref{schwarzshildflat})
la acci\'on finita (usando el m\'etodo de sustracci\'on background)\footnote{Mas adelante aclaramos que la extensi\'on del m'etodo de contrat\'erminos a espacio-tiempo asint\'oticamente planos fue extendido.} da $I^{E}=\frac{2\pi\beta r_{+}}{\kappa}$. 
Las energ\'ia libres para el agujero negro de Schwarzschild y Schwarzschild-AdS son
\begin{equation}
F_{flat}=\frac{2\pi r_{+}}{\kappa}~,\qquad 
F_{SAdS}=\frac{4\pi}{\kappa l^{2}} \bl{(}\frac{l^{2}\mu}{2}-r_{+}^{3}\br{)}~,
\end{equation} 
respectivamente. Vemos que el agujero negro de Schwarzschild 
tiene una energ\'ia libre siempre positiva $F>0$, es decir no cambia de signo por lo que no hay transiciones de fase, de hecho se puede mostrar que es inestable termodin\'amicamente, ya que la capacidad calor\'ifica es siempre $C<0$~\footnote{El hecho de que los agujeros negros de Schwarzschild tengan $C<0$, significa que el agujero negro nunca llega a un equilibrio t\'ermico, es decir, rad\'ia m\'as de lo que absorbe. Debido a que el borde es plano las part\'iculas pueden escapar al borde (es como un sistema abierto), en cambio AdS es como una caja y un agujero negro en AdS es como un sistema cerrado.}.\\
Para el caso de los agujeros negros de SAdS la energ\'ia libre
puede cambiar de signo y en este caso podemos hablar de agujeros negros grandes $r_{+}>l$ y de agujeros negros peque\~nos $r_{+}<l$ respecto del radio AdS, $l$. La energ\'ia (masa del agujero negro) se calcula 
mediante la siguiente relaci\'on termodin\'amica, donde $I^{E}=\beta F$
\begin{equation}
E=-T^{2}\frac{\partial I^{E}}{\partial T}=\frac{\mu}{2G}.
\end{equation}
En la figura \ref{trans}, la capacidad calor\'ifica $C=\pa E/\pa T$ es la pendiente de la curva de la izquierda donde los agujeros negros grandes (de energ\'ia o masa grande) tienen capacidad calor\'ifica positiva y son termodin\'amicamente estables, al contrario de los agujeros negros peque\~nos (C<0) que son inestables. En la derecha de la figura \ref{trans}, la energ\'ia libre puede cambiar de signo, donde el punto cr\'itico ($F=0$) esta dado por la temperatura de Hawking-Page, $T_{HP}=1/\pi l$. Esto es signo de transiciones de fase de primer orden~\footnote{Una transici\'on de fase de $1^{er}$ orden es tal que que la energ\'ia libre $F=0$ y la entrop\'ia camb\'ia de forma discont\'inua. En t\'erminos de agujeros negros, eso significa que la formaci\'on de un agujero negro es una transici\'on de fase de primer orden.}. En la regi\'on donde $F<0$, los agujeros negros grandes ($r_{+}>l$) son preferidos y estables en cambio en la regi\'on donde $F>0$,
el espacio AdS (global $\mu=0$, a temperatura finita) es preferido ($r_{+}<l$) pero inestable\footnote{De hecho en esta regi\'on hay agujeros negros pequen\~os ($r_{+}<l$), pero se evaporan r\'apidamente y es como si el espacio fuese AdS puro.}.  
M\'as adelante mostramos que las temperaturas de los agujeros negros de Schwarzschild y Schwarzschild-AdS son
respectivamente\footnote{Recuerde que el horizonte lo designamos con $r_{h}$, pero la ecuaci\'on para el horizonte de Schwarzschild-AdS es cuadr\'atica, por lo que hay dos ra\'ices. En ese caso, consideramos a la ra\'iz mayor como el horizonte $r_{h}=r_{+}$.}
\begin{equation}
T_{flat}=\frac{1}{4\pi r_{h}}~, \qquad T_{SAdS}=\frac{1}{4\pi r_{+}}\bl{(}1+\frac{3r_{+}^{2}}{l^{2}}\br{)}~.
\end{equation}
%
\section{Formalismo de Brown-York}
\label{BrownYork}
En el contexto de la dualidad AdS/CFT, el gravit\'on
AdS se acopla al tensor de stress energ\'ia del CFT \cite{Gubser:1997yh,Gubser:1997se}:
\begin{equation}
\int_{\partial\mathcal{M}} d^{3}x\, h^{ab}\,T_{ab}~.
\end{equation}
Entonces, desde el punto de vista hologr\'afico,
el tensor de stress de Brown-York se interpreta
como el tensor de stress energ\'ia de la teor\'ia 
dual del campo. 
En esta secci\'on, trabajamos en el sistema de coordenadas
can\'onicas $(t,r,\Sigma_{k})$ para el cual, la m\'etrica fue dada en (\ref{Ansatz1}).
Para la geometr\'ia del bulk, usaremos
la siguiente foliaci\'on con superficies
$r=R=constant$ donde la m\'etrica inducida es
\begin{equation}
ds^{2}=h_{ab}dx^{a}dx^{b}=-N(R)dt^{2}+S(R)d\Sigma_{k}^{2}~. \label{k1induced}%
\end{equation}
El tensor de stress cuasi-local de Brown-York
es definido como
\cite{Brown:1992br}
\begin{equation}
\tau^{ab}\equiv\frac{2}{\sqrt{-h}}\frac{\delta I}{\delta h_{ab}}~,%
\end{equation}
donde $I$ es la acci\'on total, debidamente suplementada con 
los contrat\'erminos~\footnote{Una de las conclusiones m\'as importantes en esta investigaci\'on es que una acci\'on debidamente regularizada, tal que el princ\'ipio variacional este bien definido $\delta I=0$, se obtiene un tensor cuasi-local $\tau^{ab}$ libre de divergencias. Mostramos que de este tensor cuasi-local podemos obtener la energ\'ia.}. 
Ya que la m\'etrica donde la teor\'ia dual vive
esta relacionada con la m\'etrica del borde
por un factor conforme, es muy importante enfatizar
que el tensor de stress de CFT esta relacionado
con el tensor de stress de Brown-York salvo un
factor conforme.
Veamos un ejemplo, donde mostramos los detalles de este m\'etodo 
para el agujero negro de Schwarzschild-AdS---
seguiremos el an\'alisis de \cite{Myers:1999psa}. Sea la m\'etrica del agujero negro de Schwarzschild-AdS
\begin{equation}
ds^{2}=-\biggl{(}1-\frac{\mu}{r}+\frac{r^{2}}{l^{2}}\biggr{)}dt^{2}%
+\biggl{(}1-\frac{\mu}{r}+\frac{r^{2}}{l^{2}}\biggr{)}^{-1}dr^{2}+r^{2}%
d\Omega^{2}~.%
\end{equation}
Si consideramos la foliaci\'on, $r=R$,
la m\'etrica inducida $h_{ab}$ de cualquier 
slice~\footnote{En espa\~nol se dice c\'ascara o rebanada. Dejamos la expresi\'on en Ingles.}, es
\begin{equation}
ds^{2}=-\biggl{(}1-\frac{\mu}{R}+\frac{R^{2}}{l^{2}}\biggr{)}dt^{2}+R^{2}%
d\Omega^{2}~,%
\end{equation}
El tensor de stress hologr\'afico se obtiene variando la 
acci\'on debidamente suplementada con 
los contrat\'erminos gravitacionales, $I_{g}$
 \begin{equation}
I=\frac{1}{2\kappa}\int_{M}{d^{4}x\sqrt{-g}(R-2\Lambda)}+\frac{1}{\kappa}\int_{\partial M}{d^{3}x\sqrt{-h}K}-\frac{1}{\kappa}\int_{\partial M}{d^{3}x\sqrt{-h}\biggl{(}\frac{2}{l}+\frac{l\mathcal{R}}{2}\biggr{)}}~,
\end{equation} 
de donde
\begin{equation}
\tau_{ab}=-\frac{1}{8\pi G}\bl{(}K_{ab}-h_{ab}K-\frac{2}{l}h_{ab}+lG_{ab}\br{)}~.
\end{equation}
Como ya puntualizamos anteriormente, la m\'etrica 
del borde es
\begin{equation}
ds^{2}_{borde}=\frac{R^{2}}{l^{2}}(-dt^{2}+l^{2}d\Omega^{2})~.
\end{equation}
pero la m\'etrica del background donde la teor\'ia
cu\'antica dual vive es $\gamma_{ab}$ definida 
de la siguiente forma
\begin{equation}
ds_{dual}^{2}=\gamma_{ab}dx^{a}dx^{b}=-dt^{2}+l^{2}d\Omega^{2}%
\end{equation}
La m\'etrica $\gamma_{ab}$ no es din\'amica 
y esta relacionada por un factor conforme
con la m\'etrica del borde. El correspondiente
tensor de stress dual es
\begin{equation}
\langle\tau_{ab}^{dual}\rangle=\lim_{R\rightarrow\infty}\frac{R}{l}\tau
_{ab}=\frac{\mu}{16\pi G l^{2}}(3\delta_{a}^{0}\delta_{b}^{0}+\gamma_{ab})~.
\end{equation}
Escrita en esta forma \cite{Myers:1999psa}, esta 
corresponde al tensor de stress de un gas t\'ermico
de part\'iculas sin masa y como ya se esperaba, debido a la simetr\'ia conforme su traza debe desaparecer $\langle\tau^{dual}\rangle=\langle\tau_{ab}^{dual}\rangle\gamma^{ab}=0$.
%
\section{Formalismo Hamiltoniano}
En espacios-tiempo AdS existen diferentes
m\'etodos para calcular la masa 
gravitacional y es importante 
compararlas ---para la condici\'on
de borde de Dirichlet esto fue hecho
en gran detalle \cite{Hollands:2005wt}
y salvo algunas ambig\"uedades relacionadas
a t\'erminos constantes en el borde,
el formalismo Hamiltoniano, la masa AMD, 
y el m\'etodo hologr\'afico producen el mismo
resultado. Aunque, como ya fue aclarado en 
\cite{Anabalon:2014fla}, cuando la simetr\'ia conforme 
se rompe en el borde la masa AMD no es la masa 
f\'isica correcta y se debe calcular la masa Hamiltoniana
del sistema. En esta secci\'on mostraremos algunos detalles del
c\'alculo de la masa Hamiltoniana y mostraremos que coincide
con la masa hologr\'afica aun cuando la simetr\'ia conforme en el borde se rompe.

Consideramos el m\'etodo de Regge-Teitelboim \cite{Regge:1974zd} para calcular la masa de teor\'ias
de campos escalares acoplados m\'inamente a la gravedad
en espacios-tiempo asint\'oticamente locales AdS. Consideramos
la acci\'on
\begin{equation}
I[g_{\mu\nu},\phi] = \int_{\mathcal{M}}{d^{4}x\sqrt{-g}\biggl{[}\frac
	{R}{2\kappa}-\frac{1}{2}(\partial\phi)^{2}-V(\phi)\biggr{]}} + \frac{1}%
{\kappa}\int_{\partial\mathcal{M}}{d^{3}xK\sqrt{-h}}~, \label{actionHamil}%
\end{equation}
para el cual la ligaduras Hamiltonianas  $\mathcal{H}_{\bot}$ and $\mathcal{H}%
_{i}$, con $i=1,2,3$, tienen contribuciones del t\'ermino gravitacional y de la materia que en este caso corresponde 
a un campo escalar m\'inimamente acoplado con un potencial
auto-interactuante $V(\phi)$. 

Estas ligaduras son funciones de las variables can\'onicas: 
la m\'etrica 3-dimensional $g_{ij}$ y el campo escalar $\phi$,
y sus correspondientes momentos conjugados $\pi^{ij}$ y $\pi_{\phi}$.
Las ligaduras Hamiltonianas est\'an dadas por
\begin{align}
{\mathcal{H}}_{\bot}  &  =\frac{2\kappa}{\sqrt{g }}\left[  \pi_{ij}\pi
^{ij}-\frac{1}{2}\left(  \pi^{i}{}_{i}\right)  ^{2}\right]  - \frac{1}%
{2\kappa}\sqrt{g }\left.  ^{(3)}R \right. \nonumber\\
&  + \frac{1}{2}\left(  \frac{\pi_{\phi}{}^{2}}{\sqrt{g }}+\sqrt{g }g
^{ij}\phi,_{i}\phi,_{j}\right)  +\sqrt{g }V\left(  \phi\right)~, \label{hperp}\\
{\mathcal{H}}_{i}  &  =-2\pi_{i}^{j}{}_{\mid j} + \pi_{\phi}\phi,_{i}~.
\label{hi}%
\end{align}
La m\'etrica 3-dimensional $g_{ij}$ puede ser reconocida del elemento
de linea escrita en su forma ADM 
\begin{equation}
\label{ADMform}ds^{2}=-(N^{\perp})^{2}dt^{2}+g _{ij}\left(  dx^{i}%
+N^{i}dt\right)  \left(  dx^{j}+N^{j}dt\right)~,
\end{equation}
donde, $g$, $^{(3)}R$ y la barra vertical $\mid$ denotan el determinante,
la curvatura escalar, y la derivada covariante asociada a m\'etrica espacial, respectivamente.
El generador can\'onico de la simetr\'ia asint\'otica definida
por el vector $\xi^{\mu}=(\xi^{\perp},\xi^{i})$ es una combinaci\'on
lineal de las ligaduras $\mathcal{H}_{\perp}, \mathcal{H}_{i}$ m\'as
un t\'ermino de superficie $Q[\xi]$
\begin{equation}
\label{congen}H[\xi]=\int_{\partial\mathcal{M}} d^{3} x \left(  \xi^{\perp}
\mathcal{H}_{\perp}+\xi^{i}\mathcal{H}_{i}\right)  +Q[\xi]~.
\end{equation}
$Q[\xi]$ es escogido de tal forma que cancela los t\'erminos
de superficie que provienen de la variaci\'on de los generadores
respecto a las variables can\'onicas. De esta forma, el generador
$H[\xi]$ posee una derivada funcional bien definida \cite{Regge:1974zd}. La forma general de $Q[\xi]$ para el generador
(\ref{congen}) esta dado por \cite{Henneaux:2006hk}
\begin{align}
\label{de}\delta Q[\xi]  &  =\oint d^{2}S_{l}\left[  \frac{G^{ijkl}}{2 \kappa
}(\xi^{\bot}\delta g_{ij}{}_{\mid k}-{\xi^{\bot}}_{,k} \delta g_{ij})+2
\xi_{k}\delta\pi^{kl}\right. \nonumber\\
&  \left.  +(2 \xi^{k}\pi^{jl}-\xi^{l}\pi^{jk}) \delta g_{jk}- (\sqrt{g}
\xi^{\bot}g^{lj}\phi,_{j}+\xi^{l}\pi_{\phi})\delta\phi\right]~,
\end{align}
donde
\begin{equation}
G^{ijkl}\equiv\frac{1}{2}\sqrt{g}(g^{ik}g^{jl}+g^{il}g^{jk}-2g^{ij}g^{kl})~.
\end{equation}
La normal y las componentes tangenciales de la deformaci\'on 
permitida
 $(\xi^{\perp},\xi^{i})$ est\'an relacionadas con las componentes 
 del espacio-tiempo $(\xi^{\perp},{}%
^{(3)}\xi^{i})$ en la siguiente forma
\begin{equation}
\xi^{\perp} =N^{\perp} \xi^{t}~, \quad\xi^{i} ={}^{(3)}\xi^{i}+N^{i}\xi^{t}~.%
\end{equation}
El siguiente paso es notar que el generador
Hamiltoniano (\ref{congen}) se reduce al t\'ermino
de superficie $Q[\xi]$ cuando las ligaduras se mantienen.
Esto es, el valor de los generadores --- las cargas
conservadas asociadas a las simetr\'ias asint\'oticas ---
est\'an justamente dadas por $Q[\xi]$. Ya que las cargas
est\'an definidas por un t\'ermino de superficie en el borde,
estas necesitan solamente el comportamiento de las variables 
can\'onicas y simetr\'ias cercanas al borde i.e. su comportamiento
asint\'otico. 
Nos centraremos en el caso est\'atico. Por definici\'on 
hay un vector de Killing tipo-tiempo $\xi=\xi^{\mu}\pa_{\mu}=\partial_{t}$,
y la correspondiente carga conservada asociada con esta simetr\'ia --- traslaciones temporales --- es de primeros princ\'ipios la 
masa $M$~\footnote{Del teorema de Noether.}. En el caso est\'atico todos los momentos
desaparecen y la expresi\'on (\ref{de}),
evaluado para $\xi= \partial_{t}$, se reduce a 
\begin{equation}
\label{qt}\delta M\equiv\delta Q[\partial_{t}]=\oint d^{2}S_{l} \left[
\frac{G^{ijkl}}{2 \kappa}(\xi^{\bot}\delta g_{ij}{}_{\mid k}-{\xi^{\bot}}_{,k}
\delta g_{ij})- \sqrt{g} \xi^{\bot}g^{lj}\phi,_{j}\delta\phi\right]~.
\end{equation}
Notamos que hay una contribuci\'on expl\'icita del campo escalar
en la masa. En general esta cantidad es distinta de cero. 
Con el fin de conseguir un mejor entendimiento
de esta contribuci\'on, es conveniente separarla de
la contribuci\'on usual gravitacional escribiendo 
$\delta M$ como  
\begin{equation}
\delta M=\delta M_{G}+\delta M_{\phi}~,%
\end{equation}
donde
\begin{equation}
\label{eq:Q_G}\delta M_{G}=\oint d^{2}S_{l} \frac{G^{ijkl}}{2 \kappa}%
(\xi^{\bot}\delta g_{ij}{}_{\mid k}-{\xi^{\bot}}_{,k} \delta g_{ij})
\end{equation}
y
\begin{equation}
\label{eq:Q_phi}\delta M_{\phi}=-\oint d^{2}S_{l} \sqrt{g} \xi^{\bot}%
g^{lj}\phi,_{j}\delta\phi~.
\end{equation}
Como ya mencionamos anteriormente, la variaci\'on de la
masa, dada por la integral de superficie (\ref{qt}), 
necesita justamente el comportamiento asint\'otico
de las variables can\'onicas y simetr\'ias. Sin embargo,
esta variaci\'on usualmente requiere m\'as informaci\'on
para ser integrada, y las condiciones de borde deben ser
impuestas. La masa de un sistema queda bien definida
despu\'es de imponer las condiciones de borde. El efecto
del fall-off lento del campo escalar en la masa de espacios-tiempo
asint\'oticamente con pelo fueron estudiados en \cite{Henneaux:2006hk,Henneaux:2002wm, Henneaux:2004zi} usando el
formalismo Hamiltoniano descrito anteriormente. Otras aproximaciones y m\'etodos pueden ser encontrados en
\cite{Hertog:2004dr,Barnich:2002pi,Gegenberg:2003jr,Banados:2005hm,Amsel:2006uf}.

Un paso m\'as all\'a se hizo en \cite{Anabalon:2014fla} donde
los c\'alculos de la masa de estas configuraciones con pelo fueron
hechas considerando informaci\'on adicional obtenidas por las
ecuaciones de campo. Para este trabajo, nos centraremos en el an\'alisis de una clase de potenciales de masa conforme $m^{2}=-2l^{-2}$ en cuatro dimensiones. La construcci\'on 
de la masa Hamiltoniana nos di\'o una \'util intuici\'on
para construir los contrat\'erminos.


%% file: capitulo5.tex
                             \chapter{Gravedad dise\~nada}
El t\'ermino de Gravedad dise\~nada (Designer gravity)
fue acu\~nado en \cite{Hertog:2004ns}. Son aquellos
espacios-tiempo asint\'oticamente AdS tales 
que para la misma acci\'on hay varias posibles 
condiciones de borde para el campo escalar y que alterando estas
condiciones, las propiedades de la teor\'ia cambian.
\section{Condiciones de borde y deformaciones multitraza $AdS_{4}$}
\label{sec2}
En esta secci\'on, revisamos el rol
de las condiciones de borde de AdS en el 
contexto de la dualidad AdS/CFT \cite{Maldacena:1997re}.
De acuerdo al diccionario hologr\'afico, imponer condiciones
de borde mixtas en el campo escalar corresponde a 
perturbar la teor\'ia del borde $N$-largo por una 
deformaci\'on multi-traza relevante, irrelevante o marginal  
\cite{Witten:2001ua}~\footnote{Ver la secci\'on \ref{defor233}}.

Vamos a empezar por exhibir algunos hechos conocidos acerca de la dualidad AdS/CFT \cite{Maldacena:1997re}. Nos gustar\'ia describir qu\'e tipo de condiciones de borde preservan la simetr\'ia conforme de la teor\'ia dual del campo e interpretarlas en el contexto de la dualidad AdS/CFT \cite{Hertog:2004dr, Henneaux:2006hk,Witten:2001ua}.
\newpage
En primer lugar, describimos el espacio-tiempo AdS$_{4}$
y explicamos como las simetr\'ias de las dos teor\'ias duales coinciden. Como explicamos anteriormente en la secci\'on (\ref{seccion2}), el grupo de isometr\'ia $SO(3,2)$ de AdS$_{4}$ act\'ua en el borde (conforme) como el grupo conforme
\footnote{El grupo conforme del espacio-tiempo de Minkowski es el grupo que deja invariante el cono de luz, en otras palabras, todas las transformaciones que dejan $ds^{2}=0$ invariante} 
actuando en el espacio-tiempo de Minkowski.

El espacio-tiempo AdS tiene el n\'umero m\'aximo de isometr\'ias en todas las dimensiones. Por lo tanto,
tiene una forma simple en un gran n\'umero de sistemas
de coordenadas (ver, p.ej., \cite{Emparan:1999pm} para una 
discusi\'on en el contexto de la dualidad AdS/CFT). Dependiendo de la elecci\'on de la coordenada radial, las foliaciones a radio constante
pueden tener una geometr\'ia diferente o incluso una topolog\'ia diferente. Por ejemplo, uno puede foliar AdS$_{4}$ 
de la siguiente forma:   
\begin{equation}
d\bar{s}^{2}=\bar{g}_{\mu\nu}dx^{\mu}dx^{\nu}=-\left(  k+\frac{r^{2}}{l^{2}}
\right)  dt^{2} + \frac{dr^{2}}{k+\frac{r^{2}}{l^{2}}}+r^{2} d\Sigma^{2}_{k}~,
\label{metricAdS}%
\end{equation}
donde $k=\{+1, 0, -1\}$ para las foliaciones esf\'erica ($d\Sigma_{1}^{2}=d\Omega^{2}$),
plana ($d\Sigma_{0}^{2} = dx^{2}+dy^{2}$), e hiperb\'olica ($d\Sigma_{-1} =
dH^{2}$), respectivamente. Aqu\'i, $d\Omega^{2}$ y $dH^{2}$ 
son la m\'etricas unitarias de una esfera e 
hiperboloide 2-dimensional, respectivamente. 
El radio $l$ de AdS$_{4}$ esta relacionada con la constante
cosmol\'ogica por $\Lambda=-3/l^{2}$.
El borde conforme est\'a en $r \rightarrow\infty$,
para el cual la m\'etrica inducida es 
\begin{equation}
h_{ab}dx^{a}dx^{b}=\frac{r^{2}}{l^{2}}(-dt^{2}+l^{2}d\Sigma_{k}^{2})~.
\end{equation}
Ahora est\'a claro que la geometr\'ia background donde vive la teor\'ia del campo, est\'a relacionada con la geometr\'ia del borde por una transformaci\'on conforme. Por lo tanto, la m\'etrica del bulk est\'a asociada con una estructura conforme en el infinito. El factor conforme juega un papel importante cuando se calcula el tensor de strees del borde.

Aun cuando las diferentes foliaciones de AdS$_{4}$ est\'an relacionadas por transformaciones de coordenadas locales, las correspondientes teor\'ias gauge duales no son f\'isicamente equivalentes (por ejemplo, en el caso $k=1$ existen transiciones de fase Hawking-Page, pero no para $k=0$). Esto es debido al hecho de que diferentes foliaciones tipo espacio de la geometr\'ia del background conducen a diferentes definiciones de la coordenada temporal (y del Hamiltoniano) del sistema cu\'antico dual.

Las coordenadas de Poincar\'e (\ref{poincare2}), cubren s\'olo una parte del espacio-tiempo AdS$_{4}$, el espacio-tiempo de Minkowski aparece de forma natural como el borde conforme. Las isometr\'ias finitas de AdS$_{4}$ mapean el borde $z=0$ a s\'i mismas y, por otra parte, act\'uan como transformaciones conformes en el borde. En particular, la transformaci\'on $(z, t, x, y)\leftarrow\lambda (z,t,x,y)$, la cual deja la m\'etrica (\ref{poincare2}) invariante, act\'ua como la dilataci\'on (transformaci\'on de escala) en el borde. Dado que el espacio-tiempo AdS no es globalmente hiperb\'olico, uno tiene que imponer las condiciones de borde. Dentro de la dualidad AdS/CFT, varias deformaciones de las condiciones de borde de AdS se interpretan como duales a las deformaciones de la CFT. Es bien sabido \cite{BF, Ishibashi:2004wx} que un escalar de la masa arbitraria en AdS puede tener ambos modos normalizable y no normalizable. Se ha demostrado en \cite{Balasubramanian:1998sn, Balasubramanian:1998de} que los modos normalizables describen las fluctuaciones en el bulk y los modos no normalizables corresponden a operadores de inserci\'on en la teor\'ia dual del campo en el borde. 
Estamos interesados en el caso en que ambos modos son normalizables:
\begin{equation}
m^{2}_{BF} + \frac{1}{l^{2}}> m^{2} \geq m^{2}_{BF} \, ,
\,\,\,\,\,\,\,\,\,\,\,\,\,\,\,\,\  m^{2}_{BF}= -\frac{9}{4l^{2}}%
\label{BFi}
\end{equation}
donde $m^{2}_{BF}$ is la cota de BF (\ref{BFi}) en cuatro dimensiones.
En lo que sigue mostramos una breve revisi\'on de las condiciones de borde que se adaptan a un campo escalar con masa conforme. Estamos interesados en la acci\'on (\ref{actionHamil}),
donde $V(\phi)$ es el potencial escalar, $\kappa=8\pi G $ con
$G$ la constante gravitacional de Newton, y el \'ultimo t\'ermino
es el t\'ermino de borde de Gibbons-Hawking. Aqu\'i, $h$
es el determinante de la m\'etrica del borde y $K$ es la traza
de la curvatura extr\'inseca. Las ecuaciones de movimiento
para el campo escalar y la m\'etrica fuer\'on dadas en (\ref{dil}) y (\ref{eqmotion}),
donde el tensor de stress del campo escalar es (\ref{stressmatter}).
Trabajamos con el ansatz general
\begin{equation}
ds^{2}=-N(r)dt^{2}+H(r)dr^{2}+S(r)d\Sigma_{k}^{2}~. \label{Ansatz1}%
\end{equation}
Como se demostr\'o por primera vez en tres dimensiones \cite{Henneaux:2002wm}, y luego se generaliz\'o a cuatro y m\'as dimensiones \cite{Hertog:2004dr, Henneaux:2006hk,Henneaux:2004zi, Amsel:2006uf}, en presencia de campos escalares las condiciones de borde est\'andares AdS son modificadas. Uno puede obtener el fall-off
consistente para la componente m\'etrica $g_{rr}$ considerando las 
ecuaciones de movimiento y el fall-off del campo escalar. 
Una discusi\'on general para cualquier masa del campo
escalar en el rango (\ref{BFi}) se puede encontrar en \cite{Henneaux:2006hk},
pero en este trabajo nos concentraremos en el caso concreto 
de masa conforme en cuatro dimensiones $m^{2}=-2l^{-2}$. Empezamos con el potencial\footnote{Se verifica inmediatamente que este potencial evade el teorema de no-pelo. Adem\'as la teor\'ia es tal que $V(0)=-3/\kappa l^{2}$, lo que significa que tiene como vac\'io a AdS.}
\begin{equation}
V(\phi)=-\frac{3}{\kappa l^{2}}-\frac{\phi^{2}}{l^{2}} +O(\phi^{4})~.
\label{vphinolog}%
\end{equation}
El fall-off del campo escalar en este caso es
\begin{equation}
\phi(r)=\frac{\alpha}{r}+\frac{\beta}{r^{2}}+O(r^{-3})~. \label{phi}%
\end{equation}
Con el fin de acomodar agujeros negros se considera el siguiente comportamiento asint\'otico para los coeficientes m\'etricos, $N(r)$ y $S(R)$  
\begin{align}
N(r)  &  =-g_{tt}=\frac{r^{2}}{l^{2}}+k-\frac{\mu}{r}+O(r^{-2})\label{gtt}~,\\
S(r)  &  =r^{2}+O(r^{-2})~. \label{gSS}%
\end{align}

Ahora, utilizamos la combinaci\'on de las ecuaciones de movimiento (\ref{eqmotion}),
$E_{t}^{t}-E_{r}^{r}=0$, de la que obtenemos
\begin{equation}
NS^{^{\prime}2}H-2NS^{^{\prime\prime}}HS+(NH)^{^{\prime}}S^{^{\prime}%
}S-2\kappa NHS^{2}\phi^{^{\prime}2}=0
\end{equation}
y entonces
\begin{equation}
H(r)=g_{rr}=\frac{l^{2}}{r^{2}}+\frac{l^{4}}{r^{4}}\biggl{(}-k-\frac
{\alpha^{2}\kappa}{2l^{2}}\biggr{)}+\frac{l^{5}}{r^{5}}\biggl{(}\frac{\mu}%
{l}-\frac{4\kappa\alpha\beta}{3l^{3}}\biggr{)}+O(r^{-6})~.
\end{equation}
La raz\'on por la que deseamos obtener el fall-off de $g_{rr}$
de esta manera es por que la masa Hamiltoniana se puede leer 
de \'el --- si hay una contribuci\'on del campo escalar a la masa, uno debe ser capaz de identificarla en $g_{rr}$.
A partir de ahora, se utilizar\'a la notaci\'on gen\'erica para la expansi\'on de $g_{rr}$ como
\begin{equation}
g_{rr}=\frac{l^{2}}{r^{2}}+\frac{al^{4}}{r^{4}}+\frac{bl^{5}}{r^{5}}+O(r^{-6})~,
\label{grr}%
\end{equation}
donde $a=-k-\frac{\kappa\alpha^{2}}{2l^{2}}$ y $b=\frac{\mu}{l}-\frac{4\kappa\alpha\beta}{3l^{3}}$. En este punto, es interesante investigar que cuando las condiciones asint\'oticas son AdS invariantes y el Hamiltoniano est\'a bien definido. Parece que, para una relaci\'on
funcional especial de los modos del campo escalar $\alpha$ y $\beta$, ambas condiciones se cumplen. Esto se hizo de manera expl\'icita en \cite{Hertog:2004dr, Henneaux:2006hk} y aqu\'i s\'olo presentamos el resultado:

\begin{equation}
\beta=C\alpha^{2}. \label{betanon}
\end{equation}

Curiosamente, tambi\'en se puede obtener un Hamiltoniano finito cuando la simetr\'ia conforme del borde se rompe.
Un an\'alisis similar se puede hacer para la llamada \textit{rama logar\'itmica} \cite{Henneaux:2004zi}. En lo que sigue nos gustar\'ia analizar cuidadosamente este caso y presentar detalles que se van a utilizar en las siguientes secciones.

Es bien sabido que una ecuaci\'on diferencial de segundo orden tiene dos soluciones linealmente independientes. Cuando la relaci\'on de las ra\'ices de la ecuaci\'on es un n\'umero entero, la soluci\'on puede desarrollar una rama logar\'itmica. Esto es exactamente lo que ocurre cuando el campo escalar satura la cota de BF, en cuyo caso el fall-off contiene un t\'ermino logar\'itmico \cite{Henneaux:2004zi}. Sin embargo, estamos interesados en un campo escalar de masa conforme $m^{2}=-2l^{-2}$. Para obtener la rama logar\'itmica, un t\'ermino c\'ubico en la expansi\'on asint\'otica del potencial del campo escalar es necesario \cite{Henneaux:2006hk}
\begin{equation}
V(\phi)=-\frac{3}{\kappa l^{2}}-\frac{\phi^{2}}{l^{2}}+\lambda\phi^{3}%
+O(\phi^{4})~, \label{vphilog}%
\end{equation}
de manera que el fall-off del campo escalar a considerar es
\begin{equation}
\phi(r)=\frac{\alpha}{r}+\frac{\beta}{r^{2}}+\frac{\gamma\ln(r)}{r^{2}%
}+O(r^{-3})~. \label{philog}%
\end{equation}
Para obtener el fall-off de $g_{rr}$ usamos el mismo fall-off para las otras componentes de la m\'etrica y la misma combinaci\'on de las ecuaciones de movimiento como en la rama no-logar\'itmica, $E_{t}^{ t}-E_{r}^{ r}=0$. Obtenemos
\begin{equation}
H(r)=g_{rr}=\frac{l^{2}}{r^{2}}+\frac{l^{4}}{r^{4}}\biggl{(}-k-\frac
{\kappa\alpha^{2}}{2l^{2}}\biggr{)}+\frac{l^{5}}{r^{5}}\biggl{(}\frac{\mu}%
{l}-\frac{4\kappa\alpha\beta}{3l^{3}}+\frac{2\kappa\alpha\gamma}{9l^{3}%
}\biggr{)}+\frac{l^{5}\ln{r}}{r^{5}}\biggl{(}-\frac{4\kappa\alpha\gamma
}{3l^{3}}\biggr{)}+O\biggl{[}\frac{\ln{(r)^{2}}}{r^{6}}\biggr{]}~.
\label{grrlog}%
\end{equation}
Utilizando de nuevo la notaci\'on gen\'erica para la expansi\'on asint\'otica de $g_{rr}$
\begin{equation}
H(r)=\frac{l^{2}}{r^{2}}+\frac{l^{4}a}{r^{4}}+\frac{l^{5}b}{r^{5}}+\frac
{l^{5}c\ln{r}}{r^{5}}+O\biggl{[}\frac{\ln{(r)^{2}}}{r^{6}}\biggr{]}~,
\end{equation}
identificamos los coeficientes relevantes como
\begin{align}
a  &  =-k-\frac{\alpha^{2}\kappa}{2l^{2}}~, \qquad b=\frac{\mu}{l}%
-\frac{4\kappa\alpha\beta}{3l^{3}}+\frac{2\kappa\alpha\gamma}{9l^{3}}~, \qquad
c=-\frac{4\kappa\gamma\alpha}{3l^{3}}~. \label{abglog}%
\end{align}
Ahora, vamos a ver cuando el fall-off del campo escalar que hemos considerado es compatible con su ecuaci\'on de movimiento:
\begin{equation}
\partial_{r}\biggl{(}\frac{\phi^{^{\prime}}S\sqrt{N}}{\sqrt{H}}%
\biggr{)}-S\sqrt{NH}\frac{\partial V}{\partial\phi}=0~.
\end{equation}
En la regi\'on asint\'otica, $r\rightarrow\infty$, esta ecuaci\'on
viene a ser
\begin{equation}
\frac{3\alpha^{2}l^{2}\lambda+\gamma}{l^{2}}+O(r^{-1})=0~,
\end{equation}
y de esta forma, el coeficiente de $\gamma$ esta fijado por $\alpha$ as $\gamma=-3l^{2}\lambda\alpha^{2}$ (o, usando la notaci\'on que se usaremos m\'as adelante,
$\gamma=C_{\gamma}\alpha^{2}$, donde $C_{\gamma}=-3l^{2}\lambda$). 
Este resultado es importante porque, como veremos en breve, tambi\'en forma parte de las condiciones que preservan la simetr\'ia conforme del borde. El \'ultimo paso en nuestra derivaci\'on es investigar cuando las condiciones de borde se preservan dentro de la simetr\'ia asint\'otica AdS. El correspondiente vector de Killing asint\'otico $\xi^{\mu}=(\xi^{r},\xi^{m})$ es
\begin{align}
\xi^{r}  &  =r\eta^{r}(x^{m})+O(r^{-1})~,\\
\xi^{m}  &  =O(1)~,\nonumber
\end{align}
donde $\{m\}$ es un \'indice que para el tiempo y las coordenadas angulares. El fall-off del campo escalar debe ser invariante dentro de las simetr\'ias asint\'oticas AdS y as\'i obtenemos:
\begin{equation}
\label{seriesphi}\phi^{\prime}(x)=\phi(x)+\xi^{\mu}\partial_{\mu}\phi(x)=
\frac{\alpha^{^{\prime}}}{r}+\frac{\beta^{^{\prime}}}{r^{2}}+\frac
{\gamma^{^{\prime}} \ln(r)}{r^{2}}+O(r^{-3})~,
\end{equation}
donde
\begin{align}
\alpha^{^{\prime}}=  &  \alpha-\eta^{r}\alpha+\xi^{m}\partial_{m}\alpha~,\\
\beta^{^{\prime}}=  &  \beta-\eta^{r}(2\beta-\gamma)+\xi^{m}\partial_{m}%
\beta\nonumber~,\\
\gamma^{^{\prime}}=  &  \gamma-2\gamma\eta^{r}+\xi^{m}\partial_{m}%
\gamma\nonumber~.\\
\nonumber
\end{align}
Si los coeficientes de la serie (\ref{seriesphi}) est\'an funcionalmente relacionados, $\beta=\beta(\alpha,x^{m})$ y $\gamma=\gamma(\alpha,x^{m})$, la simetr\'ia conforme en el borde fija la relaci\'on funcional entre los coeficientes de modo que las ecuaciones anteriores son compatibles. Se puede realizar una expansi\'on de Taylor de $\gamma^{^{\prime}}$ y $\beta^{^{\prime}}$
para obtener:
\begin{equation}
\alpha^{^{\prime}}\frac{\partial\gamma}{\partial\alpha}-\gamma^{^{\prime}%
}=0=\alpha\frac{\partial\gamma}{\partial\alpha}-\gamma+\eta^{r}%
\biggl{(}2\gamma-\alpha\frac{\partial\gamma}{\partial\alpha}\biggr{)}+\xi
^{m}\biggl{(}\frac{\partial\alpha}{\partial x^{m}}\frac{\partial\gamma
}{\partial\alpha}-\frac{\partial\gamma}{\partial x^{m}}\biggr{)} \label{gamma}%
\end{equation}
y
\begin{equation}
\alpha^{^{\prime}}\frac{\partial\beta}{\partial\alpha}-\beta^{^{\prime}%
}=0=\alpha\frac{\partial\beta}{\partial\alpha}-\beta+\eta^{r}\biggl{(}2\beta
-\gamma-\alpha\frac{\partial\beta}{\partial\alpha}\biggr{)}+\xi^{m}%
\biggl{(}\frac{\partial\alpha}{\partial x^{m}}\frac{\partial\beta}%
{\partial\alpha}-\frac{\partial\beta}{\partial x^{m}}\biggr{)} \label{beta}%
\end{equation}
Considerando el hecho de que $\eta^{r}$ y $\xi^{m}$ son independientes, conseguimos de (\ref{gamma}) que $2\gamma=\alpha\frac{\partial\gamma}{\partial\alpha}$,
lo que implica que $\gamma=C_{\gamma}\alpha^{2}$.
Este es el resultado obtenido antes de la ecuaci\'on de movimiento para el campo escalar. De la integraci\'on (\ref{beta}) obtenemos
\begin{equation}
\beta(\alpha)=(-C_{\gamma}\ln(\alpha)+C)\alpha^{2}~.%
\end{equation}
Si $C_{\gamma}=0$, el resultado coincide con la condici\'on que se encuentra para la rama no logar\'itmica (\ref{betanon}). Una vez m\'as, se puede obtener un Hamiltoniano finito incluso si la invarianza conforme se rompe.
Una formulaci\'on precisa de la dualidad AdS/CFT \cite{Maldacena:1997re} se propuso en \cite{Witten:1998qj,Gubser:1998bc} y fue desarrollado para deformaciones multi-traza en \cite{Witten:2001ua}. Las observables en el lado de la teor\'ia del campo de la dualidad son las funciones de correlaci\'on de los operadores invariantes gauge, que est\'an compuestos de campos elementales. Cualquier campo en supergravedad $\phi$ corresponde a un operador $O$ en la teor\'ia del campo en el borde. La dualidad relaciona la funcional generadora (generatriz) de funciones de correlaci\'on del operador $O$ con la funci\'on de partici\'on de cuerdas/gravedad en el espacio AdS con las condiciones de borde que se imponen a las excitaciones en el bulk. En nuestro caso, los campos relevantes en el bulk son el gravit\'on (perturbaciones m\'etricas) y el campo escalar. Los operadores correspondientes en la teor\'ia del campo dual son el tensor de stress de energ\'ia $T_{\mu\nu}$ de la teor\'ia dual del campo y un operador escalar de dimensi\'on $\Delta$, respectivamente.
Consideremos un campo escalar masivo. Al resolver la ecuaci\'on de movimiento cerca de la frontera, se obtiene:
\begin{equation}
\phi(r)=\frac{\alpha}{r^{\Delta_{-}}}+\frac{\beta}{r^{\Delta_{+}}}+...
\end{equation}
donde $\alpha$ y $\beta$ son las componentes 
relevantes y sub-relevantes de la expansi\'on
asint\'otica del campo escalar y $\Delta_{\pm}=\frac{3}{2}\pm
\sqrt{\frac{9}{4}+ m^{2}l^{2}}$.
Dependiendo del valor de la masa, los dos modos (en la secci\'on de Lorentz) puede ser divergente o finito. Por ejemplo, para una masa cuadrado-positiva $m^{2}>0$ el modo de $\beta$ es divergente en el interior y finito en el borde y el modo $\alpha$ es divergente en el borde pero finito en el interior. Entonces, el modo $\beta$ corresponde a la fuente de las corrientes en la teor\'ia dual del borde. Por otro lado, activando el modo de $\alpha$, la geometr\'ia del bulk es modificada, mientras la estructura AdS cerca del borde puede ser preservada --- este es el tipo de deformaci\'on que nos interesa en este trabajo. Puesto la soluci\'on de gravedad en el bulk cambia, uno tiene que realizar un an\'alisis linealizado alrededor del \textit{nuevo} background para calcular las funciones de correlaci\'on. Esto es exactamente lo que sucede cuando se cambia el vac\'io en torno al cual se expande para obtener las cantidades f\'isicas. Entonces, en la teor\'ia dual, se da una situaci\'on similar: la teor\'ia del campo dual se expande alrededor de un vac\'io con valores expectaci\'on de vac\'io no triviales (VEV) para los operadores apropiados. De hecho, en en el diccionario est'andar AdS/CFT \cite{Balasubramanian:1998sn, Balasubramanian:1998de}, una soluci\'on para la gravedad en el bulk con un dilat\'on no trivial corresponde en la teor\'ia del campo dual a la inserci\'on de una fuente para un operador de dimensi\'on conforme $\Delta_{-} $, VEV $ \alpha$, y la corriente $\beta=J(x)$.

El espectro de los operadores de la teor\'ia del campo dual incluye todas las cantidades invariantes gauge, es decir, el producto de las trazas de los productos de los campos (o la suma de estos productos). Los operadores de una sola traza en la teor\'ia del campo pueden ser identificados con los estados de una sola part\'icula en AdS, mientras que los operadores multi-traza corresponden a los estados de m\'ultiples part\'iculas. La importancia de las deformaciones multi-traza desde un punto de vista del lado de la gravedad se investig\'o en \cite {Witten:2001ua, Hertog:2004dr}. Las condiciones de borde mixtas juegan un papel importante, ya que corresponden a una deformaci\'on de la acci\'on de la teor\'ia del campo dada por~\footnote{Este es un ejemplo concreto de deformaci\'on, ver secci\'on \ref{defor233}}
\begin{equation}
I_{CFT}\rightarrow I_{CFT} - \int d^{3}xW[\mathcal{O}{(x)}] \label{triple}%
\end{equation}
donde $\beta(x)=\frac{dW}{d\alpha(x)}$, y $W$ es fijada por las condiciones de borde del lado de la teor\'ia de cuerdas.   
 
\section{Contrat\'erminos y acci\'on regularizada}
\label{counterregu}
En la secci\'on \ref{backsubtrac1} mostramos
a manera de ejemplo, el an\'alisis y los c\'alculo para el agujero 
negro de Schawrzschild-AdS. En esta secci\'on
mostramos una generalizaci\'on al caso de 
campos escalares m\'inimamente acoplados a la 
gravedad. En el ap\'endice \ref{onshellapendice}
se muestran algunos detalles de los c\'alculos en $D$-dimensiones del espacio-tiempo, los cuales
se reducen a este caso especial fijando $D=3+1$.

 \subsection{Principio variacional}
 Nuestro objetivo es construir contrat\'erminos
 (t\'erminos de borde) que regularizan la acci\'on
de forma que el princ\'ipio variacional este bien definida.
Los t\'erminos de borde no cambian las ecuaciones de movimiento
y de esta forma estos pueden ser incorporados en la acci\'on.
Para conseguirlo, primero consideramos
la acci\'on (\ref{actionHamil}) cuando el campo escalar
esta apagado. En este caso, la acci\'on tiene dos t\'erminos:
la acci\'on del bulk y t\'ermino de superficie de Gibbons-Hawking necesario para asegurar que la variaci\'on de
Euler-Lagrange este bien definida. La acci\'on gravitacional 
calculada en esta forma (a\'un a tree-level) contiene divergencias que surgen por integrar sobre el volumen infinito
del espacio-tiempo. En el contexto de AdS/CFT, la divergencias 
infrarrojas (IR) de la gravedad son interpretadas como las divergencias ultravioletas (UV) de dual CFT. Es bi\'en
conocido, que la forma de calcular la acci\'on del bulk
sin introducir un background es adicionar contrat\'erminos 
locales a la acci\'on, los cuales eliminan todas las divergencias, dejando una acci\'on finita que corresponde 
a la funci\'on de partici\'on del CFT. Para gravedad AdS 
puro en cuatro dimensiones, la acci\'on debe ser suplementada con el siguiente contrat\'ermino \cite{Balasubramanian:1999re}~\footnote{Incluso para el agujero negro de Schwarzschild-AdS, ver la secci\'on \ref{backsubtrac1}}:      
  
 \begin{equation}
 I^{ct}_{g}=-\frac{1}{\kappa}\int_{\partial\mathcal{M}}{d^{3}x\sqrt
 	{-h}\biggl{(}\frac{2}{l}+\frac{\mathcal{R}l}{2}\biggr{)}}~.%
 \end{equation}
 Donde, $h_{ab}$ es la m\'etrica inducida en el borde y $\mathcal{R}$ es el escalar de Ricci.
 En presencia del campo escalar, este contrat\'ermino
 no es suficiente para cancelar las divergencias en la acci\'on.
 Para este caso, un t\'ermino de borde adicional que depende del campo escalar es necesario, concretamente $I_{\phi}$.
 Estudiaremos el principio variacional de la siguiente acci\'on: 
 \begin{equation}
 I=\int{d^{4}x\sqrt{-g}\biggl{(}\frac{R}{2\kappa}-\frac{(\partial\phi)^{2}}%
 	{2}-V(\phi)\biggr{)}}+\frac{1}{\kappa}\int_{\partial\mathcal{M}}{d^{3}%
 	x\sqrt{-h}K}-\frac{1}{\kappa}\int_{\partial\mathcal{M}}{d^{3}x\sqrt
 	{-h}\biggl{(}\frac{2}{l}+\frac{\mathcal{R}l}{2}\biggr{)}} + I_{\phi}
 \label{Icomplete}%
 \end{equation}
 para un campo escalar de masa conforme $m=-2l^{-2}$. En algunos
 trabajos previos (por ejemplo, ver \cite{Lu:2013ura, Lu:2014maa}), el siguiente contrat\'ermino que produce una acci\'on finita para la rama no-logar\'itmica fue propuesto:
 \begin{equation}
 I_{\phi}^{ct} = \frac{1}{6\kappa}\int_{\partial\mathcal{M}}{d^{3}x\sqrt
 	{-h}\biggl{(}\phi n^{\nu}\partial_{\nu}\phi-\frac{\phi^{2}}{2l}\biggr{)}}~.
 \label{phict}%
 \end{equation}
 Sin embargo, esta es problem\'atica porque no es intr\'inseco 
 al borde y tambi\'en, para condiciones de borde mixtas, el principio variacional no se satisface. 
 En su reemplazo, proponemos nuevos contrat\'erminos
 para las ramas, logar\'itmica y no-logar\'itmica, de forma
 que la acci\'on es finita y el principio variacional
 esta bien definido $\delta I=0$. Los contrat\'erminos 
 intr\'insecos son construidos de tal forma que sean
 compatibles con el m\'etodo Hamiltoniano, en el sentido
 en el que los resultados concuerdan para cualquier condici\'on de borde. Comenzaremos con la rama no-logar\'itmica con el t\'ermino de borde asociado al campo escalar dado por    
 \begin{equation}
 I_{\phi}=-\int_{\partial\mathcal{M}}{d^{3}x\sqrt{-h}\biggl{[}\frac{\phi^{2}%
 	}{2l}+\frac{W(\alpha)}{l\alpha^{3}}\phi^{3}\biggr{]}}~. \label{ctphi}%
 \end{equation}
 Entonces, usando la expansi\'on en el borde de la m\'etrica
 y el campo escalar, la variaci\'on de la acci\'on da un t\'ermino de borde evaluada en el cut-off $r$:
 \begin{equation}
 \delta I=\int{d^{3}x\sqrt{-h}\biggl{[}\frac{1}{r}\biggl{(}-\sqrt{g^{rr}}%
 	\phi^{^{\prime}}-\frac{\phi}{l}-\frac{3W(\alpha)\phi^{2}}{l\alpha^{3}%
 	}\biggr{)}\biggl{(}1+\frac{1}{r}\frac{d^{2} W(\alpha)}{d\alpha^{2}%
 }\biggr{)}+\biggl{(}\frac{3W(\alpha)}{\alpha}-\beta\biggr{)}\frac{\phi^{3}%
}{l\alpha^{3}}\biggr{]}}\delta\alpha.
\end{equation}
Es f\'acil mostrar que el princ\'ipio variacional 
esta bien definido cuando el cut-off tiende al infinito:
\begin{equation}
\lim_{r\to\infty}\delta I=0~.
\end{equation}
Para la rama logar\'itmica debemos trabajar con el siguiente
contrat\'ermino para el campo escalar:
\begin{equation}
I_{\phi}=-\int_{\partial\mathcal{M}}{d^{3}x\sqrt{-h}\biggl{[}\frac{\phi^{2}%
	}{2l}+\frac{\phi^{3}}{l\alpha^{3}}\biggl{(}W-\frac{\alpha\gamma}%
	{3}\biggr{)}-\frac{\phi^{3}C_{\gamma}}{3l}\ln\biggl{(}\frac{\phi}{\alpha
	}\biggr{)}\biggr{]}}~.%
\end{equation}
Usando la expansi\'on asint\'otica de la m\'etrica y el campo
escalar, tambi\'en se muestra que el principio variacional
esta bien definido para condiciones de borde arbitrarias
cuando el cut-off en el t\'ermino de superficie tiende a 
infinito.

 \subsection{Acci\'on regularizada y variables termodin\'amicas}
 Evaluando la acci\'on nos da resultados formalmente divergentes.
 nos gustar\'a mostrar que, de hecho, que todas la divergencias pueden
 ser eliminadas usando los contrat\'erminos propuestos en la 
 secci\'on previa, de tal forma que la acci\'on queda finita. Usaremos
 la t\'ecnica est'andar de la rotaci\'on de Wick de la 
 direcci\'on temporal $t=i\tau$. Entonces, la temperatura
 esta relacionada a la periodicidad del tiempo Eucl\'ideo 
 $\tau$ ($\Delta\tau=\beta=1/T$) y la contribuci\'on relevante
 a la energ\'ia libre esta determinada por la evaluaci\'on
 de la acci\'on Eucl\'idea. 
 La acci\'on tiene cuatro t\'erminos,
 la parte del bulk $I^{E}_{bulk}$, el t\'ermino de superficie
 Gibbons-Hawking $I^{E}_{GH}$, y dos 
 contrat\'erminos del borde ($I_{g}^{ct}$,
 $I^{ct}_{\phi}$): $I= I^{E}_{bulk} + I^{E}_{GH} + I_{g}^{ct} + I_{\phi}^{ct}$. Primero calculamos estas contribuciones a la 
 rama no-logar\'itmica.
 
 Ya que estudiaremos la propiedades de la familia de soluciones exactas de agujeros negros con 
 pelo\footnote{Ver la secci\'on \ref{soluhair}}, comenzamos con el 
 siguiente ansatz para la m\'etrica 
 \begin{equation}
 ds^{2}=\Omega(x)\left[  -f(x)dt^{2}+\frac{\eta^{2}dx^{2}}{f(x)}+d\Sigma
 _{k}^{2}\right]~.  \label{Ansatz}%
 \end{equation}
 Expresiones concretas para las funciones $\Omega(x)$ y $f(x)$ fueron presentadas en la secci\'on \ref{soluhair}.
 Los c\'alculos en el $(t,x,\Sigma)$ sistema de coordenadas (\ref{Ansatz}) est'an relacionados por una 
 simple transformaci\'on de coordenadas al sistema 
  (\ref{Ansatz1}). En lo que sigue, $x_{b}$ y $r_{b}$ 
  denotan el borde y $x_{+}$, $r_{+}$ el horizonte. La 
  acci\'on Euclidea on-shell del bulk puede ser escrita
  como  
 \begin{equation}
 I_{bulk}^{E}=\int_{0}^{1/T}{d\tau}\int_{x_{+}}^{x_{b}}{d^{3}x\sqrt{g^{E}}%
 	V}\left(  \phi\right)  =\frac{\sigma_{k}}{2\eta\kappa T}\frac{d(\Omega f)}%
 {dx}\biggr{\vert}_{x_{+}}^{x_{b}}~,%
 \end{equation}
 donde $\sigma_{k}$ es el \'area de $\Sigma_{k}$ (por ejemplo, para $k=1$, $\sigma_{1}=4\pi$)\footnote{Para los casos de $k=-1,1$ se trabaja con densidades, por ejemplo la densidad de energ\'ia.} y $g^{E}$ es la 
 m\'etrica en la secc\'on Euclidea. Los dos sistemas
 de coordenadas $(t,x,\Sigma_{k})$ y $(t,r,\Sigma_{k})$
 estan relacionadas por 
 \begin{equation}
 \Omega(x)\rightarrow S(r)~,\qquad f(x)\rightarrow\frac{N(r)}{S(r)}~,\qquad
 dx\rightarrow\frac{\sqrt{NH}}{\eta S}dr\label{Changeco}%
 \end{equation}
 y de esta forma, podemos reescribir la integral 
del bulk en el sistema de coordenadas $(t,r,\Sigma_{k})$
como
 \begin{equation}
 I_{bulk}^{E}=\frac{\sigma_{k}}{2\kappa T}\frac{S}{\sqrt{NH}}\frac{dN}%
 {dr}\biggr{\vert}_{r_{+}}^{r_{b}}~.%
 \end{equation}
 Ahora calculamos el termino de Gibbons-Hawking. 
 Consideramos una hiper-superficie tipo-tiempo (timelike) 
 $x=x_{0}$. Entonces, la m\'etrica inducida $h^{\mu\nu}=g^{\mu\nu}-n^{\mu}n^{\nu}$, normal, curvatura
 extr\'inseca, y su traza $K=h^{\mu\nu}K_{\mu\nu}$ son
  \begin{equation}
 ds^{2}=h_{ab}dx^{a}dx^{b}=\Omega(x_{0})\biggl{[}-f(x_{0})dt^{2}+d\Sigma
 _{k}\biggr{]}~,
 \end{equation}%
 \begin{equation}
 n_{a}=\frac{\delta_{a}^{x}}{\sqrt{g^{xx}}}\biggr{\vert}_{x=x_{0}}~,\qquad
 K_{ab}=\frac{\sqrt{g^{xx}}}{2}\partial_{x}g_{ab}\biggr{\vert}_{x=x_{0}}~,\qquad
 K=\frac{1}{2\eta}\biggl{(}\frac{f}{\Omega}\biggr{)}^{1/2}\biggl{[}\frac
 {(\Omega f)^{^{\prime}}}{\Omega f}+\frac{2\Omega^{^{\prime}}}{\Omega
 }\biggr{]}\biggr{\vert}_{x_{0}}.\label{nkk}%
 \end{equation}
 Usando las ecuaciones de transformaci\'on (\ref{Changeco}), la contribuci\'on de este
 t\'ermino puede ser reescrita como 
  \begin{equation}
 I_{GH}^{E}=-\frac{\sigma_{k}}{\kappa T}\frac{\Omega f}{2\eta}\biggl{[}\frac
 {(\Omega f)^{^{\prime}}}{\Omega f}+\frac{2\Omega^{^{\prime}}}{\Omega
 }\biggr{]}\biggr{\vert}_{x_{b}}=-\frac{\sigma_{k}}{2T\kappa}\biggl{(}\frac
 {S}{\sqrt{NH}}\frac{dN}{dr}+\frac{2N}{\sqrt{NH}}\frac{dS}{dr}%
 \biggr{)}\biggr{\vert}_{r_{b}}~.%
 \end{equation}
 La contribuci\'on del contrat\'ermino gravitacional
 es
 \begin{equation}
 I_{g}^{ct}=\frac{2\sigma_{k}}{\kappa Tl}\biggl{(}\Omega^{3/2}f^{1/2}%
 +\frac{l^{2}k}{2}f^{1/2}\Omega^{1/2}\biggr{)}\biggr{\vert}_{x_{b}}%
 =\frac{2\sigma_{k}}{\kappa Tl}S\sqrt{N}\biggl{(}1+\frac{l^{2}k}{2S}%
 \biggr{)}\biggr{\vert}_{r_{b}}~.%
 \end{equation}
 Usando la f\'ormula general para la temperatura
 \begin{equation}
 T=\frac{N^{^{\prime}}}{4\pi\sqrt{NH}}\biggr{\vert}_{r_{+}}~,%
 \end{equation}
 uno puede escribir la suma de estas contribuciones
 en la acci\'on total como
  \begin{equation}
 I_{bulk}^{E}+I_{GH}^{E}+I^{ct}_{g}=-\frac{1}{T}\biggl{[}\frac{\sigma
 	_{k}S(r_{+})T}{4G}\biggr{]}-\frac{\sigma_{k}}{2\kappa T}\biggl{[}\frac
 {2N}{\sqrt{NH}}\frac{dS}{dr}-\frac{4}{l}S\sqrt{N}\biggl{(}1+\frac{l^{2}k}%
 {2S}\biggr{)}\biggr{]}\biggr{\vert}_{r_{b}}.~ \label{III}%
 \end{equation}
 Para el fall-off del campo escalar de masa conforme 
 $m^{2}=-2l^{-2}$ dada en (\ref{phi}) y con el fall-off 
 de la m\'etrica (\ref{gtt}) y (\ref{grr}), la acci\'on Euclidea viene a ser
 \begin{equation}
 I_{bulk}^{E}+I_{GH}^{E}+I^{ct}_{g}=-\frac{\mathcal{A}}{4G }-\frac{\sigma_{k}}{T}\biggl{(}-\frac{\mu}{\kappa}%
 +\frac{4\alpha\beta}{3l^{2}}+\frac{r\alpha^{2}}{2l^{2}}%
 \biggr{)}\biggr{\vert}_{r_{b}}~.%
 \end{equation}
 Aqui, $\mathcal{A}=\sigma_{k}S(r_{+})$ es el
 \'area del horizonte.
 Ahora es claro que el contrat\'ermino gravitacional
 no es suficiente para eliminar las divergencias en el borde $r_{b}\rightarrow\infty$, pero esta nueva divergencia lineal puede ser regularizada con el 
 siguiente contrat\'ermino que depende del campo
 escalar:
 \begin{equation}
 I_{\phi}^{ct}=\int_{\mathcal{\partial M}}{d^{3}x\sqrt{h^{E}}\biggl{[}\frac
 	{\phi^{2}}{2l}+\frac{W(\alpha)}{l\alpha^{3}}\phi^{3}\biggr{]}}=\frac
 {\sigma_{k}}{T}\biggl{(}\frac{W}{l^{2}}+\frac{\alpha\beta}{l^{2}}%
 +\frac{r\alpha^{2}}{2l^{2}}\biggr{)}\biggr{\vert}_{r_{\infty}}~. \label{Iphi}%
 \end{equation}
 La acci\'on Euclidea renormalizada puede ser reescrita
 usando $\beta=dW/d\alpha$ como
 \begin{equation}
 I^{E}=I_{bulk}^{E}+I_{GH}^{E}+I_{g}^{ct}+I_{\phi}^{ct} =-\frac{\mathcal{A}%
 }{4G} + \frac{\sigma_{k}}{T}\biggl{[}\frac{\mu}{\kappa}+\frac{1}{l^{2}%
}\biggl{(}W-\frac{\alpha}{3}\frac{dW}{d\alpha}\biggr{)}\biggr{]}
\label{Fnonlog}%
\end{equation}
y de esta forma, la energ\'ia libre viene a ser 
\begin{equation}
F=I^{E}T=M - TS~.
\end{equation}
Las relaciones termodin\'amicas nos dan la misma 
masa y entrop\'ia para los agujeros negros:
\begin{equation}
M=-T^{2}\frac{\partial I^{E}}{\partial T}=\sigma_{k}\biggl{[}\frac{\mu}%
{\kappa}+\frac{1}{l^{2}}\biggl{(}W-\frac{\alpha}{3}\frac{dW}{d\alpha
}\biggr{)}\biggr{]}
\end{equation}
y
\begin{equation}
S=-\frac{\partial(I^{E}T)}{\partial T}=\frac{\mathcal{A}}{4G}~.%
\end{equation}
Un c\'alculo similar puede ser realizado para la 
rama logar\'itmica. Trabajaremos de nuevo
con un campo escalar de masa conforme $m^{2}=-2l^{-2}$
con un fall-off (\ref{philog}) para el cual
la m\'etrica fall-off es (\ref{grrlog}).
Si trabajamos con el contrat\'ermino (\ref{Iphi}),
obtenemos
\begin{equation}
I_{bulk}^{E}+I_{GH}^{E}+I^{ct}_{g}+I_{\phi}^{ct}=-\frac{\mathcal{A}}{4G}
+\frac{\sigma_{k}}{T}\biggl{\lbrace}\frac{\mu}{\kappa}+\frac{1}{l^{2}%
}\biggl{[}W(\alpha)-\frac{\alpha}{3}\frac{dW}{d\alpha}+\frac{2\alpha\gamma}%
{9}-\frac{\alpha\gamma}{3}\ln{r}\biggr{]}\biggr{\rbrace}
\end{equation}
y vemos que a\'un hay una divergencia logar\'itmica.
Por consiguiente, uno debe considerar una nueva 
contribuci\'on del campo escalar que cancele 
esa divergencia:
\begin{equation}
\bar{I}^{\,\,ct}_{\phi}=\int_{\partial\mathcal{M}}{d^{3}x\sqrt{h^{E}%
	}\biggl{\lbrace}\frac{\phi^{3}\gamma}{3\alpha^{2}l}\biggl{[}\ln\biggl{(}\frac
	{\alpha}{\phi}\biggr{)}-1\biggr{]}\biggr{\rbrace}}=\frac{\sigma_{k}}%
{T}\biggl{[}-\frac{\alpha\gamma}{3l^{2}}+\frac{\alpha\gamma\ln{r}}{3l^{2}%
}+O(r^{-1}\ln{r})\biggr{]}~. \label{newlogct}%
\end{equation}
Tambi\'en obtenemos una acci\'on finita para la
rama logar\'itmica 
\begin{equation}
I^{E} =I_{bulk}^{E}+I_{GH}^{E}+I^{ct}_{g}+I_{\phi}^{ct}+\bar{I}^{\,\,ct}%
_{\phi} = -\frac{\mathcal{A}}{4G}+\frac{\sigma_{k}}{T}\biggl{[}\frac{\mu
}{\kappa}+\frac{1}{l^{2}}\biggl{(}W-\frac{\alpha}{3}\frac{dW}{d\alpha}%
-\frac{\alpha\gamma}{9}\biggr{)}\biggr{]}~, \label{Flog}%
\end{equation}
donde $\gamma=C_{\gamma}\alpha^{2}$ y
$C_{\gamma}=-3l^{2}\lambda$.
Las relaciones termodin\'amicas dan resultados
correctos:
\begin{equation}
M=-T^{2}\frac{\partial I^{E}}{\partial T}=\sigma_{k}\biggl{[}\frac{\mu}{\kappa}+\frac
{1}{l^{2}}\biggl{(}W-\frac{\alpha}{3}\frac{dW}{d\alpha}-\frac{\alpha\gamma}%
{9}\biggr{)}\biggr{]}
\end{equation}
y
\begin{equation}
S=-\frac{\partial(I^{E}T)}{\partial T}=\frac{\mathcal{A}}{4G}~.%
\end{equation}
La simetr\'ia conforme del borde es preservada 
cuando
 $W(\alpha)=\alpha
^{3}(C+l^{2}\lambda\ln{\alpha})$.
\section{Tensor de stress Brown-York}
En la secci\'on \ref{BrownYork}, presentamos
el formalismo de Brown-York y sus interpretaciones en la teor\'ia dual. A modo de ejemplo, mostramos
los c\'alculos para el agujero negro Schwarzschild-AdS.
En esta secci\'on estudiaremos la generalizaci\'on de 
cuando los campos escalares est\'an presentes.

En la secci\'on anterior vimos que el campo
escalar tiene contribuci\'on a la acci\'on
la misma que nos da un buen principio
variacional $\delta I=0$. De esta acci\'on
suplementada con los t\'erminos de borde del campo
escalar, construiremos el 
tensor de stress de Borwn-York, la cual
tiene contribuci\'on del campo escalar. 
 
Un procedimiento similar puede ser usado para los agujeros
negros con pelo, pero uno de be a\~nadir contrat\'erminos 
del borde relacionados con el campo escalar. En el caso
de la rama no-logar\'itmica, la acci\'on completa es (\ref{Icomplete}) 
y el contrat\'ermino escalar fue dada en (\ref{phict}), donde
 $G_{ab}$ es el tensor de Einstein para la folicaci\'on   
 (\ref{k1induced}) dado por $G_{ab}=\delta_{a}
 ^{t}\delta_{b}^{t}Nk/S$. El tensor de stress regularizado
 es 
 \begin{equation}
 \tau_{ab}=-\frac{1}{\kappa}\biggl{(}K_{ab}-h_{ab}K+\frac{2}{l}h_{ab}%
 -lG_{ab}\biggr{)}-\frac{h_{ab}}{l}\biggl{[}\frac{\phi^{2}}{2}+\frac{W(\alpha
 	)}{\alpha^{3}}\phi^{3}\biggr{]}~.\label{BY1}%
 \end{equation}
 Entonces, las componentes del tensor de stress son 
 \begin{align}
 \tau_{tt} &  =\frac{l}{R}\biggl{[}\frac{\mu}{8\pi G l^{2}}+\frac{1}{l^{4}%
 }\biggl{(}W-\frac{\alpha\beta}{3}\biggr{)}\biggr{]}+O(R^{-2})~,\\
 \tau_{\theta\theta} &  =\frac{l}{R}\biggl{[}\frac{\mu}{16\pi G}-\frac
 {1}{l^{2}}\biggl{(}W-\frac{\alpha\beta}{3}\biggr{)}\biggr{]}+O(R^{-2}%
 )~,\nonumber\\
 \tau_{\phi\phi} &  =\frac{l\sin^{2}{\theta}}{R}\biggl{[}\frac{\mu}{16\pi
 	G}-\frac{1}{l^{2}}\biggl{(}W-\frac{\alpha\beta}{3}%
 \biggr{)}\biggr{]}+O(R^{-2})~.\nonumber
 \end{align}
 El tensor de stress de la teor\'ia dual del campo puede ser puesta
 en la forma similar a la del agujero negro de Schwarzschild-AdS:  
 \begin{equation}
 \langle\tau_{ab}^{dual}\rangle=\frac{3\mu}{16\pi G l^{2}}\delta_{a}%
 ^{0}\delta_{b}^{0}+\frac{\gamma_{ab}}{l^{2}}\biggl{[}\frac{\mu}{16\pi G %
 }-\frac{1}{l^{2}}\left(  W(\alpha)-\frac{\alpha\beta}{3}\right)  \biggr{]}~.
 \end{equation}
 La traza puede ser f\'acilmente calculada y obtenemos
 \begin{equation}
 \langle\tau^{dual}\rangle=-\frac{3}{l^{4}}\biggl{[}W(\alpha)-\frac{\alpha
 	\beta}{3}\biggr{]}~.
 \end{equation}
A diferencia del agujero negro de Schwarzschild-AdS, para los agujeros negros con pelo existen diferentes tipos de condiciones de borde, 
concretamente esta preserva o no la simetr\'ia conforme. Como se esperaba, cuando la simetr\'ia conforme es preservada, $W=C\alpha^{3}$,
la traza del tensor de stress desaparece $\langle
\tau^{dual}\rangle=0$ y no hay anomal\'ia conforme. 
Un procedimiento similar pero m\'as complicado puede ser aplicado 
a la rama lograr\'itmica. La acci\'on (\ref{Icomplete}) tiene
una nueva contribuci\'on (\ref{newlogct}) que cancela la divergencia 
logar\'itmica, y el nuevo tensor de stress cuasilocal  regularizado es 
 \begin{equation}
 \tau_{ab}=-\frac{1}{\kappa}\biggl{(}K_{ab}-h_{ab}K+\frac{2}{l}h_{ab}%
 -lG_{ab}\biggr{)}-\frac{h_{ab}}{l}\biggl{[}\frac{\phi^{2}}{2}+\frac{\phi^{3}%
 }{\alpha^{3}}\biggl{(}W-\frac{\alpha\gamma}{3}\biggr{)}+\frac{\phi^{3}\gamma}{3\alpha^{2}}\ln{\biggl{(}\frac{\alpha}{\phi}\biggr{)}}\biggr{]}~,\label{BY2}%
 \end{equation}
cuyas componentes son
 \begin{align}
 \tau_{tt} &  =\frac{l}{R}\biggl{[}\frac{\mu}{8\pi G l^{2}}+\frac{1}{l^{4}%
 }\biggl{(}W-\frac{\alpha\beta}{3}-\frac{\alpha\gamma}{9}%
 \biggr{)}\biggr{]}+O\biggl{[}\frac{(\ln{R})^{3}}{R^{2}}\biggr{]}~,\\
 \tau_{\theta\theta} &  =\frac{l}{R}\biggl{[}\frac{\mu}{16\pi G }-\frac
 {1}{l^{2}}\biggl{(}W-\frac{\alpha\beta}{3}-\frac{\alpha\gamma}{9}%
 \biggr{)}\biggr{]}+O\biggl{[}\frac{(\ln{R})^{3}}{R^{2}}\biggr{]}~,\nonumber\\
 \tau_{\phi\phi} &  =\frac{l\sin^{2}{\theta}}{R}\biggl{[}\frac{\mu}{16\pi
 	G }-\frac{1}{l^{2}}\biggl{(}W-\frac{\alpha\beta}{3}-\frac{\alpha\gamma}%
 {9}\biggr{)}\biggr{]}+O\biggl{[}\frac{(\ln{R})^{3}}{R^{2}}\biggr{]}.\nonumber
 \end{align}
 El tensor de stress de la teor\'ia dual viene a ser
 \begin{equation}
 \langle\tau_{ab}^{dual}\rangle=\frac{3\mu}{16\pi G l^{2}}\delta_{a}%
 ^{0}\delta_{b}^{0}+\frac{\gamma_{ab}}{l^{2}}\biggl{[}\frac{\mu}{16\pi G %
 }-\frac{1}{l^{2}}\left(  W(\alpha)-\frac{\alpha\beta}{3}-\frac{\alpha\gamma}{9}\right)  \biggr{]}~.
 \end{equation}
 Su traza es
 \begin{equation}
 \langle\tau^{dual}\rangle=-\frac{3}{l^{4}}\biggl{(}W-\frac{\alpha\beta}%
 {3}-\frac{\alpha\gamma}{9}\biggr{)}
 \end{equation}
 y, como se esperaba, esta desaparece para las condiciones de borde
 que preserva la condiciones de borde que preserva la simetr\'ia conforme:
 \begin{equation}
 \langle\tau^{dual}\rangle=0\Rightarrow\gamma=-3l^{2}\lambda\alpha^{2}~,\qquad
 W(\alpha)=\alpha^{3}[C+l^{2}\lambda\ln{\alpha}]~.
 \end{equation} 
  
\section{Masa Hamiltoniana}
Mediante el formalismo Hamiltoniano
calculamos la masa o energ\'ia 
del agujero negro y mostraremos
la contribuci\'on no-trivial del
campo escalar. Nos centramos en el 
caso concreto de teor\'ias campos escalares
m\'inimamente acoplados a la gravedad
y de masa conforme: $m^{2}=-2/l^{2}$.
Esta masa conforme 
es mayor que la masa de BF (Breitenlohner-Freedman)
$m_{BF}^{2}=-9/4l^{2}$ y esta por debajo de la cota
unitaria: $m_{BF}^{2}+1/l^{2}=-5/4l^{2}$,

\begin{equation}
-5/4l^{2}>m^{2}\geqslant -9/4l^{2}~.
\end{equation}  

De esta forma ambos modos de la expansi\'on (en la vecindad
del borde $r\rightarrow\infty$) del 
campo escalar son normalizables

\begin{equation}
\phi(r)=\frac{\alpha}{r}+\frac{\beta}{r^{2}}+O(r^{-3})~.
\end{equation}

Mostramos la equivalencia de la masa
hologr\'afica (obtenida mediante el m\'etodo
de renormalizaci\'on hologr\'afica) y la masa Hamiltoniana.
Finalmente vemos una aplicaci\'on concreta a la soluci\'on exacta de agujero negro con pelo.

\newpage
 \subsection{Rama logar\'itmica y no-logar\'itmica}

En la expansi\'on del potencial como una serie de potencias en torno a $\phi=0$, se demostr\'o en \cite{Henneaux:2006hk} la ausencia de ramas logar\'itmicas en el comportamiento asint\'otico de la m\'etrica y del campo escalar, es debido a que la serie no contiene un t\'ermino c\'ubico. Este conjunto de condiciones asint\'oticas permite soluciones exactas de agujeros negros escalares \cite{Acena:2013jya,Anabalon:2013eaa,Martinez:2004nb}. El fall-off del campo escalar y la m\'etrica en el infinito se obtuvo en la secci\'on \ref{sec2}.

Ahora, evaluamos las expresiones generales (\ref{eq:Q_G}) y (\ref{eq:Q_phi}) para configuraciones est\'aticas, usando
las anteriores condiciones asint\'oticas. Consideramos
el borde localizado en $r=\infty$. Integrando las 
`coordenadas angulares', obtenemos la contribuci\'on 
gravitacional
\begin{equation}
\delta M_{G}=\frac{\sigma_{k}}{\kappa}[r\delta a+l\delta b+O(1/r)]
\label{eq:delta_mg}%
\end{equation}
y la contribuci\'on del campo escalar
\begin{equation}
\delta M_{\phi}=\frac{\sigma_{k}}{l^{2}}[r\alpha\delta\alpha+\alpha\delta
\beta+2\beta\delta\alpha+O(1/r)]~. \label{eq:delta_mphi}%
\end{equation}
Mediante la adici\'on de ambas contribuciones tenemos la
variaci\'on de la masa
\begin{equation}
\delta M=\frac{\sigma_{k}}{\kappa l^{2}}[r(l^{2}\delta a+\kappa\alpha
\delta\alpha)+l^{3}\delta b+\kappa(\alpha\delta\beta+2\beta\delta
\alpha)+O(1/r)]~. \label{varmass}%
\end{equation}
Es importante recordar que esta expresi\'on para $\delta M$ tiene sentido s\'olo en el caso  
cuando se cumplen las ligaduras Hamiltonianas. En el caso
est\'atico, solo hay una ligadura no trivial, $H_{\perp}=0$,
el cual para las condiciones asint\'oticas dadas anteriormente 
obtenemos
\begin{equation}
\frac{k+a}{\kappa}+\frac{\alpha^{2}}{2l^{2}}=0~. \label{a}%
\end{equation}
La divergencia lineal en (\ref{varmass}) es removida
por el reemplazo de (\ref{a}) en (\ref{varmass}).
Entonces, la variaci\'on asint\'otica de la masa
queda finita
\begin{equation}
\delta M=\frac{\sigma_{k}}{\kappa l^{2}}[l^{3}\delta b+\kappa(\alpha
\delta\beta+2\beta\delta\alpha)]~. \label{varmassfin}%
\end{equation}
Para integrar las variaciones en (\ref{varmassfin}), las 
condiciones de borde para el campo escalar son necesarias.
En particular, la integraci\'on de (\ref{varmassfin})
requiere una relaci\'on funcional entre $\alpha$ y $\beta$.
 Si definimos $\beta=dW(\alpha)/d\alpha$, la masa del 
 espacio tiempo est\'a dada por
\begin{equation}
\label{eq:mass1}M=\sigma_{k}\left[  \frac{l b}{\kappa}+\frac{1}{l^{2}}\left(
\alpha\frac{dW(\alpha)}{d\alpha}+W(\alpha)\right)  \right]~.
\end{equation}
Observamos que en la masa (\ref{eq:mass1}) se define salvo una constante. Esta constante se fija a cero con el fin de fijar una masa cero para el espacio-tiempo AdS a nivel local debido a que en cuatro dimensiones no hay energ\'ia Casimir.

Para obtener la rama logar\'itmica, es necesario usar el 
potencial auto-interactuante (\ref{vphilog}) de tal
forma que el campo escalar
a considerar es (\ref{philog}). La ligadura Hamiltoniana
$H_{\perp}=0$ se satisface si (\ref{a}) y
\begin{equation}
\frac{lc}{\kappa}-4\alpha^{3}\lambda=0 \label{d}%
\end{equation}
se cumplen.
Ahora, evaluamos (\ref{eq:Q_G}) y (\ref{eq:Q_phi}). En 
este caso encontramos
\begin{equation}
\label{mgl}\delta M_{G}=\biggr{\lbrace} \frac{l\delta b}{\kappa}+\frac{\delta
	a}{\kappa} r+\frac{l\delta c}{\kappa} \ln(r)+O\left(  \frac{\ln(r)^{2}}%
{r}\right)  \biggl{\rbrace} \sigma_{k}%
\end{equation}
y
\begin{align}
\label{mpl}\delta M_{\phi}  &  =\left[  \frac{\alpha\delta\beta+2\beta
	\delta\alpha+3\alpha^{2}l^{2}\lambda\delta\alpha}{l^{2}}+r\frac{\alpha
	\delta\alpha}{l^{2}}\right. \nonumber\\
&  \left.  -12\lambda\alpha^{2}\delta\alpha\ln(r)+O\left(  \frac{\ln(r)^{2}%
}{r}\right)  \right]  \sigma_{k}~.%
\end{align}
Ambas contribuciones contienen divergencias lineales 
y logar\'itmicas. Sumando (\ref{mgl}) y (\ref{mpl}),
las divergencias lineales se cancelan en virtud de (\ref{a})
 y las divergencias logar\'itmicas desaparecen considerando 
 (\ref{d}). De este modo, obtenemos una expresi\'on finita para la variaci\'on de la masa,
\begin{equation}
\delta M=\left[  \frac{l\delta b}{\kappa}+\frac{\alpha\delta\beta+2\beta
	\delta\alpha+3\alpha^{2}l^{2}\lambda\delta\alpha}{l^{2}}\right]  \sigma_{k}~.%
\end{equation}
Nuevamente, necesitamos condiciones de borde, 
una relaci\'on funcional entre $\alpha$ y $\beta$,
 para integrar $\delta M$. Consideramos
 la relaci\'on general $\beta=\frac{dW}{d\alpha}$,
 de forma que la masa Hamiltoniana esta dada por
\begin{equation}
M=\left[  \frac{lb}{\kappa}+\frac{1}{l^{2}}\left(  \alpha\frac{dW}{d\alpha
}+W(\alpha)+\alpha^{3}l^{2}\lambda\right)  \right]  \sigma_{k}~.%
\end{equation}
La masa puede estar relacionada con el primer t\'ermino
sub-relevante de $g_{tt}$ usando (\ref{abglog}). 
As\'i, la masa se puede ser escrita como
\begin{equation}
M=\left[  \frac{\mu}{\kappa}+\frac{1}{l^{2}}\left(  W(\alpha)-\frac{1}%
{3}\alpha\frac{dW}{d\alpha}+\frac{1}{3}\alpha^{3}l^{2}\lambda\right)  \right]
\sigma_{k}~.%
\end{equation}
Por lo tanto, la expresi\'on $M=\mu\sigma_{k}\kappa^{-1}$
es obtenida s\'olo para $\alpha=0$ o 
\begin{equation}
W(\alpha)=\alpha^{3}\left[  C+l^{2}\lambda\ln(\alpha)\right]~,
\end{equation}
que corresponden con las condiciones de borde AdS invariantes
\cite{Henneaux:2006hk}.

\subsection{Equivalencia entre masa hologr\'afica y masa Hamiltoniana}

Armado con el formalismo de Brown-York suplementado con contrat\'erminos, se puede obtener la energ\'ia de agujeros negros con pelo. La m\'etrica del borde se puede escribir, al menos localmente, en forma similar a ADM. Siempre y cuando la geometr\'ia del borde tiene una isometr\'ia generada por el vector de Killing $\xi^{a}=(\partial_{t})^{a}$, la energ\'ia es, de forma usual, la carga conservada.
En concreto, vamos a utilizar las coordenadas $(t,r,\sigma_{k})$ con la m\'etrica (\ref{Ansatz1}) y la foliaci\'on (\ref{k1induced}) parametrizada como
\begin{equation}
d\Sigma_{k}^{2}=\frac{dy^{2}}{1-ky^{2}}+(1-ky^{2})d\phi^{2}~.%
\end{equation}
La energ\'ia
\begin{equation}
E=\int{d\sigma^{i}\tau_{ij}\xi^{j}}=\int{dyd\phi Su^{i}\tau_{ij}\xi^{j}}~,%
\end{equation}
est\'a asociada con la superficie $t=constante$, para el cual la m\'etrica inducida es
\begin{equation}
ds^{2}=\sigma_{ij}dx^{i}dx^{j}=Sd\Sigma_{k}^{2}~,%
\end{equation}
con el vector normal $u^{a}=N^{-1/2}(\partial_{t})^{a}$.
Para la rama no-logar\'itmica, usando el tensor de stress quasilocal
(\ref{BY1}), se obtiene
\begin{equation}
E=\sigma_{k}\biggl{[}\frac{\mu}{\kappa}+\frac{1}{l^{2}}\biggl{(}W-\frac
{\alpha}{3}\frac{dW}{d\alpha}\biggr{)}\biggr{]}~.
\end{equation}
Con un c\'alculo similar para la rama logar\'itmica, pero con el tensor de stress quasilocal (\ref {BY2}), se obtiene la siguiente energ\'ia del agujero negro con pelo:
\begin{equation}
E=\sigma_{k}\biggl{[}\frac{\mu}{\kappa}+\frac{1}{l^{2}}\biggl{(}W-\frac{1}%
{3}\alpha\frac{dW}{d\alpha}-\frac{\alpha\gamma}{9}\biggr{)}\biggr{]} =
\sigma_{k}\biggl{[}\frac{\mu}{\kappa}+\frac{1}{l^{2}}\biggl{(}W-\frac{1}%
{3}\alpha\frac{dW}{d\alpha}-\frac{\alpha^{3}C_{\gamma}}{9}\biggr{)}\biggr{]}~.
\end{equation}
Esto muestra un perfecto acuerdo con la masa Hamiltoniana incluso si la simetr\'ia conforme se rompe en el borde --- con ambos m\'etodos es posible obtener una energ\'ia finita y los resultados coinciden. 
Sin embargo, la prescripci\'on AMD \cite{Ashtekar:1999jx} para el c\'alculo de la masa de un espacio-tiempo con pelo no es adecuado cuando el campo escalar rompe la invarianza asint\'otica anti-de Sitter \cite{Anabalon:2014fla}.

\subsection*{Ejemplo para una soluci\'on exacta}
Como ejemplo concreto, discutimos las condiciones de borde y algunas propiedades hologr\'aficas de las soluciones exactas de \cite{Acena:2013jya, Anabalon:2013eaa}. Consideramos el siguiente potencial escalar mostrada anteriormente en la secci\'on \ref{soluhair}, y para evitar confusiones cambiamos de notaci\'on $\alpha\rightarrow\Upsilon$. Para algunos valores particulares del par\'ametro $\Upsilon$, se convierte en una de las truncaciones de $\omega$-deformed gauged $\mathcal{N}$=8 supergravedad \cite{Anabalon:2013eaa, Guarino:2013gsa, Tarrio:2013qga}:
\begin{align}
V(\phi)  &  =\frac{\Lambda(\nu^{2}-4)}{6\kappa\nu^{2}}\biggl{[}\frac{\nu
	-1}{\nu+2}e^{-\phi l_{\nu}(\nu+1)}+\frac{\nu+1}{\nu-2}e^{\phi l_{\nu}(\nu
	-1)}+4\frac{\nu^{2}-1}{\nu^{2}-4}e^{-\phi l_{\nu}}\biggr{]}\\
&  +\frac{\Upsilon}{\kappa\nu^{2}}\biggl{[}\frac{\nu-1}{\nu+2}\sinh{\phi
	l_{\nu}(\nu+1)}-\frac{\nu+1}{\nu-2}\sinh{\phi l_{\nu}(\nu-1)}+4\frac{\nu
	^{2}-1}{\nu^{2}-4}\sinh{\phi l_{\nu}}\biggr{]}~.\nonumber
\end{align}
Utilizando el ansatz para la m\'etrica (\ref{Ansatz}), las ecuaciones de movimiento pueden ser integradas para el factor conforme (\ref{omegaandres}) donde $\Upsilon$, $\nu$, $\kappa$ y $\Lambda=-3l^{-2}$ son par\'ametros del potencial y $\eta$ es una constante de integraci\'on. Todos ellos caracterizan la soluci\'on con pelo. Con esta elecci\'on del factor de conforme, es sencillo de obtener las expresiones para el campo escalar~\footnote{Estas fueron mostradas anteriormente en la secci\'on \ref{soluhair}}
\begin{equation}
\phi(x)=l_{\nu}^{-1}\ln{x}%
\end{equation}
y la funci\'on m\'etrica
\begin{equation}
f(x)=\frac{1}{l^{2}}+\Upsilon\biggl{[}\frac{1}{\nu^{2}-4}-\frac{x^{2}}{\nu
	^{2}}\biggl{(}1+\frac{x^{-\nu}}{\nu-2}-\frac{x^{\nu}}{\nu+2}%
\biggr{)}\biggr{]}+\frac{x}{\Omega(x)}%
\end{equation}
donde $l_{\nu}^{-1}=\sqrt{(\nu^{2}-1)/2\kappa}$.

Para comparar con los resultados presentados en la secci\'on anterior, debemos trabajar con las coordenadas can\'onicas de AdS. Vamos a discutir la rama $x\in(1,\infty)$ para el cual el campo escalar se define positivamente. Cambiamos la coordenada $r$ de forma que la funci\'on delante de la secci\'on transversal, $d\Sigma_ {k}$, tiene la siguiente fall-off:
\begin{equation}
\Omega(x)=r^{2}+O(r^{-3})~.
\end{equation}
Esta elecci\'on est\'a motivada por el hecho de que el t\'ermino $O(r^{-2})$ genera un t\'ermino lineal en el fall-off de $\Omega$. Los tres primeros t\'erminos sub-relevantes son
\begin{equation}
x=1+\frac{1}{\eta r}+\frac{m}{r^{3}}+\frac{n}{r^{4}}+\frac{p}{r^{5}}+O(r^{-6})
\end{equation}
y pueden ser calculados teniendo en cuenta la expansi\'on alrededor de
$r=\infty$:
\begin{equation}
\Omega(x)=r^{2}-\frac{24m\eta^{3}+\nu^{2}-1}{12\eta^{2}}-\frac{24n\eta^{4}%
	-\nu^{2}+1}{12\eta^{3} r}+\frac{720m^{2}\eta^{6}-480p\eta^{5}+\nu^{4}%
	-20\nu^{2}+19}{240\eta^{4} r^{2}}+O(r^{-3})~.
\end{equation}
Despu\'es de un c\'alculo sencillo obtenemos
\begin{equation}
x=1+\frac{1}{\eta r}-\frac{(\nu^{2}-1)}{24\eta^{3}r^{3}}\biggl{[}1-\frac
{1}{\eta r}- \frac{9(\nu^{2}-9)}{80\eta^{2}r^{2}}\biggr{]}+O(r^{-6}).
\end{equation}
Las siguientes expansiones asint\'oticas para las funciones m\'etricas:
\begin{equation}
-g_{tt}=f(x)\Omega(x)=\frac{r^{2}}{l^{2}}+1+\frac{\Upsilon+3\eta^{2} }%
{3\eta^{3} r}+O(r^{-3})~,
\end{equation}
\begin{equation}
g_{rr}=\frac{\Omega(x)\eta^{2}}{f(x)}\biggl{(}\frac{dx}{dr}\biggr{)}=\frac
{l^{2}}{r^{2}}- \frac{l^{4}}{r^{4}}-\frac{l^{2}(\nu^{2}-1)}{4\eta^{2} r^{4}}-
\frac{l^{2}(3\eta^{2} l^{2}+\Upsilon l^{2}-\nu^{2}+1)}{3\eta^{3} r^{5}%
}+O(r^{-6})~.
\end{equation}
La expansi\'on asint\'otica del campo escalar en estas coordenadas es
\begin{equation}
\phi(x)= l_{\nu}^{-1}\ln{x}= \frac{1}{l_{\nu}\eta r}-\frac{1}{2l_{\nu}\eta
	^{2}r^{2}}-\frac{\nu^{2}-9}{24\eta^{3}r^{3}}+O(r^{-4})
\end{equation}
y entonces, en la notaci\'on est'andar, obtenemos $\alpha=1/l_{\nu}\eta$,
$\beta=-1/2l_{\nu}\eta^{2}$. Ambos modos son normalizables y, ya que 
$\beta=C\alpha^{2}$ con $C=-l_{\nu}/2$, la simetr\'ia conforme en el borde se preserva. Ahora, podemos calcular f\'acilmente la masa Hamiltoniana del sistema como se propuso en \cite{Anabalon:2014fla} 
\begin{equation}
M=\sigma\biggl{[}\frac{\mu}{\kappa}+\frac{1}{l^{2}}\biggl{(}W-\frac{\alpha}{3}\frac
{dW}{d\alpha}\biggr{)}\biggr{]}
\end{equation}
y considerando $W=-l_{\nu}\alpha^{3}/6$, $\sigma=4\pi$, y $l_{\nu}%
^{-1}=\sqrt{(\nu^{2}-1)/2\kappa}$ obtenemos
\begin{equation}
M=-\frac{\sigma}{\kappa}\biggl{(}\frac{3\eta^{2}+\Upsilon}{3\eta^{3}}\biggr{)}~,
\end{equation}
que coincide con la masa hologr\'afica.
Finalizamos con la interpretaci\'on de estas soluciones con pelo dentro de la dualidad AdS/CFT. Es decir, puesto que $W=-l_{\nu}\alpha^{3}/6$, 
corresponde a la adici\'on de una deformaci\'on de triple traza a la acci\'on del borde como es (\ref{triple}) (ejemplos similares se pueden encontrar en \cite{Hertog:2004dr, Hertog:2004rz}):
\begin{equation}
I_{CFT}\rightarrow I_{CFT} +\frac{l_{\nu}}{6} \int d^{3}x\mathcal{O}^{3}~.%
\end{equation}
Para diferentes agujeros negros con pelo, los cuales est\'an caracterizados por el par\'ametro $\nu$, la relaci\'on entre $\alpha$ y $\beta$ no cambia por lo que hay deformaciones triple traza, pero con diferentes acoplamientos.
\section{Discusi\'on}
Ya que esta secci\'on contiene 
c\'alculos detallados e interpretaciones, 
presentamos algunas conclusiones generales.

El m\'etodo de contrat\'erminos 
\cite{Balasubramanian:1999re}, el cual 
fue obtenido en el contexto de la dualidad AdS/CFT  inicialmente fue propuesto
para soluciones asint\'oticamente AdS 
\cite{deHaro:2000xn,Bianchi:2001kw,Skenderis:2002wp} y entonces fue generalizado a soluciones
asint\'oticamente planas \cite{Astefanesei:2005ad,Mann:2005yr,Mann:2006bd,Astefanesei:2006zd,
	Marolf:2006bk,Astefanesei:2009mc,Astefanesei:2009wi,Astefanesei:2010bm} e incluso
a soluciones dS \cite{Balasubramanian:2001nb,Ghezelbash:2001vs,Ghezelbash:2002vz,
	Astefanesei:2003gw}, aunque en los 
\'ultimos dos casos la interpretaci\'on
hologr\'afica no es clara y generalmente
es poco aceptada. Interesantemente,
este m\'etodo da el tensor de stress cuasilocal y las cargas conservadas, en una forma muy similar con la bien comprendida
holograf\'ia de espacios-tiempo asint\'oticamente AdS.

Cuando la teor\'ia contiene campos escalares,
existe una diversidad de condiciones de
borde mixtas que pueden ser impuestas,
en particular las condiciones de borde
que rompen la simetr\'ia conforme 
del borde. El m\'etodo de renormalizaci\'on hologr\'afica  \cite{deHaro:2000xn,Bianchi:2001kw,Skenderis:2002wp,Skenderis:2000in}   
que usa la expansi\'on de Fefferman-Graham 
fue generalizado para condiciones de borde mixtas que corresponden a la rama no-logar\'itmica de las soluciones en 
\cite{Papadimitriou:2007sj}.

En este trabajo construimos contrat\'erminos expl\'icitamente covariantes que son similares a las propuestas por Balasubramanian y Kraus
\cite{Balasubramanian:1999re} y generalizadas 
para teor\'ias (potenciales moduli) que 
contienen adem\'as soluciones de rama logar\'itmica. Para construir estos contrat\'erminos nos guiamos por el m\'etodo
Hamiltoniano que provee las condiciones de borde correctas (en particular, el fall-off del campo escalar) de tal forma que las cargas 
conservadas son finitas. Tambi\'en verificamos que 
el principio variacional para la acci\'on 
gravitacional este bien definida. Puede que
no sea una sorpresa de que la masa hologr\'afica 
concuerda con la masa Hamiltoniana para todas 
las condiciones de borde. Sin embargo, 
cuando comparamos con el formalismo AMD
hay un c\'ambio dr\'astico cuando las condiciones
de borde no preservan la simetr\'ia conforme y,
como fue mostrado en \cite{Anabalon:2014fla}, la
masa AMD no es adecuada para este caso. 

Como futuras direcciones, nos 
gustar\'ia considerar contrat\'erminos
para otras masas conformes del campo escalar
y para otras teor\'ias en mayores dimensiones.
Ser\'ia \'util, si fuese posible, plantear
un algoritmo general para construir contrat\'erminos
usando el m\'etodo Hamiltoniano --- no es del todo
claro si eso es posible para soluciones
de gravedad que son asint\'oticamente dS. 
Para agujeros negros extremos, existen
diferentes m\'etodos para calcular las cantidades
conservadas, el formalismo de la funci\'on entrop\'ia de Sen \cite{Sen:2005wa,Sen:2005iz,Sen:2007qy} (para agujeros negros rotantes fue generalizado en \cite{Astefanesei:2006dd} y en el contexto de la dualidad AdS/CFT, ver por ejemplo \cite{Astefanesei:2007vh,Astefanesei:2008wz,Morales:2006gm,Astefanesei:2010dk,Astefanesei:2011pz}).
Sin embargo, este m\'etodo nos da las cargas usando 
los datos de la geometr\'ia cerca del horizonte y,
cuando hay un flujo RG no-trivial en teor\'ias con
escalares encendidos, es interesante comparar
las cargas conservadas calculadas en el horizonte 
con las obtenidas en el borde por el m\'etodo de contrat\'erminos. El m\'etodo de contrat\'erminos 
tambi\'en fue usado en \cite{Anabalon:2015ija} para estudiar los diagramas de fase de una clase general de agujeros negros con pelo de horizonte esf\'erico.

Una perspectiva diferente naturalmente 
surge cuando los campos escalares y la gravedad interactuan. De hecho, se demostr\'o anteriormente que en espacios-tiempo asint\'oticamente planos y cuando el potencial escalar es convexo, 
el \'unico agujero negro esf\'ericamente sim\'etrico es la soluci\'on de Schwarzschild \cite{Bekenstein:1972ny, Bekenstein:1972ky}, 
el cual fue generalizado a auto-interacciones
no negativas \cite{Heusler:1992ss,
	Sudarsky:1995zg}, para un reciente 
review ver \cite{Herdeiro:2015waa}.
Se esperaba que los teoremas de no-pelo 
se preservaran cuando, asint\'oticamente 
existe una constante cosmol\'ogica no trivial.
La existencia num\'erica de agujeros negros AdS
fueron verificadas en varios art\'iculos \cite{Hertog:2004dr,Sudarsky:2002mk,
	Torii:2001pg}  (un gran n\'umero de agujeros negros con pelo ex\'actos fuer\'on encontrados cuando el campo escalar tiene una masa $m^{2}=-2/l^{2}$ 
\cite{Acena:2013jya,Anabalon:2013eaa,Martinez:2004nb,Kolyvaris:2009pc, Gonzalez:2013aca,Feng:2013tza,Anabalon:2012sn, Anabalon:2012ta}).
Algunos de esos agujeros negros son linealmente estables \cite{Torii:2001pg,Anabalon:2015vda}.
Otra interesante direcci\'on es sobre
soluciones de estrellas bos\'onicas y la relaci\'on 
con las inestabilidades de algunas soluciones AdS (y el propio AdS) \cite{Astefanesei:2003qy,Bizon:2011gg,Buchel:2013uba,Astefanesei:2003rw,Buchel:2015rwa}%
.

%% file: capitulo6.tex
       \chapter{Termodin\'amica de agujeros negros}
Los agujeros negros asint\'oticamente 
AdS juegan un rol importante en el entendimiento
de la din\'amica y termodin\'amica de las teor\'ias hologr\'aficas duales del campo
v'ia la dualidad AdS/CFT \cite{Maldacena:1997re}.
En particular, estos agujeros negros son 
duales a los estados t\'ermicos de la 
teor\'ia del campo en el borde. Las transiciones
de fase de primer orden en el bulk pueden
ser relacionadas a las transiciones de fase 
confinaminamiento/deconfinamiento en la teor\'ia
del campo dual \cite{Witten:1998zw}. 
Ya que los campos escalares aparecen como
el moduli en teor\'ia de cuerdas, 
es importante entender la termodin\'amica
y sus propiedades gen\'ericas de los agujeros
negros con pelo. 

Motivados por esas consideraciones, en esta secci\'on estudiaremos en detalle la termodin\'amica de una clase general de
agujeros negros con pelo escalar $4$-dimensional
\cite{Anabalon:2013eaa,Anabalon:2012ta,Anabalon:2013sra} (ejemplos en otras dimensiones
o para diferentes topolog\'ias del horizonte
pueden ser encontradas en \cite{Acena:2013jya,Acena:2012mr,Anabalon:2013qua, Lu:2013ura,Lu:2014maa,Feng:2013tza, Xu:2014uka, Wen:2015xea}). El potencial escalar 
esta caracterizado por dos par\'ametros 
y la soluci\'on del agujero negros tiene una 
constante de integraci\'on que esta relacionada
con su masa. Para algunos valores particulares
de los par\'ametros en el potencial, las 
soluciones pueden ser inmersas en supergravedad
\cite{Anabalon:2013eaa, Feng:2013tza}. El potencial del campo escalar contiene como
casos especiales todas las soluciones exactas est\'aticas no cargadas discutidas ampliamente
en la literatura \cite{Martinez:2004nb, Kolyvaris:2009pc,
	Gonzalez:2013aca} (para los detalles, ver \cite{Anabalon:2012dw}).
Esas configuraciones est\'aticas fueron 
extendidas a soluciones de agujeros negros din\'amicos \cite{Zhang:2014dfa, Zhang:2014sta}.

Hay algunas sutilezas en la definici\'on
de la masa de los agujeros negros con pelo
\cite{Hertog:2004dr,Anabalon:2014fla,Henneaux:2006hk,Barnich:2002pi, Amsel:2006uf}. En \cite{Anabalon:2014fla}, un m\'etodo 
concreto del calculo de la masa de un agujero
negro asint\'oticamente AdS fue propuesto.
Este m\'etodo es muy \'util desde un punto
de vista pr\'actico porque esta usa justamente
la expansi\'on de las funciones m\'etricas 
en el borde. Sobre todo, esta puede ser usada
para agujeros negros con pelo que preservan o no 
la simetr\'ia conforme (las isometr\'ias AdS) 
del borde. Usaremos este m\'etodo, el cual esta basada en el formalismo Hamiltoniano
\cite{Regge:1974zd}, para calcular la masa de las soluciones de agujeros negros. 

Sin embargo, basados en la f\'isica de
la dualidad AdS/CFT, un m\'etodo diferente
fue desarrollado, y se conoce como Renormalizaci\'on Hologr\'afica 
\cite{Henningson:1998gx} (ver, tambi\'en, \cite{Balasubramanian:1999re,Skenderis:2002wp,
	Skenderis:2000in,de Haro:2000xn, Papadimitriou:2004ap}%
\footnote{Un m\'etodo similar para espacios-tiempo asint\'oticamente AdS fue desarrollado en
	\cite{Astefanesei:2005ad,Mann:2005yr} y algunas aplicaciones concretas fueron presentadas en \cite{Astefanesei:2009wi, Astefanesei:2010bm}.}) --- para condiciones
de borde mixtas del campo escalar, este
m\'etodo fue desarrollado y promovido en \cite{Papadimitriou:2007sj}. La principal idea
detr\'as de este m\'etodo es que, considerando
la holograf\'ia, las divergencias infrarrojas
(IR) que aparecen en el lado de la gravedad 
son equivalentes con las divergencias ultravioletas de la teor\'ia dual del campo.
Entonces, para curar estas divergencias, uno 
necesita adicionar contrat\'erminos 
que son locales y dependen en la geometr\'ia 
intr\'inseca del borde. De esta forma, uno
puede usar el formalismo cuasi-local de 
Brown y York \cite{Brown:1992br} suplementadas con estos contrat\'erminos para calcular
la acci\'on Euclidea regularizada y el
stress tensor del borde. La energ\'ia
es la carga asociada con el vector de Killing
$\partial_{t}$ y esta puede ser obtenida del
stress tensor del borde.

Con estos resultados, uno puede investigar
la termodin\'amica y los diagramas de fase de 
las soluciones de agujeros negros con pelo.
En particular, mostramos que existen
transiciones de fase de primer orden que llevan a una discontinuidad en en la entrop\'ia. 
Resultados similares fueron obtenidas para
una clase general de soluciones de agujeros negros en una teor\'ia con un campo escalar
invariante conforme \cite{Giribet:2014fla}.      

Este cap\'itulo esta organizado de la siguiente manera:
En la primera parte describimos un m\'etodo desarrollado
en \cite{Astefanesei:2009wi} para calcular la temperatura de cualquier
agujero negro estacionario. En la siguiente parte
mostramos las transiciones de fase los agujeros negros 
con horizonte esf\'erico y planar. Presentamos una discusi\'on con el resumen para cada caso.

 \section{Espacio-tiempo de Rindler}
 
 En el espacio-tiempo de Minkowski se puede definir un n\'umero infinito de observadores (sistemas de coordenadas) inerciales y no-inerciales. Los observadores inerciales son invariantes bajo el grupo de Lorentz $SO(1,3)$ que dejan
 invariante la m\'etrica
  \begin{equation}
 ds^{2}=-dt^{2}+dx^{2}+dy^{2}+dz^{2}.
 \label{minko11}
 \end{equation}
  En espacio-tiempos curvos el principio de equivalencia asegura que el espacio-tiempo es localmente de Minkowski. Es decir, localmente no es posible saber
  si acceleraci\'on es debido a la gravedad o a una fuerza externa. Consideremos observadores no-inerciales en el espacio-tiempo de Minkowski. Esto puede conseguirse mediante la siguiente transformaci\'on de coordenadas
 \begin{equation}
 t=\rho\sinh{(a\tau)}, \qquad x=\rho\cosh{(a\tau)}
 \end{equation}
 que mapea la m\'etrica (\ref{minko11}) a
 \begin{equation}
 ds^{2}=-a^{2}\rho^{2}d\tau^{2}+d\rho^{2}+dy^{2}+dz^{2}~,
 \label{rindler1}
 \end{equation}
 donde los dominios de la aceleraci\'on y el par\'ametro temporal son respectivamente, $a^{\mu}a_{\mu}=a^{2}$,  $0\leq a\leq\infty$, $-\infty\leq\tau\leq\infty$.
 Estos son los observadores de Rindler y tienen acceso  s\'olo a una parte del
 espacio-tiempo de Minkowski, el resto es inaccesible causalmente, oculto por el horizonte de Rindler. Es importante recordar que si 
 el observador no-inercial deja de acelerar, este horizonte desaparece.
 Este tipo de observadores pueden encontrarse en las cercan\'ias
 del horizonte de sucesos del agujero negro de Schwarzschild, pero a diferencia del espacio-tiempo de Rindler, el horizonte del agujero negro no puede
 desaparecer bajo un cambio global de coordenadas.
 Entonces el campo gravitacional del agujero negro de Schwarzschild es equivalente a la aceleraci\'on que mide un observador no-inercial en un espacio-tiempo de
 Minkowski. Este es el principio de equivalencia.
 El espacio-tiempo de Rindler en la secci\'on Euclidea $\tau\rightarrow -i\tau^{E}$, donde $\tau^{E}\in(0,2\pi)$, presenta una singularidad c\'onica en $\rho=0$: $ds^{2}=\rho^{2}d(a\tau^{E})^{2}+d\rho^{2}..$~, la que puede ser evitada
 considerando la periodicidad $\beta=\frac{2\pi}{a}$. Esta periodicidad
 esta relacionada con la temperatura mediante $T=\frac{1}{\beta}=\frac{a}{2\pi}$. Entonces, un observador de Rindler mide una 
 temperatura distinta de cero, y que por el principio de equivalencia un observador en las cercan\'ias del agujero negro de Schwarzschild (localmente) mide una temperatura. Basados en esta idea en \cite{Astefanesei:2009wi}
 mostraron un m\'etodo general para determinar la temperatura de agujeros negros estacionarios.  
 
 \section{Temperatura de agujeros negros estacionarios}
 \label{formulatemp}
 
 En cap\'itulos anteriores mostramos que en la secci\'on Euclidea podemos obtener la funci\'on de partici\'on y, de esta, las cantidades termodin\'amicas. Sin embargo hay casos en los cuales este m'etodo no funciona correctamente, tal es el caso de agujeros negros en $5$ dimensiones como las soluciones de anillos negros (black rings) debido a que la m\'etrica en la secci\'on Euclidea 
 no es una m\'etrica para una variedad real\footnote{Las componentes angulares son funciones complejas}.\\ En nuestro caso, el ansatz est\'atico (\ref{anzats2}) en la secci\'on Euclidea $t\rightarrow -it_{E}$ es:
 \begin{equation}
 ds^{2}=N(r)dt_{E}^{2}+H(r)dr^{2}+S(r)d\Sigma_{k}^{2}.
 \end{equation}      
 Vemos que la m\'etrica resultante es real, pero cuando 
 consideramos soluciones estacionarias\footnote{Por ejemplo 
 la soluci\'on de Kerr} donde el t\'ermino 
 $g_{t\varphi}dtd \varphi$ esta presente en la m\'etrica. En la secci\'on Euclidea este nuevo t\'ermino toma la forma $-ig_{t\varphi}dt_{E}d\varphi$, entonces la m\'etrica queda compleja. Estas configuraciones con geometr\'ia 
 compleja y acci\'on real (funci\'on de partici\'on) funcionan adecuadamente y el procedimiento se conoce con el nombre de \textit{cuasi-Euclidean method} \cite{Brown:1990di}\footnote{En este m\'etodo se realiza una transformaci\'on de Wick a las variables intensivas, para el caso de black ring, $\varphi\rightarrow -i\varphi$ }.\\ Aqu\'i nos concentraremos
 en el caso est\'atico de cuatro dimensiones. Realizando el cambio $\rho=\sqrt{N}$, en la secci\'on Eucl\'idea y reordenando se puede identificar la singularidad c\'onica en el sector $(\tau_{E},r)$
 \begin{equation}
 ds^{2}=g_{rr}\frac{4N}{[(N)^{'}]^{2}}\bl{[}\rho^{2}\frac{[(N)^{'}]^{2}}{4Ng_{rr}}d\tau_{E}^{2}+d\rho^{2}\br{]}~.
 \end{equation}
 Recordando el caso de Rindler, identificamos la periodicidad $\Delta\tau_{E}=\beta$ con la temperatura, 
 \begin{equation}
 T=\frac{1}{\beta}=\frac{(N^{2})^{'}}{4\pi\sqrt{N^{2}g_{rr}}}\br{\vert}_{H}~.
 \label{generalT}
 \end{equation}
 Cuando $g_{t\varphi}=0$, la m\'etrica 
 invariante bajo la inversi\'on temporal  $t\rightarrow -t$, se dice que es
 el caso est\'atico. Son las soluciones est\'aticas
 las que estudiamos a lo largo de la presente tesis.\\
  
 Usando esta formula podemos calcular la temperatura de los agujeros negros de
 Schwarzschild y Reissner-Nordstr\"om, asint\'oticamente planos:
 \begin{equation}
 T_{flat}=\frac{1}{4\pi r_{h}}~, \qquad T_{RN-flat}=\frac{1}{4\pi r_{+}}\bl{(}
 1-\frac{q^{2}}{4r^{2}_{+}}\br{)}~.
 \label{schRN}
 \end{equation}
 De igual forma la temperatura de los agujeros negros de 
 Schawazschild-AdS y RN-AdS:
 \begin{equation}
 T_{Sch-AdS}=\frac{1}{4\pi r_+} \left(1+\frac{3r_+^{~2}}{l^2}\right)~, \qquad T_{RN-AdS}=\frac{1}{4\pi r_+} \left(1+\frac{3r_+^{~2}}{l^2}-\frac{q^2}{4r_+^{~2}}\right)~.
 \end{equation}
 Es interesante notar que las temperaturas de los agujeros negros de Schwarzschild-AdS y RN-AdS,
 cuando el horizonte $r_{+}$ es peque\~no $r_{+}\ll l$, se aproximan a las temperaturas de los agujeros negros 
 asint\'oticamente planos~\footnote{Esto es equivalente a decir que el radio AdS es muy grande $l\rightarrow\infty$, tomando ese l\'imite se puede ver inmediatamente que las expresiones para la temperatura se reducen a (\ref{schRN})}.
 Finalmente, calculamos la temperatura de los agujeros negros con pelo escalar (\ref{Ansatz2}) y (\ref{Ansatz3}) de secciones transversales $k=0,1,-1$:
 \begin{equation}
 ds^{2}=\Omega(x)\left[  -f(x)dt^{2}+\frac{\eta^{2}dx^{2}}{f(x)}+d\theta^{2}+d\Sigma_{k}^{2}\right]~,  \label{Ansatzk1}%
 \end{equation}
 con la funci\'on m\'etrica
 \begin{equation}
 f(x)=\frac{1}{l^{2}}+\alpha\bl{[}\frac{1}{\nu^{2}-4}-\frac{x^{2}}{\nu^{2}}\bl{(}1+\frac{x^{-\nu}}{\nu-2}-\frac{x^{\nu}}{\nu+2}\br{)}\br{]}+\frac{kx}{\Omega (x)}~.
 \label{Ansatk2}
 \end{equation} 
 Una identidad importante es

 \begin{equation}
 f^{'}\Omega(x)=\frac{\alpha}{\eta^{2}}+2k+k\nu\frac{x^{\nu}+1}{x^{\nu}-1}~.
 \end{equation}  
 Entonces la temperatura de acuerdo a la f\'ormula (\ref{generalT})
 \begin{equation}
 T=\frac{f^{'}}{4\pi\eta}\br{\vert}_{x_{h}}=\frac{1}{4\pi\eta\Omega(x_{h})}\bl{(}\frac{\alpha}{\eta^{2}}+2k+k\nu\frac{x_{h}^{\nu}+1}{x_{h}^{\nu}-1}\br{)}~,
 \end{equation}
 donde la ecuaci\'on que define el horizonte es $f(x_{h},\eta)=0$.
 \section{Transiciones de fase de agujeros negros esf\'ericos}
 
 Como ejemplo, trabajaremos 
 con el agujero negro de Schwarzschild-AdS (SAdS) en coordenadas $(t,x,\theta,\varphi)$, el que puede ser obtenido fijando el par\'ametro
 hairy $\nu=1$
 \begin{equation}
 ds^2=\Omega(x) \bl{(}-f(x)dt^{2}+\frac{\eta^{2}dx^{2}}{f(x)}+d\theta^{2}+\sin^{2}{\theta}d\varphi^{2}\br{)}~,
 \end{equation}
 \begin{equation}
 \Omega(x)=\frac{1}{\eta^{2}(x-1)^{2}}~,\qquad f(x)=\frac{1}{l^{2}}+\frac
 {1}{3}\alpha(x-1)^{3}+\eta^{2}x(x-1)^{2}~. \label{schw}%
 \end{equation}
 Para obtener el agujero negro de SAdS
 en su forma can\'onica, necesitamos el 
 siguiente cambio de coordenadas
 \begin{equation}
 x=1+\frac{1}{\eta r}~, \qquad x=1-\frac{1}{\eta r}~. \label{change}%
 \end{equation}
 Tenemos dos ramas que corresponden a 
 $x\in\lbrack0,1)$ y $x\in\lbrack1,\infty)$.
 Ya que hay algunas sutilezas en el c\'alculo de la acci\'on para la rama $x\in\lbrack 1,\infty)$ (por ejemplo, la curvatura extr\'inseca cambia de signo debido al cambio de la normal para la foliaci\'on $x=constante$) en lo que sigue trabajaremos expl\'icitamente en la rama $x\in[0,1]$. Usando el cambio de coordenadas (\ref{change}) obtenemos el agujero negro de SAdS en coordenadas can\'onicas:
 \begin{equation}
 \Omega(x)f(x)=F(r)=1-\frac{\mu}{r}+\frac{r^{2}}{l^{2}}\qquad,\qquad\mu
 =\frac{\alpha+3\eta^{2}}{3\eta^{3}}~.%
 \end{equation}
 En este caso, el potencial del campo escalar se reduce a $V=\frac{\Lambda}{\kappa}$, y el par\'ametro de la teor\'ia $\alpha$ pasa a ser una constante de integraci\'on. 
 Es bien conocido que la acci\'on tiene divergencias a\'un a tree level debido a la integraci\'on en un volumen infinito. Para regularizar la acci\'on, usaremos los contraterminos \cite{Balasubramanian:1999re}:
 \begin{equation}
 I[g_{\mu\nu}]=I_{bulk}+I_{GH}-\frac{1}{\kappa}\int_{\partial\mathcal{M}}%
 {d^{3}x\sqrt{-h}\biggl{(}\frac{2}{l}+\frac{\mathcal{R}l}{2}\biggr{)}}~,%
 \end{equation}
 donde $\mathcal{R}$ es el escalar de Ricci de la m\'etrica del borde $h_{ab}$.
 Ahora calcularemos la acci\'on del $bulk$. En este caso, siendo que el campo escalar desaparece, el potencial viene a ser la constante cosmol\'ogica: $V=\frac{\Lambda
 }{\kappa}=-\frac{3}{l^{2}\kappa}$. Usamos la traza del tensor de Einstein y la siguiente combinaci\'on de las ecuaciones de movimiento
 \begin{align}
 E_{t}^{t}-E_{\phi}^{\phi}=0\Rightarrow0  &  =f^{^{\prime\prime}}+\frac
 {\Omega^{^{\prime}}f^{^{\prime}}}{\Omega}+2\eta^{2}~,\\
 E_{t}^{t}+E_{\phi}^{\phi}=0\Rightarrow2\kappa V(\phi)  &  =-\frac
 {(f\Omega^{^{\prime\prime}}+f^{^{\prime}}\Omega^{^{\prime}})}{\Omega^{2}%
 	\eta^{2}}+\frac{2}{\Omega}~,\nonumber
 \end{align}
 para obtener
 \begin{equation}
 I_{bulk}^{E}=\frac{4\pi\beta}{\eta^{3}\kappa l^{2}}\biggl{[}-\frac{1}%
 {(x_{b}-1)^{3}}+\frac{1}{(x_{h}-1)^{3}}\biggr{]}=\frac{4\pi\beta}{\kappa
 	l^{2}}(r_{b}^{3}-r_{h}^{3})~.
 \end{equation}
 Aqu\'i, $x_{b}$ y $x_{h}$ son las localizaciones del borde y del horizonte, y $\beta$
 es la periodicidad del tiempo Euclideo que esta relacionado con la temperatura por $\beta=T^{-1}$.\\
 El t\'ermino de superficie de Gibbons-Hawking
 puede ser calculado si elegimos la foliaci'on
 $x=constante$ con la m\'etrica inducida
 \begin{equation}
ds^{2}=h_{ab}dx^{a}dx^{b}=\Omega(x)\left[
 -f(x)dt^{2}+d\theta^{2}+\sin^{2}\theta d\phi^{2}\right]~. 
 \end{equation}
 La normal y la curvatura extr\'inseca son
 \begin{equation}
 n_{a}=\frac{\delta_{a}^{x}}{\sqrt{g^{xx}}}\,\,,\qquad K_{ab}=\frac
 {\sqrt{g^{xx}}}{2}\partial_{x}h_{ab}%
 \end{equation}
 y la contribuci\'on de Gibbons-Hawking a la acci\'on es
 \begin{equation}
 I_{GH}^{E}=-\frac{2\pi\beta}{\kappa}\biggl{[}-\frac{6}{l^{2}\eta^{3}(x-1)^{3}%
 }-\frac{4}{\eta(x-1)}-\biggl{(}\frac{\alpha+3\eta^{2}}{\eta^{3}}%
 \biggr{)}\biggr{]}\biggr{\vert}_{x_{b}}=-\frac{2\pi\beta}{\kappa
 }\biggl{(}\frac{6r_{b}^{3}}{l^{2}}+4r_{b}-3\mu\biggr{)}~.
 \end{equation}
 La \'ultima contribuci\'on esta dada por el contrat\'ermino gravitacional, el cual es un t\'ermino intr\'inseco de superficie que depende s\'olo de la geometr\'ia del borde
 \begin{equation}
 I_{g}^{E}=\frac{2\pi\beta}{\kappa}\biggl{[}\frac{4}{l^{2}\eta^{3}%
 	(x_{b}-1)^{3}}+\frac{4}{\eta(x_{b}-1)}-2\mu\biggr{]}=\frac{2\pi\beta}{\kappa
 }\biggl{(}\frac{4r_{b}^{3}}{l^{2}}+4r_{b}-2\mu\biggr{)}~.
 \end{equation}
 Podemos ver que las divergencias proporcionales a $r_{b}$ y $r_{b}^{3}$ se cancelan, y la acci\'on regularizada es
 \begin{equation}
 I^{E}=I_{bulk}^{E}+I_{GH}^{E}+I_{g}^{E}=\frac{4\pi\beta}{\kappa l^{2}%
 }\biggl{[}\frac{1}{\eta^{3}(x_{h}-1)^{3}}+\frac{\mu l^{2}}{2}\biggr{]}=\frac
 {4\pi\beta}{\kappa l^{2}}\biggl{(}-r_{h}^{3}+\frac{\mu l^{2}}{2}\biggr{)}~.
 \end{equation}
 Los c\'alculos para el agujero negro general con pelo (\ref{Ansatz}),
 (\ref{omega}), (\ref{f}) son muy similares, pero debemos agregar un contrat\'ermino que dependa del campo escalar \cite{Skenderis:2002wp, Papadimitriou:2007sj,Henningson:1998gx}. Nosotros trabajamos con un contrat\'ermino que es intr\'inseco a la geometr\'ia del borde (no depende de la normal al borde o de las derivadas del campo escalar)
 \cite{Papadimitriou:2007sj}:
 \begin{equation}
 I_{\phi}^{E}=\int_{\partial\mathcal{M}}{d^{3}x^{E}\sqrt{h^{E}}\biggl{(}\frac
 	{\phi^{2}}{2l}-\frac{l_{\nu}}{6l}\phi^{3}\biggr{)}}=\frac{4\pi\beta}{\kappa
 }\biggl{[}-\frac{\nu^{2}-1}{4l^{2}\eta^{3}(x_{b}-1)}+\frac{\nu^{2}-1}%
 {3l^{2}\eta^{3}}\biggr{]}~. \label{cap4Iphi}%
 \end{equation}
 La suma de los otros tres primeros t\'erminos de la acci\'on es\footnote{En el ap\'endice estan las expresiones en D-dimensiones de $I_{bulk}^{E}, I_{surf}^{E}, I_{ct}^{E}$ asi como la suma de estos tres primeros t\'erminos}
 \begin{equation}
 I_{bulk}^{E}+I_{surf}^{E}+I_{g}^{E}=-\frac{1}{T}\biggl{(}\frac{AT}%
 {4G}\biggr{)}+\frac{4\pi\beta}{\kappa}\biggl{[}\frac{\nu^{2}-1}{4l^{2}\eta
 	^{3}(x_{b}-1)}+\frac{12\eta^{2}l^{2}+4\alpha l^{2}-4\nu^{2}+4}{12l^{2}\eta
 	^{3}}\biggr{]}~,
 \end{equation}
 donde $\mathcal{A}=4\pi\Omega(x_{h})$ es el \'area del horizonte. Es importante mencionar que el contrat\'ermino gravitacional \cite{Balasubramanian:1999re} no es suficiente para cancelar la divergencia en la acci\'on (aun queda un t\'ermino divergente proporcional a $(x_{b}-1)^{-1}$) pero cuando agregamos el contrat\'ermino  (\ref{cap4Iphi})
 obtenemos la acci\'on finita:
 \begin{equation}
 I^{E}=\beta\biggr{(}-\frac{AT}{4G}+\frac{4\pi}{\kappa}\frac{3\eta^{2}+\alpha
 }{3\eta^{3}}\biggr{)}~.
 \end{equation}
 Como ya vimos, en el l\'imite cl\'asico, la acci\'on esta relacionada con el potencial termodin\'amico (la energ\'ia libre $F$ en este caso), el cual es $F=I^{E}/\beta=M-TS$. Usando las relaciones termodin\'amicas podemos mostrar que la masa es (o comparando con la definici\'on de la energ\'ia libre):
 \begin{equation}
 M=\frac{1}{2G}\left(  \frac{\alpha+3\eta^{2}}{3\eta^{3}}\right)~.
 \end{equation}
 Usando las siguientes expresiones para la temperatura y la entrop\'ia 
 \begin{equation}
 T=\frac{f^{^{\prime}}(x)}{4\pi\eta}\biggr{\vert}_{x=x_{h}}=\frac{1}{4\pi
 	\eta\Omega(x_{h})}\biggl{[}\frac{\alpha}{\eta^{2}}+2+\nu\frac{x_{h}^{\nu}%
 	+1}{x_{h}^{\nu}-1}\biggr{]}\,\,,\qquad S=\frac{A}{4G}=\frac{4\pi\Omega(x_{h}%
 	)}{4G}~,\label{temperature}%
 \end{equation}
 podemos verificar que la primera ley $dM=TdS$ se satisface, considerando la ecuaci\'on del horizonte $f(x_{h},\eta)=0$ 
 \begin{equation}
 \frac{\pa M}{\pa\eta}\frac{d\eta}{dx_{h}}=T\bl{(}\frac{\pa S}{\pa x_{h}}+\frac{\pa S}{\pa\eta}\frac{d\eta}{dx_{h}}\br{)}~.
 \end{equation}
 Como ya mencionamos anteriormente, hay dos tipos de soluciones. Para la familia con campo escalar positivo $x\in [0,\infty)$ la masa es 
 \begin{equation}
 M=-\frac{1}{2G}\biggl{(}\frac{\alpha+3\eta^{2}}{3\eta^{3}}\biggr{)}
 \end{equation}
y la temperatura
 \begin{equation}
 T=-\frac{1}{4\pi\eta\Omega(x_{h})}\biggl{[}\frac{\alpha}{\eta^{2}}+2+\nu
 \frac{x_{h}^{\nu}+1}{x_{h}^{\nu}-1}\biggr{]}~,
 \end{equation}
con la entrop\'ia dada por la ley de \'area.
  
 \newpage 
 \subsection{Transici\'on de fase a la Hawking y Don Page}
 
 El espacio-tiempo AdS en la regi\'on asint\'otica puede ser interpretado 
 como una pared de potencial y esta se comporta como una caja infinita
 (tiene borde conforme)\footnote{Ver la secci\'on \ref{geoAdS}}. 
 Dado que no es un espacio-tiempo globalmente hiperb\'olico es necesario imponer condiciones de borde.  
 
 
 El campo escalar satisface diferentes condiciones de borde dependiendo si este es positivo o negativo, el cual corresponde a dos familias de soluciones mencionadas anteriormente\footnote{Donde el campo escalar es $\phi=l_{\nu}^{-1}\ln{x}$, en el intervalo $x\in[0,1]$ el campo escalar es $\phi<0$. En el intervalo $x\in [0,\infty)$ el campo escalar es $\phi>0$}. Una teor\'ia cl\'asica del campo es completamente definida cuando las condiciones de borde est'an pre-escritas. Para cualesquiera condiciones de borde en el campo escalar, la configuraci\'on de vac\'io dado por la soluci\'on del agujero negro-SAdS debe ser incluido como un estado permitido de la teor\'ia. Por lo tanto, en el ensemble can\'onico, su energ\'ia libre puede ser comparada con la del agujero negro a una temperatura dada. Figura $1(a)$ muestra que, para una familia con un campo escalar positivo, SAdS es siempre m\'as favorable que la configuraci\'on hairy. Figura $(b)$ muestra el mismo fen\'omeno para la familia con un campo escalar negativo. Encontramos que, para valores valores gen\'ericos de $\alpha$ el comportamiento cualitativo de los diagramas de fase no cambia. 
\begin{figure}[h!]
	\centering
	\begin{subfigure}[b]{0.4\textwidth}
		\includegraphics[width=\textwidth]{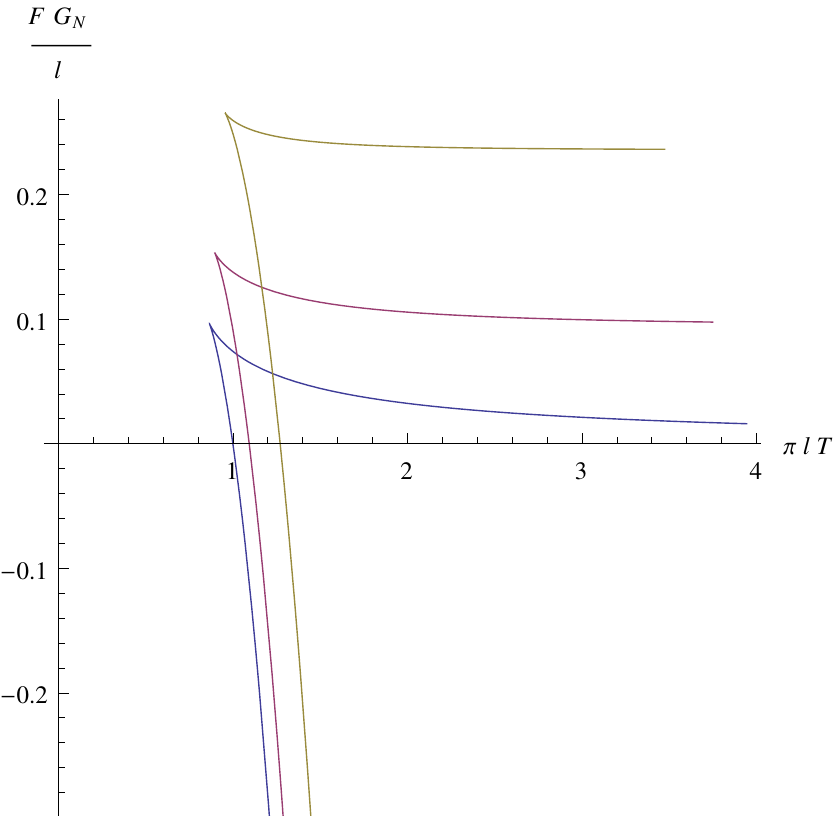}
		\caption{(a)}
	\end{subfigure}
	\begin{subfigure}[b]{0.4\textwidth}
		\includegraphics[width=\textwidth]{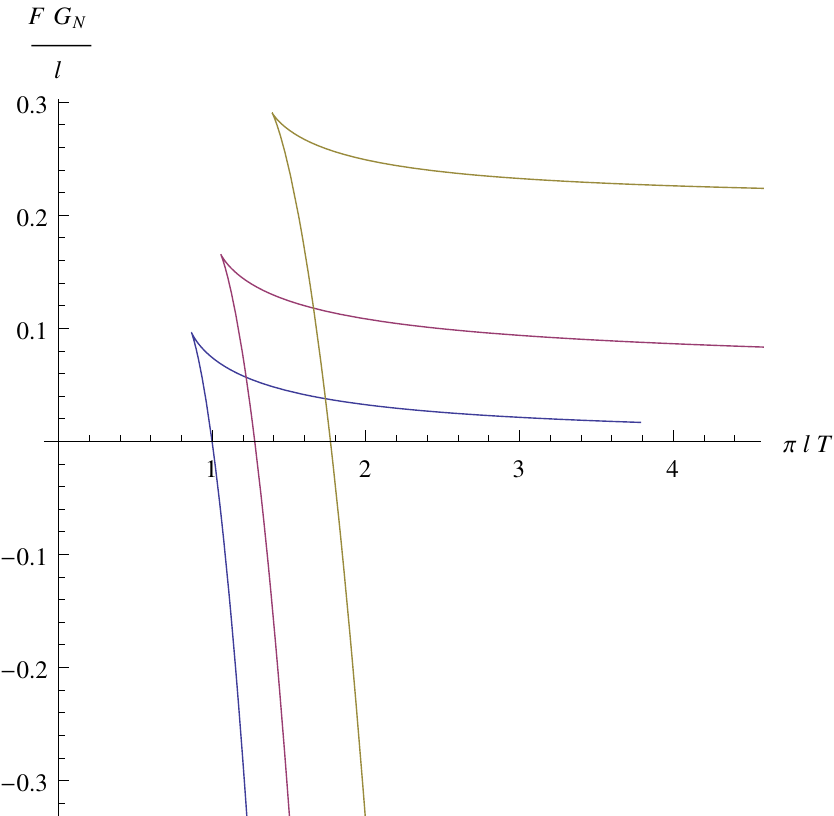}
		\caption{(b)}
	\end{subfigure}
	\hfill\caption{(a) Energ\'ia libre adimensional versus temperatura adimensional,
		para diferentes valores de $\nu$ y  $\alpha=-10 l^{-2}$ y campo escalar positivo. Los gr\'aficos son para $\nu=1,\nu=1.9$ y $\nu=3$ (de abajo acia arriba).
		La energ\'ia libre de Schwarzschild AdS  ($\nu=1$) tiende a cero cuando T tiende a infinito. La energ\'ia libre de los agujeros negros con pelo tienden a una constante a temperatura infinita. (b) Energ\'ia libre adimensional versus temperatura adimensional, para diferentes valores de $\nu$,  $\alpha=10 l^{-2}$ y campo escalar negativo. Los gr\'aficos son para $\nu=1,\nu=1.9$ y $\nu=3$ (de abajo hacia arriba).}
\end{figure}
%

 Como en el caso SAdS, aqu\'i hay dos ramas consistentes de agujeros negros grandes y peque\~nos. Figuras $2(a)$ y $2(b)$ muestran la masa versus temperatura para familias con campo escalar positivo y negativo respectivamente.

\begin{figure}[h!]
	\centering
	\begin{subfigure}[b]{0.4\textwidth}
		\includegraphics[width=\textwidth]{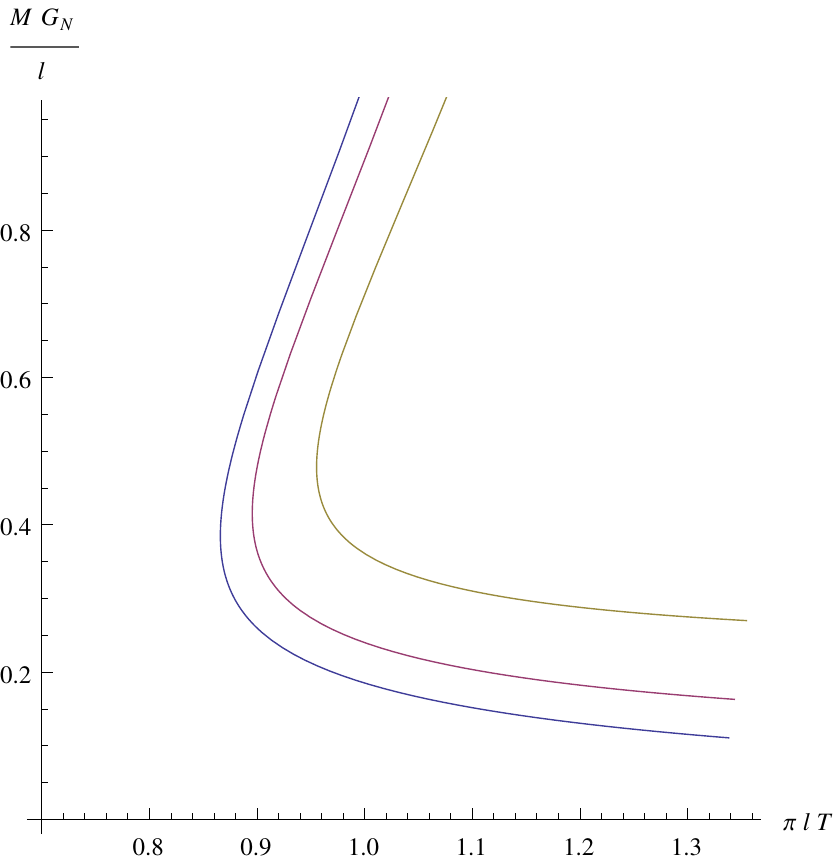}
		\caption{(c)}
	\end{subfigure}
	\begin{subfigure}[b]{0.4\textwidth}
		\includegraphics[width=\textwidth]{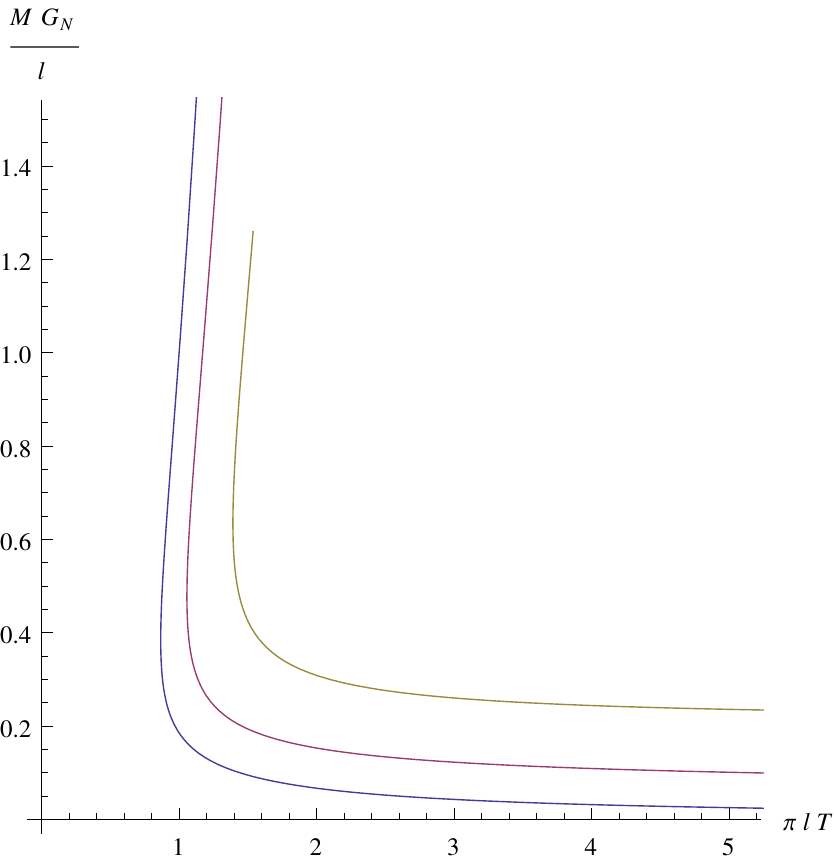}
		\caption{(d)}
	\end{subfigure}
	\caption{(c) Masa adimensional versus temperatura adimensional, para diferentes valores de $\nu$, $\alpha=10 l^{-2}$ y campo escalar negativo. Los graficos son para  $\nu=1,\nu=1.9$ y $\nu=3$ (de izquierda a derecha). Aqu\'i es posible ver la existencia de peque\~nos y grandes agujeros negros, ex\'actamente similar a Schwarzschild AdS.\\
		(d) Masa adimensional versus temperatura adimensional, para diferentes valores de $\nu$, $\alpha=-10 l^{-2}$ y campo escalar positivo. Los gr'aficos son para  $\nu=1,\nu=1.9$ y $\nu=3$ (de izquierda a derecha). Aqu\'i es posible ver la existencia de peque\~nos y grandes agujeros negros, ex\'actamente similar a Schwarzschild AdS.}
\end{figure}
 %
 Estos gr\'aficos nos dan informaci\'on sobre el calor espec\'ifico
 \begin{equation}
 C=\frac{\partial M}{\partial T},%
 \end{equation}
 que es interpretado como un cambio en la pendiente.
 
  \begin{figure}[ptbh]
  	\centering
  	\begin{subfigure}[b]{0.4\textwidth}
  		\includegraphics[width=\textwidth]{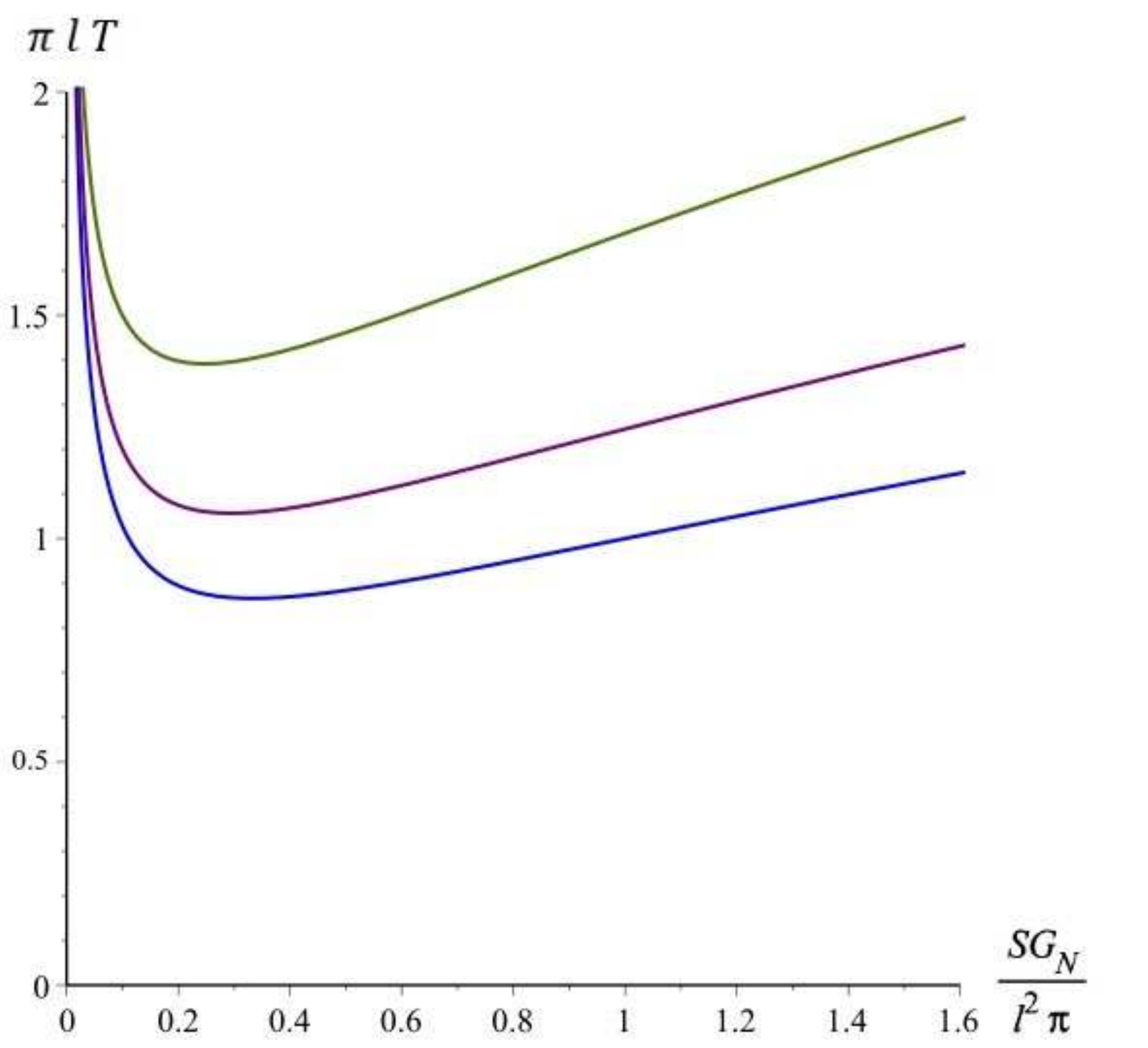}
  		\caption{(e)}
  	\end{subfigure}
  	\begin{subfigure}[b]{0.4\textwidth}
  		\includegraphics[width=\textwidth]{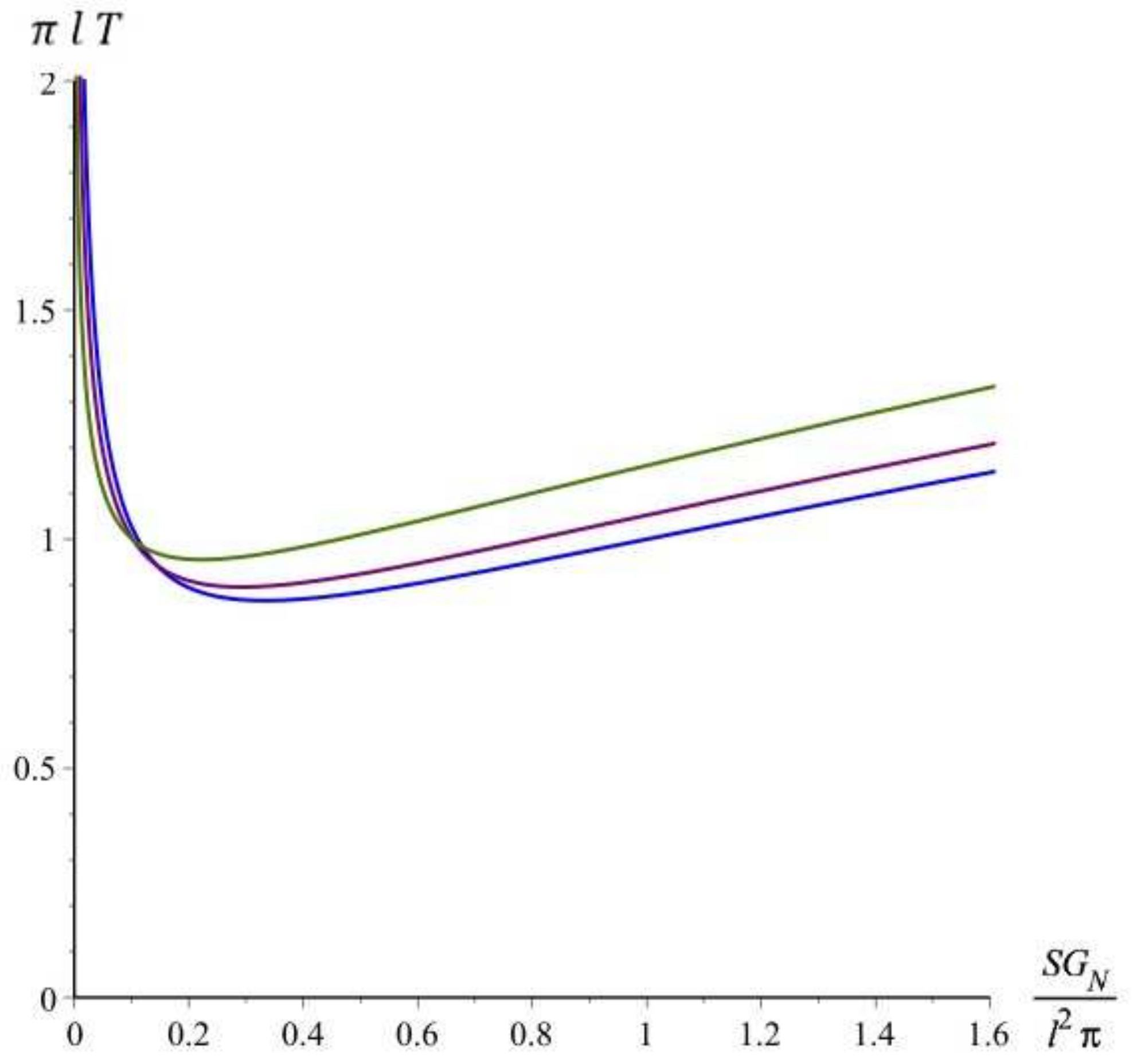}
  		\caption{(f)}
  	\end{subfigure}
  	\caption{(e) Temperatura adimensional versus entrop\'ia adimensional, para diferentes valores de $\nu$, $\alpha=10 l^{-2}$ y campo escalar negativo. Los gr'aficos son para  $\nu=1,\nu=1.9$ y $\nu=3$ (de izquierda a derecha). Aqu\'i es posible ver la existencia de peque\~nos y grandes agujeros negros, ex\'actamente similar a Schwarzschild AdS.\\
  		(f) Masa adimensional versus temperatura adimensional, para diferentes valores de $\nu$, $\alpha=-10 l^{-2}$ y campo escalar positivo. Los gr'aficos son para  $\nu=1,\nu=1.9$ y $\nu=3$ (de izquierda a derecha). Aqu\'i es posible ver la existencia de peque\~nos y grandes agujeros negros, ex\'actamente similar a Schwarzschild AdS.}
  \end{figure}
 La rama entera de peque\~nos agujeros negros (para ambas familias) es inestable termodin\'amicamente y tiene energ\'ia libre positiva, mientras que los agujeros negros grandes son estables termodin\'amicamente y la energ\'ia libre negativa para temperaturas $T>T_{c}$. 
 
 \newpage
 
 A diferencia de los agujeros negros planares, para el cual no existen transiciones de fase de primer orden respecto AdS. Las soluciones de  agujeros negros con pelo escalar y horizonte esf\'erico, existen transiciones de fase de primer orden respecto a AdS a temperatura finita --- las soluciones de los agujeros negros grandes que tienen energ\'ia libre negativa respecto a AdS son claramente preferidas. 
 
 \subsection{Discusi\'on}
 
 Desde el punto de vista de la
 dualidad AdS/CFT, el estudio de la 
 termodin\'amica de agujeros negros
 asint\'oticamente AdS es relevante para
 entender los diagramas de fase de algunas
 teor\'ias duales del campo hologr\'aficas.
 Investigamos la termodin\'amica de una clase
 general de agujeros negros con pelo con condiciones
 de borde para el campo escalar de masa conforme $m^{2}=-2/l^{2}$, los cuales preservan las
 isometr\'ias de AdS. Es importante remarcar
 la cercana similitud que observamos con la estructura
 familiar del agujero negro SAdS. Los agujeros negros
 grandes son termodin\'amicamente estables, y los peque\~nos
 tienen calor espec\'ifico negativo.
 
 Calculamos la acci\'on Euclidea (y el potencial termodin\'amico) usando el formalismo cuasilocal
 suplementado con contrat\'erminos. Usando esos resultados,
 mostramos que existen transiciones de fase de primer orden
 entre AdS t\'ermico y el agujero negro con pelo. Por
 otro lado, comparando la energ\'ia libre del agujero negro
 con pelo con el de la soluci\'on SAdS, 
 se muestra que el agujero negro SAdS es siempre
 preferido.

 \section{Transiciones de fase de agujeros negros planares}

 En esta secci\'on construimos soluciones neutras de solitones con pelo escalar en un espacio-tiempo asint\'oticamente AdS \cite{Maldacena:1997re}. Este an\'alisis es importante en el contexto de la dualidad AdS/CFT porque las soluciones del bulk
 corresponden a fases de la teor\'ia dual del campo \cite{Witten:1998zw}.
 
 Consideramos teor\'ias de gravedad
 acoplados m'inimamente a un campo escalar con 
 un potencial $V(\phi)$. Siendo que para la misma auto-interacci\'on existen distintas condiciones de borde para el campo escalar (que pueden o no romper la simetr\'ia conforme), uno puede especificar la teor\'ia del campo \cite{Hertog:2004ns} con un potencial efectivo  \cite{Hertog:2004ns,Hertog:2004rz, Anabalon:2015vda}.  
 
 Diferentes foliaciones del espacio-tiempo AdS llevan a diferentes definiciones del tiempo y por lo tanto a distintos Hamiltonianos  de la
 teor\'ia del campo dual. Ya que el cl\'asico 
 background de (super)gravedad, con posibles correcciones $\alpha^{'}$, es equivalente a 
 una teor\'ia cu\'antica gauge en el correspondiente 
 cascar\'on\footnote{Se refiere a una de las hiper-superficies, definidas por una folicaci\'on tipo-tiempo ($r=R=constante$).}, uno espera diferentes teor\'ias gauge f\'isicamente distintas para distintas foliaciones. De hecho cuando la topolog\'ia del horizonte es Ricci-plano y sin direcciones compactas, no existen transiciones de fase de primer orden similares a las transiciones de fase de Hawking-Page \cite{Hawking} que existen para agujeros negros esf\'ericamente sim\'etricos. 
  
 Sin embargo, cuando algunas de las direcciones
 espaciales son compactificadas asint\'oticamente en un c\'irculo, uno
 espera la existencia de una energ\'ia negativa
 de Casimir de una teor\'ia del campo
 no-supersim\'etrica que vive en la topolog\'ia
 correspondiente. Horowitz y Myers mostrar\'on
 en \cite{Horowitz:1998ha} que, de hecho, existe una soluci\'on (bulk) gravitacional apodado como \textsl{AdS solit\'on}\footnote{Como veremos m\'as adelante, una de las direcciones de la m\'etrica del solit'on es peri\'odica por lo que puede interpretarse como un n\'umero topol\'ogico que caracteriza la m\'etrica del solit\'on. Adem\'as esta tiene una energ\'ia. Esto encaja dentro del concepto de solit\'on en f\'isica no lineal y teor\'ia cu\'antica de campos.} con una energ\'ia mas baja que el de AdS. Esta soluci\'on fue
 obtenida por una doble continuaci\'on anal\'itica (una en la coordenada temporal y
 la otra en una de las direcciones angulares compactificadas) del agujero negro planar. Esta encaja de una manera muy bonita con la propuesta
 de Witten \cite{Witten:1998zw} donde una teor\'ia gauge no-supersim\'etrica de Yang-Mills
 puede ser descrita dentro de la dualidad AdS/CFT
 compactificando una de las direcciones e imponiendo condiciones de borde antiperi\'odicas para los fermiones alredor del c\'irculo.
 
 Los solitones AdS neutros con pelo escalar
 fueron previamente analizados (ver, e.g. \cite{Brihaye:2013tra, Ogawa:2011fw, Shi:2016bxz, Kleihaus:2013tba,
 	Brihaye:2012ww, Smolic:2015txa, Cadoni:2011yj}), aunque varios de esos
 estudios usan m'etodos num\'ericos. Por lo tanto, es interesante encontrar ejemplos de
 solitones AdS con pelo anal\'iticos e investigar sus propiedades gen\'ericas. En 
 a\~nos recientes, estas soluciones anal\'iticas
 fueron construidas, por ejemplo en \cite{Acena:2013jya, Acena:2012mr,Lu:2013ura, Feng:2013tza,
 Wen:2015xea, Fan:2015tua, Fan:2015ykb}. De all\'i que la construcci\'on de soluciones anal\'iticas si es posible. Usamos una de las soluciones exactas particulares de agujeros negros obtenidas en      
 \cite{Acena:2013jya,Acena:2012mr} y obtendremos
 los correspondientes solitones usando la doble
 continuaci\'on anal\'itica como en \cite{Horowitz:1998ha}. Los solitones con 
 pelo AdS son los candidatos del estado ground
 de la teor\'ia \cite{Woolgar:2016axs}.
 
 Ya que el solit\'on AdS es la soluci\'on
 con la energ\'ia m\'inima dentro de 
 condiciones de borde dadas \cite{Galloway:2001uv,Galloway:2002ai},
 es natural investigar la existencia de 
 transiciones de fase respecto a este 
 background t\'ermico. En un bonito trabajo
 \cite{Surya:2001vj}, mostrar\'on que existe 
 transiciones de fase de primer orden entre
 los agujeros negros planares y el solit\'on
 AdS. Construimos el solit\'on AdS con pelo
 y calculamos su masa por el m'etodo de
 Balasubramanian y Kraus \cite{Balasubramanian:1999re} suplementado
 con contrat\'erminos extra para el campo escalar
 como fue propuesto en \cite{Anabalon:2015xvl}.
 Investigamos la existencia de transiciones de fase de primer orden respecto al solit\'on AdS
 con pelo y discutimos el efecto del 
 pelo en el comportamiento termodin\'amico.
 
 \subsection{AdS solit\'on}

 Comenzaremos con una peque\~na 
 revisi\'on de \cite{Surya:2001vj}.
 Sin embargo, para conectar este an\'alisis
 con el resto de la secci\'on, los c\'alculos
 se har\'an usando el m'etodo de contrat\'erminos de Balasubramanian y Krauss \cite{Balasubramanian:1999re}.  
 Consideramos la acci\'on usual
 para la gravedad suplementado con el 
 contrat\'ermino gravitacional\footnote{Aqu\'i el t\'ermino de la curvatura escalar de Ricci es nulo $\mathcal{R}=0$. Debido a que la m\'etrica del borde ($r=\infty$) es plano i.e. $\mathcal{R}^{a}_{bcd}=0$.} propuesto en
 \cite{Balasubramanian:1999re}
  \begin{equation}
 I[g_{\mu\nu}]=\int_{\mathcal{M}}d^{4}x\left(  R-2\Lambda\right)  \sqrt
 {-g}+2\int_{\partial\mathcal{M}}{d^{3}x~K\sqrt{-h}}-\int_{\partial\mathcal{M}%
 }{d^{3}x~\frac{4}{l}\sqrt{-h}}~, \label{actionSchw}%
 \end{equation}
 donde $\Lambda=-3/l^{2}$ es la constante cosmol\'ogica ($l$ es el radio de AdS), $16\pi G=1$ con $G$ la constante gravitacional de Newton, el segundo t\'ermino
 es el t\'ermino de borde de Gibbons-Hawking,
 y el \'ultimo t\'ermino es el contrat\'ermino
 gravitacional. Aqu\'i, $h$ es el determinante
 de la m\'etrica inducida del borde y $K$ 
 es la traza de la curvatura extr\'inseca.
 La soluci\'on del agujero negro planar es
 \begin{equation}
 ds^{2}=-\biggl{(}-\frac{\mu_{b}}{r}+\frac{r^{2}}{l^{2}}\biggr{)}dt^{2}%
 +\biggl{(}-\frac{\mu_{b}}{r}+\frac{r^{2}}{l^{2}}\biggr{)}^{-1}dr^{2}%
 +\frac{r^{2}}{l^{2}}(dx_{1}^{2}+dx_{2}^{2})~, \label{bh}%
 \end{equation}
 donde $\mu_{b}$ es el par\'ametro de masa
 y consideraremos las coordenadas compactificadas $0\leq x_{1}\leq L_{b}$ and $0\leq x_{2}\leq L$~\footnote{El horizonte es un toro: $S^{1}\times S^{1}$.}. La normalizaci\'on es tal
 que la coordenada temporal y las coordenadas $x_{1}$ y $x_{2}$ tienen la misma dimensi\'on 
 y de esta manera la continuaci\'on anal\'itica
 para obtener el soliton AdS produce la misma
 geometr\'ia del borde.\\
 El rol de contrat\'ermino es cancelar la divergencia infraroja de la acci\'on de esta forma el resultado final es 
 finito\footnote{Este c\'alculo se puede hacer usando directamente el m\'etodo propuesto en la secci\'on \ref{counterregu}}: 
 \begin{equation}
 I^{E}_{b}=\frac{2L L_{b}\beta_{b}}{l^{4}}\biggl{(}-r^{3}_{h}+\frac{\mu
 	_{b}l^{2}}{2}\biggr{)}=-\frac{L L_{b}\beta_{b}r^{3}_{h}}{l^{4}}~.
 \label{Schwact}%
 \end{equation}
 El radio del horizonte es denotado por
 $r_{h}$ y $\beta_{b}$ es la periodicidad
 del tiempo Euclideo que esta relacionada
 a la temperatura del agujero negro por:
 \begin{equation}
 T = \beta_{b}^{-1}=\frac{(-g_{tt})^{^{\prime}}}{4\pi}\biggr{\vert}_{r=r_{h}%
 }=\frac{3r_{h}}{4\pi l^{2}}~. \label{schtemp}%
 \end{equation}
 Usando las relaciones termodin\'amicas usuales
 y la energ\'ia libre $F = I^{E}_{b}/
 \beta_{b}$, obtenemos la energ\'ia y entrop\'ia 
 del agujero negro planar:
 \begin{equation}
 \label{enT}E=-T^{2}\frac{\partial I_{b}^{E}}{\partial T}=\frac{2LL_{b}\mu_{b}%
 }{l^{2}}~,%
 \end{equation}
 \begin{equation}
 S=-\frac{\partial(I_{b}^{E}T)}{\partial T}=\frac{LL_{b}r_{h}^{2}}{4l^{2}G%
 }=\frac{\mathcal{A}}{4G}~.%
 \end{equation}
 
 Por completitud, presentaremos el cuasi-local stress tensor
 de Brown and York \cite{Brown:1992br}. Este tensor
 esta relacionado con el stress tensor de la teor\'ia del campo dual, salvo un factor conforme. Considerando la folicaci\'on $r=R=constante$. La m'etrica del bulk esta asociada con una estructura conforme en el borde y la geometr\'ia donde la teor\'ia del campo esta definida se relaciona con la geometr\'ia del borde mediante una trasformaci\'on conforme:    
 \begin{equation}
 ds_{dual}^{2} = \frac{l^2}{R^2}ds^2 = \gamma_{ab}dx^{a}dx^{b}=-dt^{2}+dx_{1}^{2}+dx_{2}^{2}~.
 \label{cftschw}
 \end{equation}
 El stress tensor dual es \cite{Myers:1999psa} 
 \begin{equation}
 \langle\tau_{ab}^{dual}\rangle=\lim_{R\rightarrow\infty}\frac{R}{l}\tau_{ab}=\frac{\mu_{b}}{16\pi G_{N}l^{2}}[3\delta_{a}^{0}\delta_{b}^{0}+\gamma_{ab}]~,
 \end{equation}
 donde $\tau_{ab}$ es el Brown-York stress tensor. Vemos que el stress tensor dual es similar a la de un gas t\'ermico de part'iculas sin masa, y como es de esperarse, su traza es cero 
 $\langle\tau^{dual}\rangle=\langle\tau_{ab}^{dual}\rangle\gamma^{ab}=0$.
 La energ\'ia es la carga conservada asociada con la isometr\'ia (de la m\'etrica del borde) generada por el vector de Killing $\xi^{a}=(\pa_{t})^{a}$.  Obtenemos  
 \begin{equation}
 E=Q_{\xi_{t}}=\int{d\Sigma^{i}\tau_{ij}\xi^{j}} = \frac{L L_{b}}{l^{2}\kappa}\bl{[}\mu_{b}+\frac{l^{2}}{4R}+O(R^{-2})\br{]}, 
 \label{charge}     
 \end{equation}
 que esta en acuerdo con (\ref{enT}).
 El solit\'on AdS fue obtenido en \cite{Horowitz:1998ha}
 \begin{equation}
 ds^{2}=-\frac{r^{2}}{l^{2}}d\tau^{2}+\biggl{(}-\frac{\mu_{s}}{r}+\frac{r^{2}%
 }{l^{2}}\biggr{)}^{-1}dr^{2}+\biggl{(}-\frac{\mu_{s}}{r}+\frac{r^{2}}{l^{2}%
 }\biggr{)}d\theta^{2}+\frac{r^{2}}{l^{2}}dx_{2}^{2}~, \label{Schwsol}%
 \end{equation}
usando una doble continuaci\'on anal\'itica 
$t\rightarrow i\theta$, $x_{1}%
\rightarrow i\tau$ de la m\'etrica agujero negro planar (\ref{bh}). Para distinguir de la 
soluci\'on del agujero negro, denotamos por $\mu_{s}$ el par\'ametro de masa
del solit\'on AdS y, en la secci\'on Euclidea
($\tau\rightarrow i\tau_{E}$), la periodicidad
es $0\leq\tau_{E}\leq\beta_{s}$. Para obtener
una soluci\'on Lorentziana regular, la coordenada
$r$ esta restringida a $r_{s}\leq r$, donde
\begin{equation}
-\frac{\mu_{s}}{r_{s}}+\frac{r_{s}^{2}}{l^{2}}=0.
\end{equation}
Para evitar la singularidad c\'onica en el
plano $(r,\theta)$, imponemos la siguiente
periodicidad\footnote{Esta f\'ormula se demostr\'o en la secci\'on \ref{formulatemp}} para $\theta$:
\begin{equation}
\label{periosol}L_{s}=\frac{4\pi\sqrt{g_{\theta\theta}g_{rr}}}{(g_{\theta
		\theta})^{^{\prime}}}\biggr{\vert}_{r=r_{s}}=\frac{4\pi l^{2}}{3r_{s}}~.%
\end{equation}
La acci\'on on-shell Euclidea y la masa del solit\'on
AdS pueden ser obtenidas de forma similar al del agujero negro (pero no 
presentaremos los detalles aqu'i):
\begin{equation}
I^{E}_{s} = -\frac{LL_{s}\beta_{s}\mu_{s}}{l^{2}} \label{Schwactsol}%
\end{equation}
y la masa puede ser obtenida usando las relaciones 
termodin\'amicas con la energ\'ia libre
$F=I^{E}_{s}/\beta_{s}=M$ (o del tensor de stress cuasilocal)
y el resultado es
\begin{equation}
M=-\frac{LL_{s}\mu_{s}}{l^{2}}~.%
\end{equation}
La masa del solit\'on AdS corresponde a la energ\'ia Casimir asociada a las direcciones 
compactas de la teor\'ia dual en el borde y es negativa \cite{Horowitz:1998ha}. 
Las energ\'ias libres (acci\'on on-shell) del agujero negro y del solit\'on respectivamente (\ref{Schwact}),(\ref{Schwactsol}) son negativas de all\'i que son termodin\'amicamente estables (no hay transiciones de fase) respecto al background AdS.

En \cite{Surya:2001vj} se muestra que existen
transiciones de fase de primer orden entre agujeros negros
y el solit\'on AdS. Con toda la informaci\'on obtenida hasta ahora, es sencillo verificar la existencia de las transiciones de fase de primer orden\footnote{La transici\'on de fase de primer orden plantea que la derivada de la energ\'ia libre es discont\'inua, para nuestro caso $S=-\pa F/\pa T$. Entonces quiere decir que la entrop\'ia es discontinua en el punto cr\'itico
definido por $F=0$ y que respecto a ese punto cr\'itico la energ\'ia libre es $F>0$ (t\'ermicamente inestable) o $F<0$ (t\'ermicamente estable).\\
Para el caso de transiciones de fase de segundo orden,  la primera derivada de $F$ es continua pero la segunda derivada es discontinua.}.
Para comparar las soluciones Euclideas, 
debemos imponer las mismas condiciones de periodicidad, 
las cuales en el borde ($r\rightarrow \infty$) son: $\beta_{b}=\beta_{s}$ y $L_{s}=L_{b}$. Comparemos ahora 
las acciones on-shell (\ref{Schwact}),(\ref{Schwactsol}), 
esto equivale a comparar las energ\'ias libres:
\begin{equation}
\Delta I=I_{b}^{E}-I_{s}^{E}=\frac{L}{2\kappa l^{4}}
\bl{(}\frac{4\pi l^{2}}{3}\br{)}^{3}L_{b}\beta_{b}(L_{s}^{-3}-\beta_{b}^{-3})=\frac{L}{2\kappa l^{4}}
\bl{(}\frac{4\pi l^{2}}{3}\br{)}^{3}L_{b}\beta_{b}\bl{(}\frac{1}{L_{s}^{3}}-T^{3}\br{)}~.
\label{bhsolfree}
\end{equation}
El cambio de signo es indicaci\'on de una transici\'on de
fase a primer orden entre el agujero negro planar
y el solit\'on AdS. Vemos que a diferencia de la transici\'on
de fase Hawking-Page para agujeros negros esf\'ericos AdS, 
esta es controlada por la proporci\'on $r_{h}/l$ (par\'ametro de orden), en el caso planar la proporci\'on $r_{h}/r_{s}$ es relevante\footnote{Recuerde que $r_{h}$ es radio del  horizonte del agujero negro y $r_{s}$ es el radio m\'inimo que la m\'etrica del solit\'on puede describir $r_{s}\leq r$ .}. Por consiguiente debemos comparar la periodicidad del tiempo Euclideo (el inverso de la temperatura) con la periodicidad de las coordenadas compactas obtenidas por la doble continuaci\'on anal\'itica.    

Para los agujeros negros esf\'ericos AdS \cite{Witten:1998zw,Hawking}, cuando $r_h<l$
el espacio-tiempo AdS a temperatura finita domina y cuando 
$r_h>l$ los agujeros negros grandes son t\'ermicamente estables. 
Para el caso planar, hay cambio dr\'astico \cite{Surya:2001vj}. Esto es, la estabilidad del agujero negro planar no depende s\'olo de la temperatura, tambien depende de su tama\~no. Escribiendo el \'area del agujero negro $\mathcal{A}=LL_{b}r_{h}^{2}/l^{2}$ en t\'erminos de $T$ y $L_{s}$ como en \cite{Surya:2001vj}, encontramos la proporci\'on     
\begin{equation}
\frac{\mathcal{A}}{Tl^{3}}=\frac{L}{l}\bl{(}\frac{4\pi}{3}\br{)}^{2}L_{s}T~.
\end{equation}
Si $\mathcal{A}l^{-2}<<Tl$ (y $L$ del mismo orden del radio AdS $l$) entonces $L_{s}T<<1$, y usando (\ref{schtemp}) y (\ref{periosol}) se verifica que $r_{h}<<r_{s}$ y $F>0$, 
lo que significa que los agujeros negros peque\~nos y calientes (respecto a $r_{s}$) son inestables y decaen a agujeros negros peque\~nos. Por otro lado, si $\mathcal{A}l^{-2}>>Tl$ entonces $L_{s}T>>1$, y usando (\ref{schtemp}) y (\ref{periosol}) obtenemos $r_{s}<<r_{h}$ y $F<0$, esto es una indicaci\'on de que los agujeros negros grandes y fr\'ios son estables. Cuando 
$\mathcal{A} \sim Tl^3$ el solit\'on y el agujero negro estan en equilibrio e incluye los casos cuando son fr\'ios y peque\~nos o grandes y calientes. El punto cr\'itico es cuando $\Delta I=0$, para el cual $1=L_{s}T$ y $r_{h}=r_{s}$. Note que la transici\'on
de fase esta controlada por el par\'ametro
adimensional $z=TL_{s}$.
\newpage
\subsection{Solit\'on AdS con pelo}
Consideramos las soluciones regulares exactas 
de agujeros negros con pelo escalar 
\cite{Acena:2013jya,Acena:2012mr,Anabalon:2013qua}. La acci\'on es
\begin{equation}
I[g_{\mu\nu},\phi]=\int_{\mathcal{M}}{d^{4}x\sqrt{-g}\biggl{[}R-\frac
	{(\partial\phi)^{2}}{2}-V(\phi)\biggr{]}}+2\int_{\partial\mathcal{M}}%
{d^{3}xK\sqrt{-h}} \label{action}%
\end{equation}
y estamos interesados en el siguiente potencial 
moduli:
\footnote{Para algunos de los valores particulares de los par\'ametros, viene a ser una de las truncasiones de $\omega$-deformed gauged $\mathcal{N}=8$ supergravedad \cite{Dall Agata:2012bb}
	, ver \cite{Anabalon:2013eaa, Guarino:2013gsa, Tarrio:2013qga}.}
\begin{align}
V(\phi)  &  =\frac{\Lambda(\nu^{2}-4)}{3\nu^{2}}\biggl{[}\frac{\nu-1}{\nu
	+2}e^{-\phi l_{\nu}(\nu+1)}+\frac{\nu+1}{\nu-2}e^{\phi l_{\nu}(\nu-1)}%
+4\frac{\nu^{2}-1}{\nu^{2}-4}e^{-\phi l_{\nu}}\biggr{]}\\
&  +\frac{2\alpha}{\nu^{2}}\biggl{[}\frac{\nu-1}{\nu+2}\sinh{\phi l_{\nu}%
	(\nu+1)}-\frac{\nu+1}{\nu-2}\sinh{\phi l_{\nu}(\nu-1)}+4\frac{\nu^{2}-1}%
{\nu^{2}-4}\sinh{\phi l_{\nu}}\biggr{]}~.\nonumber\\
\nonumber
\end{align}
Nos concentraremos en el caso 
concreto de $\nu=3$. Sin embargo, 
los solitones AdS con pelo para
otros valores de $\nu$ probablemente 
tambi\'en existen, pero el an\'alisis
es t\'ecnicamente m\'as complicado y 
no lo presentaremos en el presente trabajo.
En este caso el potencial del campo escalar
vienea a ser \newline  
\begin{align}
\label{potential}V(\phi)  &  =\frac{2\Lambda}{27}\biggl{(}5e^{-\phi\sqrt{2}%
}+10e^{\phi\sqrt{2}/2}+16e^{-\phi\sqrt{2}/4}\biggr{)}\\
&  +\frac{4\alpha}{45}\biggl{[}\sinh{\biggl{(}\phi\sqrt{2}\biggr{)}}%
-10\sinh{\biggl{(}\phi\sqrt{2}/2\biggr{)}}+16\sinh{\biggl{(}\phi\sqrt
	{2}/4\biggr{)}}\biggr{]}~.\nonumber\\
\nonumber
\end{align}
El potencial tiene dos partes que son controladas
por los par\'ametros $\Lambda$ y $\alpha$.
Asint\'oticamente, donde el campo escalar desaparece, justamente el par\'ametro $\Lambda$
sobrevive y esta relacionado con el radio AdS como $\Lambda=-3l^{-2}$.
Usando el ansatz para la m\'etrica dada en (\ref{Ansatz}),  
las ecuaciones de movimiento pueden ser integradas para el factor conforme
\cite{Acena:2013jya, Acena:2012mr, Anabalon:2013qua, Anabalon:2012ta}
\begin{equation}
\Omega(x)=\frac{9x^{2}}{\eta^{2}(x^{3}-1)^{2}}~. \label{omega}%
\end{equation}
Con esta elecci\'on del factor conforme,
es sencillo obtener las expresiones para
el campo escalar
\begin{equation}
\phi(x)=2\sqrt{2}\ln{x}%
\end{equation}
y la funci\'on m\'etrica%
\begin{equation}
f(x)=\frac{1}{l^{2}}+\alpha\biggl{[}\frac{1}{5}-\frac{x^{2}}{9}%
\biggl{(}1+x^{-3}-\frac{x^{3}}{5}\biggr{)}\biggr{]}. \label{f}%
\end{equation}
Donde $\eta$ es s\'olo una constante de 
integraci\'on. El par\'ametro $\alpha$ es
positivo para $x<1$ y negativo para $x>1$. 
Nos concentraremos en el caso $x<1$.
El borde conforme esta en $x=1$, donde
la m\'etrica viene a ser
\begin{equation}
ds^{2}=\frac{R^{2}}{l^{2}}\biggl{[}-dt^{2}+dx_{1}^{2}+dx_{2}^{2}\biggr{]}.
\label{Bblack}%
\end{equation}
Usaremos la siguiente notaci\'on 
para el factor conforme:
\begin{equation}
R^{2}\equiv\frac{1}{\eta^{2}(x-1)^{2}}~.%
\end{equation}
La geometr\'ia donde la teor\'ia dual
vive tiene la m\'etrica
\begin{equation}
ds_{dual}^{2} = \frac{l^{2}}{R^{2}}ds^{2} = \gamma_{ab}dx^{a}dx^{b}%
=-dt^{2}+dx_{1}^{2}+dx_{2}^{2}~. \label{cftschw}%
\end{equation}
La acci\'on Euclidea regularizada para 
estos agujeros negros fue obtenida en  
\cite{Anabalon:2015xvl,Anabalon:2017eri} (ver, tambi\'en, \cite{Anabalon:2016yfg}) (en lo que sigue usaremos la siguientes notaciones similares a la anterior secci\'on para $\beta_{b}$ y
$L_{b}$):
\begin{equation}
I^{E}_{BH}=\beta_{b}\biggr{(}-\frac{\mathcal{A}T}{4G_{N}}+\frac{2LL_{b}}%
{l^{2}}\frac{\alpha}{3\eta^{3}}\biggr{)}=-\frac{LL_{b}\alpha\beta_{b}}%
{3l^{2}\eta^{3}}~, \label{Ibh}%
\end{equation}
donde el \'area del horizonte y la temperatura
del agujero negro son
\begin{equation}
\mathcal{A}=\frac{LL_{b}\Omega(x_{h})}{l^{2}}, \qquad T=\frac{\alpha}{4\pi
	\eta^{3}\Omega}~.%
\end{equation}
La masa del agujero negro es \cite{Anabalon:2015xvl,Anabalon:2014fla}
\begin{equation}
M_{b}= \frac{2LL_{b}\mu_{b}}{l^{2}}, \qquad\mu_{b}=\frac{\alpha}{3\eta^{3}}
\label{massbh}%
\end{equation}
tambi\'en se puede verificar usando las 
relaciones termodin\'amicas usuales, y por
supuesto se verifica la primera ley.
Ahora construiremos el solit\'on AdS. 
Usando de nuevo la doble continuaci\'on 
anal\'itica $x_{1}\rightarrow i\tau$ y $t\rightarrow i \theta$ en
(\ref{Ansatz}), la m\'etrica viene a ser
\begin{equation}
ds^{2}=\Omega_{s}(x)\left[  -\frac{d\tau^{2}}{l^{2}}+\frac{\lambda^{2}dx^{2}%
}{f(x)}+f(x)d\theta^{2}+\frac{dx_{2} ^{2}}{l^{2}}\right]  . \label{anzatsol}%
\end{equation}
De manera similar al agujero negro con pelo,
el factor conforme (\ref{omega}) es
\begin{equation}
\Omega_{s}(x)=\frac{9x^{2}}{\lambda^{2}(x^{3}-1)^{2}}~, \label{omegasol}%
\end{equation}
pero ahora denotaremos la constante
de integraci\'on con $\lambda$ para
distinguir de la constante de integraci\'on
$\eta$ del agujero negro. Para evadir
la singularidad c\'onica en el plano
$(x,\theta)$, imponemos la periodicidad: 
\begin{equation}
L _{s}=\frac{4\pi\lambda}{f^{^{\prime}}}\biggr{\vert}_{x=x_{s}}=\frac
{4\pi\lambda^{3}\Omega_{s}}{\alpha}~, \label{periosoliton}%
\end{equation}
donde $x_{s}$ es el m\'inimo valor de $x$,
concr\'etamente la ra\'iz mayor de $f(x_{s})=0$.
Despu\'es de imponer la periodicidad en $\theta$
y restringir la coordenada $x$ de tal forma que la m\'etrica es Lorentziana, obtendremos una 
soluci\'on regular bien definida
Usaremos el metodo de \cite{Anabalon:2015xvl}
para calcular la acci\'on regularizada Euclidea
y el resultado es
\begin{equation}
I_{soliton}^{E}=-\frac{L\beta_{s}\Omega_{s}(x_{s})}{4l^{2}G_{N}}+\frac
{2LL_{s}\beta_{s}}{l^{2}}\frac{\alpha}{3\lambda^{3}}=-\frac{LL_{s}\beta_{s}%
}{l^{2}}\biggl{(}\frac{\alpha}{3\lambda^{3}}\biggr{)}~, \label{Isol}%
\end{equation}
de donde la masa puede ser inmediatamente le\'ida:
\begin{equation}
M_{soliton}=-\frac{LL_{s}\mu_{s}}{l^{2}}, \qquad\mu_{s}=\frac{\alpha}%
{3\lambda^{3}}~. \label{masssoliton}%
\end{equation}
Por completitud, calculamos 
el strees tensor cuasi-local de Brown y York
y calculamos la masa del solit\'on
a partir de ella. El nuevo 
contrat\'ermino para el campo escalar 
da una nueva contribuci\'on al stress tensor
\begin{equation}
I^{\phi}=-\int{d^{3}x\sqrt{-h}\bl{(}\frac{\phi^{2}}{2l}-\frac{l_{\nu}}{6l}\phi^{3}\br{)}}~, \qquad \tau_{ab}^{\phi}=-\frac{2}{\sqrt{-h}}\frac{\delta I^{\phi}}{\delta h^{ab}}~.
\end{equation}
Entonces el stress tensor renormalizado es
\begin{equation}
\tau_{ab}=-\frac{1}{\kappa}\bl{(}K_{ab}-h_{ab}K+\frac{2}{l}h_{ab}-lE_{ab}\br{)}-\frac{h_{ab}}{l}\bl{(}\frac{\phi^{2}}{2}-\frac{l_{\nu}}{6}\phi^{3}\br{)}~,
\label{stresstensor}
\end{equation}
donde las componentes son:
\begin{equation}
\tau_{\tau\tau}=\frac{\alpha(x-1)}{3\lambda^{2}l}+O[(x-1)^{2}]~,
\end{equation}
\begin{equation}
\tau_{\theta\theta}=\frac{2\alpha (x-1)}{3\lambda^{2}l}+O[(x-1)^{2}]~,
\end{equation}
\begin{equation}
\tau_{x_{2}x_{2}}=-\frac{\alpha (x-1)}{3\lambda^{2}l}+O[(x-1)^{2}]~.
\end{equation}
La m\'etrica (\ref{anzatsol}) en el borde esta 
relacionada con la m\'etrica conforme como
\begin{equation}
ds_{dual}^{2}=\frac{l^{2}}{R^{2}}ds^{2}=-d\tau^{2}+d\theta^{2}+dx_{2}^{2}~,
\end{equation}
donde $R\equiv\- 1/\eta(x-1)$, y en el borde $R\rightarrow\infty$ implica $x\rightarrow x_{b}$. Entonces el stress tensor de la teor\'ia dual al solit\'on es: 
\begin{equation}
\langle\tau_{ab}^{dual}\rangle=\lim_{R\rightarrow\infty}\frac{R}{l}\tau_{ab}=\lim_{x\rightarrow x_{b}}\bl{[}-\frac{1}{\lambda l(x-1)}\br{]}\tau_{ab}=\frac{1}{l^{2}}\bl{(}\frac{\alpha}{3\lambda^{3}}\br{)}[-3\delta_{a}^{\theta}\delta_{b}^{\theta}+\gamma_{ab}]~.
\end{equation}
La cantidad conservada (masa del solit\'on) generada 
por el vector de Killing  $\xi^{i}=(\pa_{\tau})^{i}$ es: 
\begin{equation}
M=\oint_{\Sigma}{d^{2}y\sqrt{\sigma}m^{a}\tau_{ab}\xi^{b}}=\frac{L L_{s} f^{1/2}\Omega}{\sqrt{-g_{\tau\tau}}}(\pa_{\tau})^{i}\tau_{ij}(\pa_{\tau})^{j}=-\frac{LL_{s}}{ l^{2}}\bl{[}\frac{\alpha}{3\lambda^{3}}+O(x-1)\br{]}~,
\end{equation}
donde la foliaci\'on es de tipo espacio
${\tau=constante}$, con el vector normal y unitario: $m^{i}=\frac{(\pa_{t})^{i}}{\sqrt{-g_{\tau\tau}}}$,
\begin{equation}
ds^{2}=\sigma_{ij}dx^{i}dx^{j}=\Omega(x)\left[f(x)d\theta^{2}+\frac{dx_{2}^{2}}{l^{2}}\right]~.
\end{equation}  
\subsection{Implicaciones para las transiciones de fase}

En el marco de la dualidad AdS/CFT, los agujeros negros son interpretados como estados t\'ermicos en la teor\'ia dual del campo. Mostraremos que existen transiciones de fase de primer orden 
entre el agujero negro con pelo planar y el solit\'on con pelo AdS.

Con los resultados de la secci\'on
previa, estamos listos para investigar 
la existencia de las transiciones de fase.   
\footnote{El caso $k=1$, cuando la topolog\'ia del horizonte es esf\'erico, fue estudiado en  \cite{Anabalon:2015ija}.} Nos concentraremos
en el caso $D=4$. Antes de comparar las
acciones, es importante puntualizar que de las
definiciones de $x_{s}$ y $x_{h}$ se concluye
que son iguales, $x_{s}=x_{h}$. A primer vista,
esto puede ser un poco extra\~no porque 
en general se espera estos dependan 
de los par\'ametros de masa $\lambda$ y 
$\eta$ para el solit\'on y el agujero negro.
Sin embargo, en estas coordenadas poco usuales,
$x_{s}$ y $x_{h}$ est'an definidas por (\ref{f}),
pero la verdadera \'area del horizonte y
centro del solit\'on est'an determinados por
el factor conforme en frente de la m\'etrica.
Este factor conforme depende del par\'ametro de
masa, as\'i que definimos:  

\begin{equation}
r_{b}^{2}=\frac{\Omega(x_{h},\eta)}{l^{2}},\qquad r_{s}^{2}=\frac{\Omega
	(x_{s},\lambda)}{l^{2}}~.\label{ratio}%
\end{equation}

Como antes (\ref{bhsolfree}), necesitamos
comparar las energ\'ia libres de las soluciones
en la misma teor\'ia y debemos imponer
las mismas condiciones de periodicidad
en el borde $\beta_{b}=\beta_{s}$ y $L_{s}=L_{b}$. El solit\'on AdS con pelo
tiene energ\'ia negativa (el espacio
AdS en coordenadas planares tiene masa cero)
y este es el ground state de la teor\'ia.
De all\'i que, la energ\'ia del agujero negro
con pelo debe ser calculado respecto a este
estado base y obtenemos 
\begin{equation}
E=M_{bh}-M_{soliton}=\frac{LL_{b}}{l^{2}}(2\mu_{b}+\mu_{s})~,
\end{equation}
con $\mu_{b}$ y $\mu_{s}$ definidas en (\ref{massbh}) y (\ref{masssoliton}).
Las misma periodicidad del tiempo Euclideo
implica la misma temperatura. Consideramos 
la soluci\'on del solit\'on con pelo como
el background t\'ermico:
\begin{equation}
\Delta F=\beta_{b}^{-1}(I_{BH}^{E}-I_{soliton}^{E})=\frac{TL\alpha}{3 l^{2}%
}\biggl{(}\frac{L_{s}\beta_{s}}{\lambda^{3}}-\frac{L_{b}\beta_{b}}{\eta^{3}%
}\biggr{)}~.
\end{equation}
Usando las expresiones de la temperatura
del agujero negro $T$ y la periodicidad
$L_{s}$, podemos escribir la diferencia 
de las energ\'ias libres como
\begin{equation}
\Delta F=\frac{4\pi LL_{s}}{3l^{2}}\biggl{[}\frac{\Omega(\lambda,x_{s})}%
{L_{s}}-T\Omega(\eta,x_{h})\biggr{]} = \frac{4\pi L}{3l^{2}}\Omega
(\lambda,x_{s})\biggl{(}1-\frac{r_{b}^{3}}{r_{s}^{3}}\biggr{)}~.
\end{equation}
Escrita en t\'erminos de la temperatura,
hay un c\'ambio dr\'astico comparado
con el caso sin pelo porque el factor
conforme aparece expl\'icitamente. 
Cl\'aramente, el signo de esta expresi\'on
es controlada por la raz\'on $r_{b}/r_{s}$.

A pesar de la aparici\'on del factor
conforme, el punto cr\'itico
donde $\Delta F=0$ corresponde nuevamente
a la temperatura $T_{c}=1/L_{s}$
(esto es porque cuando $\Delta F=0$, $\mu_{b}=\mu_{s}$ y entonces $\eta=\lambda$).
Esto es lo que uno espera para una teor\'ia 
del campo conforme, ya que la transici\'on
de fase debe depender de la raz\'on
de las escalas. \newline
Escribiendo el \'area del agujero negro
en t\'erminos de $\beta_{b}$ y $\beta_{s}$, 
encontramos que
\begin{equation}
\frac{\mathcal{A}}{Tl^{3}}=\frac{\alpha L}{4\pi l^{5}}\frac{\beta_{b}^{2}%
	L_{s}}{\eta^{3}}=\frac{L\mathcal{L}}{l}\biggl{(}\frac{\lambda}{\eta
}\biggr{)}\label{ratiosol}%
\end{equation}
donde
\begin{equation}
\mathcal{L}=\frac{16\pi^{2}}{\alpha^{2}l^{4}}\biggl{[}\frac{9x_{h}^{2}}%
{(x_{h}^{3}-1)^{2}}\biggr{]}^{3}%
\end{equation}
Sin embargo, siendo que $x_{h}$ satisface
$f(x_{h})=0$, esta puede ser calculada como
funci\'on del par\'ametro $\alpha$ del 
potencial moduli, lo cual implica
que el coeficiente $\mathcal{L}\left(  \alpha,l\right)$ es s\'olo una funci\'on  
de $\alpha$ y $l$. De la definici\'on
de (\ref{ratio}), uno puede f\'acilmente
obtener $r_{b}/r_{s}=\lambda/\eta$ y de all\'i
(\ref{ratiosol}) puede ser reescrita 
en esta forma \'util:    
\begin{equation}
\frac{\mathcal{A}}{Tl^{3}}=\frac{L\,\mathcal{L}\left(  \alpha,l\right)  }%
{l}\frac{r_{b}}{r_{s}}%
\end{equation}
Aqu\'i hay una importante diferencia 
en comparaci\'on con el caso sin pelo,
concr\'etamente la aparici\'on de la funci\'on
$\mathcal{L}\left(
\alpha,l\right)  $. Cuando $\alpha$ es muy
peque\~no entonces $\mathcal{L}$ tambi\'en
es peque\~no. En este caso, uno puede mantener
el r\'adio del horizonte $r_{b}$ del mismo
tama\~no que $r_{s}$. Por consiguiente,
para peque\~nos $\alpha$, no s\'olo
los agujeros negros peque\~nos, tambi\'en
los agujeros negros grandes son inestables 
y decaen a solitones AdS con pelo. Cuando el
par\'ametro $\alpha$ es grande, el comportamiento
termodin\'amico de los agujeros negros con pelo 
es similar al agujero negro planar sin pelo.
\newpage
\subsection{Discusi\'on}

Hawking  y Page mostraron que existen 
transiciones de fase entre el agujero negro
esf\'erico (Schwarzschild) AdS y el 
espacio-tiempo global ($k=1$) AdS. 
Es bien conocido que las transiciones de fase,
ambas en el lado de la gravedad y en el lado 
de la teor\'ia gauge don sensitivas a la topolog\'ia de la folicaci\'on AdS. Para
los agujeros negros AdS con geometr\'ia 
del horizonte planar, no existen transiciones
de fase Hawking-Page respecto al espacio-tiempo
AdS. En otras palabras, la fase del agujero negro planar es siempre dominante para cualquier
temperatura diferente de cero.  
Interesantemente, se mostr'o que cuando
una (o m\'as de las direcciones) son 
compactas existen tambi\'en transiciones de fase Hawking-Page entre los agujeros negros planares
y el solit'on AdS, el cual es obtenida por una 
doble continuaci\'on anal\'itica del agujero negro. Obtuvimos un comportamiento
similar para los agujeros negros con pelo,
pero ahora el ground state corresponde 
al solit'on con pelo. Una importante
diferencia con el caso sin pelo es que
la transici\'on de fase es tambi\'en controlada
por el par\'ametro $\alpha$ en el potencial
escalar. Una vez que $\alpha$ esta fija,
la teor\'ia esta fija, pero para peque\~nos
$\alpha$ la teor\'ia continen agujeros negros
calientes (peque\~nos o grandes) que son inestables y decaen a solitones AdS con pelo.
Este cambio dr\'astico esta relacionado con el 
hecho de que cuando $\alpha$ desaparece,
la soluci\'on del agujero negro con pelo
viene a ser una singularidad desnuda. 
La auto-interacci\'on del campo escalar 
es muy d'ebil y entonces las temperaturas 
grandes pueden desestabilizar el sistema 
independientemente del tama\~no del agujero 
negro. Cuando fijamos la constante cosmol\'ogica de estas soluciones
a cero, se obtienen soluciones de agujeros negros con pelo, asint\'oticamente planas que son regulares. Para futuras direcciones futuras queremos aplicar y extender el m\'etodo de contrat\'erminos para las estas soluciones asintoticamente planas\footnote{Estos proyectos fueron finalizados en fechas posteriores a la tesis \cite{Astefanesei:2018vga,Astefanesei:2019mds} y el an\'alisis de su inestabilidades fueron estudiadas en \cite{Astefanesei:2019pfq}. Para los lectores queremos mencionar que estos modelos fueron recientemente embedded en SUGRA \cite{Anabalon:2017yhv}.}


%% file: conclusiones.tex

\chapter{Conclusiones}

La dualidad AdS/CFT es una herramienta
eficaz para obtener informaci\'on de teor\'ias gauge fuertemente acopladas a partir de sistemas gravitacionales cl\'asicos. En particular los campos escalares en AdS 
corresponden a operadores en la teor\'ia cu\'antica dual. Los agujeros negros describen los estados t\'ermicos de la teor\'ia cu\'antica dual. En esta secci\'on describimos brevemente los resultados de las secciones 5 y 6 los cuales son concr\'etramente los resultados del trabajo de investigaci\'on.

En la presente tesis estudiamos las condiciones de borde de un campo escalar (m\'inimante acoplado a la gravedad) y de masa conforme $m^{2}=-2/l^{2}$, para la cual ambos modos son normalizables. Estudiamos las condiciones de borde que preservan o rompen la simetr\'ia conforme.

Propusimos nuevos contrat\'erminos locales para la rama logar\'itmica, las que nos dieron un buen princ\'ipio variacional $\delta I=0$. 
Calculamos la acci\'on on-shell y mediante f\'ormulas termodin\'amicas obtuvimos la entrop\'ia y energ\'ia gravitacional. Vimos que la masa tiene contribuci\'on del campo escalar. Esta contribuci\'on desaparece cuando el campo escalar preserva la simetr\'ia conforme.
Calculamos el tensor de stress de la CFT a partir del tensor de stress de Brown-York y mostramos que la anomal\'ia de traza desaparece cuando la simetr\'ia conforme es preservada.

Con el fin de complementar los resultados, trabajamos con el procedimiento Hamiltoniano para obtener la energ\'ia del sistema, esto debido a que la energ\'ia es de primeros principios la cantidad conservada asociada a la simetr\'ia de las traslaciones temporales.
Vimos que la masa calculada mediante este m\'etodo da el mismo resultado que el obtenido por el m\'etodo hologr\'afico, a\'un cuando la simetr\'ia conforme se rompe.
   
Finalmente estudiamos las transiciones de fase agujeros negros con pelo escalar y de horizonte esf\'erico. Mostramos que existen transiciones de fase de primer orden entre AdS t\'ermico y los agujeros negros con pelo. Usamos los m\'etodos desarrollados en las secciones anteriores para calcular la acci\'on on-shell y las cantidades termodin\'amicas.
Se mostr'o que, comparando la energ\'ia libre del agujero negro (con pelo) con la soluci\'on
SAdS, esta \'ultima es siempre m\'as estable.
Para el caso de agujeros negros planares, es bien sabido que no existen transiciones de fase, ya que la energ\'ia libre es siempre negativa. Investigaciones previas mostraron que si compactificamos una o m\'as direcciones es posible obtener transiciones de fase Hawking-Page entre agujeros negros planares y el solit\'on AdS. Construimos solitones con pelo y estudiamos las transiciones de fase, respecto a este este estado ground. En este caso el par\'ametro $\alpha$ que aparece en el potencial y que define la teor\'ia juega un rol importante en las transiciones de fase. Para peque\~nos valores de $\alpha$ (la autointeracci\'on del campo escalar es peque\~na) se tienen agujeros negros calientes, grandes y peque\~nos, que son inestables y decaen en 
solitones AdS con pelo.      

\section{Futuras direcciones}

El m\'etodo de contrat\'erminos ha demostrado 
ser una herramienta muy eficaz para determinar
las cantidades conservadas de sistemas  gravitacionales asont\'oticamente AdS, planas y en otras diversas formulaciones de la relatividad general. Actualmente, yo y dem\'as colaboradores estamos trabajando sobre transiciones de fase de agujeros negros (cargados y con pelo escalar) asint\'oticamente planos.
La termodin\'amica y los diagramas de fase 
de estos agujeros negros con pelo
\cite{Anabalon:2013qua,Herdeiro:2015waa,Herdeiro:2015gia,Herdeiro:2014goa} pueden estudiarse
con un m\'etodo similar al de los contrat\'erminos
\cite{Astefanesei:2005ad,Mann:2005yr,Astefanesei:2009wi}.
 Este at\'iculo esta pr\'oximo a ser publicado.
 
Paralelamente, en colaboraci\'on con el Prof. Edelstein estamos investigando sobre las transiciones de fase en SUGRA $\mathcal{N}=8$, y las transiciones de fase de los agujeros negros planares respecto al solit\'on con pelo, para cualquier valor del par\'ametro pelo $\nu$. 

Otra de las interesantes direcci\'ones futuras ser\'ia estudiar los diagramas de fase de la familia de agujeros negros cargados (exactos) con pelo presentados en \cite{Anabalon:2013sra}. En este caso, uno puede
estudiar ambos ensembles, can\'onico y gran can\'onico.
En el ensemble can\'onico la carga, el cual
es una variable extensiva, debe mantenerse fija. Debido a que el espacio-tiempo AdS con una carga fija no es soluci\'on de las ecuaciones de movimiento, es apropiado 
calcular la acci\'on Euclidea respecto al estado ground que es el agujero negro extremo en este caso \cite{Chamblin:1999hg}. Usando argumentos similares
como en \cite{Astefanesei:2010dk,Astefanesei:2011pz} 
se mostr'o en \cite{Anabalon:2013sra} que existen agujeros
negros extremos un horizonte finito y se espera que el ensemble can\'onico este bien definido.

Los c\'alculos de funciones de correlaci\'on en el contexto de la inflaci\'on pueden realizarse mediante procedimientos similares a los que se realizan en el contexto de la dualidad AdS/CFT. Dada la experiencia desarrollada en el contexto de la dualidad AdS/CFT, pretendemos explorar 
en los diversos m\'etodos para obtener funciones de correlaci\'on en el escenario de la inflaci\'on.

%% file: apendice.tex
\chapter{Acci\'on regularizada}
\label{apendice1}
En esta secci\'on usamos los s\'imbolos $\nu_{ij}$ y $\nu$ para designar a las componentes unitarias y el determinante de la secci\'on
transversal~\footnote{No confundir con el par\'ametro hairy $\nu$}. El n\'umero de dimensiones
del espacio-tiempo es $D=n+1$, donde $n$ es el n\'umero de dimensiones del borde tipo-tiempo, $r=\infty$ ($x=1$). 
\section{Acci\'on-bulk on-shell}
\label{onshellapendice}
Sea la acci\'on de la gravedad acoplado
minimanente a un campo escalar
\begin{equation}
I=\int{d^{n+1}x\sqrt{-g}\bl{[}\frac{R}{2\ka}-\frac{(\pa\phi)^{2}}{2}-V(\phi)\br{]}}+\frac{1}{\kappa}\int_{\pa\mathcal{M}}{d^{n}x\sqrt{-h}K}+I_{g}+I_{\phi}~.
\end{equation}
Las ecuaciones de movimiento
para la m\'etrica son
\begin{equation}
G_{\mu\nu}=\ka T_{\mu\nu}~,
\end{equation}
\begin{equation}
G_{\mu\nu}=R_{\mu\nu}-\frac{1}{2}g_{\mu\nu}R~, \qquad  T_{\mu\nu}=\pa_{\mu}\phi\pa_{\nu}\phi-g_{\mu\nu}\bl{[}\frac{(\pa\phi)^{2}}{2}+V\br{]}~,
\end{equation}
donde la traza es
\begin{equation}
G=-\frac{R(n-1)}{2}~, \qquad T=-(n-1)\bl{[}\frac{(\pa\phi)^{2}}{2}+V\frac{(n+1)}{(n-1)}\br{]}~.
\end{equation}
De donde obtenemos
\begin{equation}
G_{\mu\nu}=\ka T_{\mu\nu} \rightarrow \frac{R}{2\ka}=\frac{(\pa\phi)^{2}}{2}+V\frac{(n+1)}{(n-1)}~.
\end{equation}
Entonces, la acci\'on on-shell en el secci\'on Euclidea del bulk es 
\begin{equation}
I_{bulk}^{E}=-\frac{2}{n-1}\int{d^{n+1}x}\sqrt{g^{E}}V(\phi)~.
\end{equation}
\section{Sistema de coordenadas $(t,x,\Sigma_{k,n-1})$}
Aprovechamos el hecho de que las ecuaciones de movimiento
en este sistema de coordenadas en D-dimensiones 
fueron obtenidas en \cite{Acena:2013jya} y que el c\'alculo
de la acci\'on on-shell debidamente regularizada 
es muy sencillo en estas coordenadas.
Sea el ansatz (\ref{Ansatz})\footnote{Es importante aclarar que el ansatz (\ref{Ansatz}) es para cuatro dimensiones y que para dimensiones mayores la secci\'on transversal tiene nuevos t\'erminos}, donde la secci\'on trasversal puede ser: hiperb\'olica, plana, esf\'erica ($k=-1,0,1$)  
\begin{equation}
ds^{2}=\Om(x)\bl{[}-f(x)dt^{2}+\frac{\et^{2}dx^{2}}{f(x)}+d\Si_{k}^{2}\br{]}~.
\label{andres}
\end{equation}
Las ecuaciones de movimiento en este sistema de 
coordenadas son
\begin{align}
E_{t}\,^{t}-E_{x}\,^{x}  &  =0\Longrightarrow 2\kappa\phi^{\prime2}=\frac{D-2}%
{2\Omega^{2}}\left[  3\left(  \Omega^{\prime}\right)  ^{2}-2\Omega
\Omega^{\prime\prime}\right]~, \label{eqphi}\\
E_{t}\,^{t}-\frac{1}{D-2}g^{ab}E_{ab}  &  =0\Longrightarrow f^{\prime\prime
}+\frac{D-2}{2\Omega}\Omega^{\prime}f^{\prime}+2k\eta^{2}%
=0\label{eqf}\\
E_{t}\,^{t}+\frac{1}{D-2}g^{ab}E_{ab}  &  =0~.\Longrightarrow 2\kappa V=-\frac
{D-2}{2\eta^{2}\Omega^{2}}\left[  f\Omega^{\prime\prime}+\frac{D-4}{2\Omega
}f\left(  \Omega^{\prime}\right)  ^{2}+\Omega^{\prime}f^{\prime}\right]
+\frac{k(D-2)}{\Omega}~. \label{eqV}%
\end{align}
Reordenamos convenientemente las ecuaciones 
(\ref{eqf}) y (\ref{eqV})
\begin{equation}
\frac{d}{dx}[\Om^{(D-2)/2}f^{'}]+2\et^{2} k\Om^{(D-2)/2}=0~,
\end{equation}   
\begin{equation}
-\frac{2\et^{2}\Om^{D/2}(2\ka V)}{D-2}=f\Om^{''}\Om^{(D-4)/2}+\Om^{'}(f\Om^{(D-4)/2})^{'}-2\et^{2}k\Om^{(D-2)/2}~,
\end{equation}
combinando las dos ecuaciones anteriores
se obtiene 
\begin{equation}
2\ka V=-\frac{D-2}{2\eta^{2}\Omega^{D/2}}[\Omega^{\frac{D-4}{2}}(f\Omega)^{'}]^{'}~.
\end{equation}
Sea la secci\'on transversal
\begin{equation}
d\Sigma_{k}^{2}=\nu_{ij}dx^{i}dx^{j}~,
\label{transversal}
\end{equation}
en donde $det(\nu_{ij})=\nu$ y 
$\sqrt{-g}=\Om^{D/2}\eta\sqrt{\nu}$. Entonces
la acci\'on on-shell para la parte del $bulk$ 
es f\'acilmente integrada entre $x_{h}$ (lugar donde se ubica el horizonte, que es la ra\'iz mayor de la ecuaci\'on $f(x_{h})=0$)  y $x_{b}$
(lugar donde se ubica el borde)
\begin{equation}
I^{E}_{bulk}=\frac{\be\si_{k,n-1}}{2\ka\et}\bl{[}\Omega^{\frac{D-4}{2}}(f\Omega)^{'}\br{]}_{x_{h}}^{x_{b}}~.
\end{equation}
Definimos el \'area unitaria (n-1)-dimensional de la secci\'on transversal (\ref{transversal}) como simbolizada por $\sigma_{k,n-1}$ (para la 2-esfera es $\sigma_{1,2}=4\pi$). Para calcular el t\'ermino de Gibbons-Hawking, elegimos la foliciaci\'on time-like a $x=constant$, con
la m\'etrica inducida $h^{\mu\nu}=g^{\mu\nu}-n^{\mu}n^{\nu}$
\begin{equation}
h_{ab}dx^{a}dx^{b}=\Omega(x)[-f(x)dt^{2}+d\Si_{k}^{2}]~,
\label{xctefoli}
\end{equation}  
la normal, curvatura extr\'inseca y su traza $K=h^{\mu\nu}K_{\mu\nu}$ son:
\begin{equation}
n_{a}=\frac{\de_{a}^{x}}{\sqrt{g^{xx}}}~,\qquad  K_{ab}=\frac{\sqrt{g^{xx}}}{2}\pa_{x}h_{ab}~, \qquad
K=\frac{1}{2\et}\bl{(}\frac{f}{\Om}\br{)}^{1/2}\bl{[}\frac{(\Om f)^{'}}{\Om f}+(D-2)\frac{\Om^{'}}{\Om}\br{]}~.
\end{equation}
Entonces el t\'ermino de Gibbons-Hawking 
en la secci\'on Euclidea es
\begin{equation}
I_{GH}^{E}=-\frac{\beta\si_{k,n-1}}{2\ka\et }\Om^{(D-2)/2}f\bl{[}\frac{(f\Om)^{'}}{f\Om}+(D-2)\frac{\Om^{'}}{\Om}\br{]}~.
\end{equation}
Los contrat\'erminos de Balasubramanian-Krauss necesarios
para eliminar las divergencias, para dimensiones $D\leq 7$ son \cite{Emparan:1999pm}:
\begin{equation}
I_{g}=-\frac{1}{\ka}\int{d^{n}x\sqrt{-h}\bl{[}\frac{n-1}{l}+\frac{l\m{R}}{2(n-2)}+\frac{l^{3}}{2(n-4)(n-2)^{2}}\bl{(}\m{R}_{ab}\m{R}^{ab}-\frac{n\m{R}^{2}}{4(n-1)}\br{)}\br{]}}~.
\end{equation}
La curvatura intr\'inseca, tensor de Ricci, la curvatura escalar y el invariante
$R_{ab}R^{ab}$ de la foliaci\'on time-like (\ref{xctefoli}) son:
\begin{equation}
\m{R}_{ij}=\frac{(n-2)k}{\Omega}\sigma_{ij}~, \qquad  \m{R}=\frac{k(n-2)(n-1)}{\Omega}~, \qquad \m{R}_{ab}\m{R}^{ab}=\frac{(n-2)^{2}(n-1)k^{2}}{\Omega^{2}}~.
\label{Rh}
\end{equation}
Donde la componente $\m{R}_{tt}=0$. Dadas las cantidades geom\'etricas calculamos el tercer t\'ermino 
\begin{equation}
\m{R}_{ab}\m{R}^{ab}-\frac{n\m{R}^{2}}{4(n-1)}=-\frac{k^{2}}{4\Omega^{2}}(n-2)^{2}(n-1)(n-4)~.
\end{equation}
Entonces el contrat\'ermino gravitacional de
Balasubramanian-Krauss en la secci\'on Eucl\'idea es
\begin{equation}
I^{E}_{g}=\frac{\beta\si_{k,n-1}}{\ka }\frac{(D-2)}{l}\sqrt{\Omega^{D-1}f }\bl{(}1+\frac{l^{2}k}{2\Omega}-\frac{l^{4}k^{2}}{8\Omega^{2}}\br{)}~.
\end{equation}
Considerando las expresiones de la temperatura y  entrop\'ia, donde el \'area del horizonte de eventos de dimensi'on $(n-1)$, es: $\mathcal{A}=\sigma_{n-1,k}\Omega^{(n-1)/2}\vert_{x_{h}}$. Procedemos a sumar las tres contribuciones
\begin{equation}
\beta^{-1}=T=\frac{f^{'}}{4\pi\eta}\br{\vert}_{x_{h}}, \qquad S=\frac{\mathcal{A}}{4G}~,
\end{equation}
\begin{eqnarray}
I^{E}_{bulk}+I^{E}_{GH}+I^{E}_{g}&=&-\frac{1}{T}\bl{(}\frac{\m{A}T}{4G}\br{)} \\ \notag
&& -\frac{\si_{D-2,k}}{2\ka T}\Omega^{(D-2)/2}(D-2)\bl{[}\frac{f\Omega^{'}}{\eta\Omega}-\frac{\sqrt{\Omega f}}{2l}\bl{(}1+\frac{l^{2}k}{2\Omega}-\frac{l^{4}k^{2}}{8\Omega^{2}}\br{)} \br{]} \br{\vert}_{x_{b}}~. \\ \notag
\end{eqnarray}
\section{Sistema de coordenadas $(t,r,\Sigma_{k,n-1})$}
Procederemos a realizar el c\'alculo en las
coordenadas usuales, una de las ventajas es que los resultados son m\'as intuitivos que el caso anterior. No es necesario volver a calcular todo, simplemente realizaremos un cambio de coordenadas,
sea el ansatz est\'atico en este nuevo sistema coordenado
\begin{equation}
ds^{2}=-N(r)dt^{2}+H(r)dr^{2}+S(r)d\Si_{k}^{2}~.
\end{equation} 
Los sistemas coordenados $(t,x,\Sigma_{n-1,k})$ y $(t,r,\Sigma_{n-1,k})$ est'an relacionados por
\begin{equation}
\Om(x) \rightarrow S(r)~, \qquad f(x)\rightarrow \frac{N(r)}{S(r)}~, \qquad \frac{\sqrt{NH}}{\et S}dr \rightarrow dx~.
\end{equation}
El potencial es
\begin{equation}
2\ka V=-\frac{D-2}{2\eta^{2}\Omega^{D/2}}[\Omega^{\frac{D-4}{2}}(f\Omega)^{'}]^{'} \rightarrow 2\kappa V=-\frac{D-2}{2S^{\frac{D-2}{2}}\sqrt{NH}}\frac{d}{dr}\bl{(}\frac{S^{\frac{D-2}{2}}}{\sqrt{NH}}\frac{dN}{dr}\br{)}~.
\end{equation}
En estas nuevas coordenadas, $r_{h}$ es el 
radio del horizonte de eventos del agujero negro
tal que es la soluci\'on mayor de $N(r_{h})=0$ y el borde $x_{b}$ es $r_{b}=\infty$. El t\'ermino del $bulk$ es
\begin{equation}
I^{E}_{bulk}=\frac{\be\si_{k,n-1}}{2\ka\et}\bl{[}\Omega^{\frac{D-4}{2}}(f\Omega)^{'}\br{]}_{x_{h}}^{x_{b}}  \rightarrow
I_{bulk}^{E}=\frac{\beta\si_{k,n-1}}{2\ka }\frac{dN}{dr}\frac{S^{(n-1)/2}}{\sqrt{NH}} \br{\vert}_{r_{h}}^{r_{b}}~.
\end{equation} 
Sea la foliaci\'on time-like $r=R=constante$, donde la m\'etrica inducida es $h^{\mu\nu}=g^{\mu\nu}-n^{\mu}n^{\nu}$ 
\begin{equation}
h_{ab}dx^{a}dx^{b}=-N(R)dt^{2}+S(R)d\Si_{k}^{2}~,
\end{equation}
la normal, la curvatura extr\'inseca y la traza $K=h^{\mu\nu}K_{\mu\nu}$ son
\begin{equation}
n_{\mu}=\frac{\de_{\mu}^{r}}{\sqrt{g^{rr}}}~,
\qquad K_{\mu\nu}=\frac{\sqrt{g^{rr}}}{2}\pa_{r}h_{\mu\nu}~,
\qquad K=\frac{1}{2\sqrt{H}}\bl{[}\frac{N^{'}}{N}+(n-1)\frac{S^{'}}{S}\br{]}~.
\end{equation}
Los t\'erminos de Gibbons-Hawking y los contrat\'erminos gravitacionales de Balasubramanian-Krauss en la secci\'on Euclidea son 
\begin{equation}
I_{GH}^{E}=-\frac{\beta\si_{k,n-1}}{2\ka\et }\Om^{\frac{D-2}{2}}f\bl{[}\frac{(f\Om)^{'}}{f\Om}+(D-2)\frac{\Om^{'}}{\Om}\br{]}\br{\vert}_{x_{b}}
\rightarrow I^{E}_{GH}=-\frac{\si_{k,n-1}}{2\ka T}\frac{S^{\frac{D-2}{2}}}{\sqrt{NH}}\bl{[}\frac{dN}{dr}+(D-2)\frac{N}{S}\frac{dS}{dr}\br{]} \br{\vert}_{r_{b}}~,
\end{equation}
\begin{equation}
I^{E}_{g}=\frac{\beta\si_{k,n-1}(n-1)}{l\ka }\Omega^{\frac{D-1}{2}}f^{\frac{1}{2}} \bl{(}1+\frac{l^{2}k}{2\Omega}-\frac{l^{4}k^{2}}{8\Omega^{2}}\br{)}_{x_{b}} \rightarrow I_{g}^{E}=\frac{\beta\si_{k,n-1}(n-1)}{l\ka }S^{\frac{D-2}{2}}N^{\frac{1}{2}}\bl{(}1+\frac{l^{2}k}{2S}-\frac{l^{4}k^{2}}{8S^{2}}\br{)}_{r_{b}}~.
\end{equation}
Considerando de nuevo las expresiones para la temperatura y entrop\'ia del agujero negro, donde el \'area del horizonte de eventos es $\mathcal{A}=\sigma_{k,n-1}S^{\frac{D-2}{2}}\vert_{r_{h}}$
\begin{equation}
\beta^{-1}=T=\frac{N^{'}}{4\pi\sqrt{NH}} \br{\vert}_{r_{h}}, \qquad S=\frac{\mathcal{A}}{4G}~,
\end{equation}
uno puede escribir la suma de las tres contribuciones en la acci\'on total como
\begin{eqnarray}
I^{E}_{bulk}+I^{E}_{GH}+I^{E}_{g}&=&-\frac{1}{T}\bl{(}\frac{\m{A} T}{4G}\br{)} \\ \notag
&& -\frac{\si_{D-2,k}}{2\ka T}S^{(D-2)/2}(D-2)\bl{[}\frac{NS^{'}}{S\sqrt{NH}}-\frac{2\sqrt{N}}{l}\bl{(}1+\frac{l^{2}k}{2S}-\frac{l^{4}k^{2}}{8S^{2}}\br{)} \br{]} \br{\vert}_{r_{b}}~. \\ \notag
\end{eqnarray}
\newpage
\section{Ecuaciones de movimiento}
Las ecuaciones de movimiento en las coordenadas, $(t,x,\Sigma_{k})$ y $(t,r,\Sigma_{k})$,
son
\begin{equation}
2\kappa\phi^{\prime2}=\frac{D-2}%
{2\Omega^{2}}\left[  3\left(  \Omega^{\prime}\right)  ^{2}-2\Omega
\Omega^{\prime\prime}\right]; \qquad \frac{2\kappa\phi^{'2}}{D-2}=\frac{1}{2S^{2}}[S^{'2}-2SS^{''}]+\frac{S^{'}}{2S}\frac{(NH)^{'}}{NH}
\end{equation}  
\begin{equation}
\frac{d}{dx}\bl{[}\Omega^{\frac{D-2}{2}}\frac{df}{dx}\br{]}=-2\eta^{2}k\Omega^{\frac{D-2}{2}}; \qquad \qquad \frac{d}{dr}\bl{[}\frac{S^{D/2}}{\sqrt{NH}}\frac{d}{dr}\bl{(}\frac{N}{S}\br{)}\br{]}=-2k\sqrt{NH}S^{\frac{D-4}{2}}
\end{equation}
\begin{equation}
2\ka V=-\frac{D-2}{2\eta^{2}\Omega^{D/2}}[\Omega^{\frac{D-4}{2}}(f\Omega)^{'}]^{'}; \qquad 2\kappa V=-\frac{D-2}{2S^{\frac{D-2}{2}}\sqrt{NH}}\frac{d}{dr}\bl{(}\frac{S^{\frac{D-2}{2}}}{\sqrt{NH}}\frac{dN}{dr}\br{)}
\end{equation}
y para el campo escalar
\begin{equation}
\pa_{x}[\Omega^{\frac{D-4}{2}}f\phi^{'}]=\eta^{2}\Omega^{D/2}\frac{\pa V}{\pa \phi}; \qquad
\pa_{r}\bl{(}S^{\frac{D-2}{2}}\phi^{'}\sqrt{\frac{N}{H}}\br{)}=\sqrt{NH}S^{\frac{D-2}{2}}\frac{\pa V}{\pa\phi}
\end{equation}
\chapter{Tensor de stress Brown-York}
Los t\'erminos de borde de la acci\'on, son lo \'unicos que contribuyen al cuasilocal stress tensor de Brown-York definido en \cite{Brown:1992br}
\begin{equation}
\tau^{ab}\equiv\frac{2}{\sqrt{-h}}\frac{\delta I}{\delta h_{ab}}~.%
\end{equation}
Para dimensiones $D=3,4,5$ son necesarios s\'olo dos contrat\'erminos gravitacionales. Nombraremos $\Psi$ a los contrat\'erminos del campo escalar, los cuales dependen expl\'icitamente de las condiciones de borde del campo escalar 
\begin{equation}
I_{GH}+I_{g}+I_{\phi}=\frac{1}{\ka}\int{d^{n}x\sqrt{-h}K}-\frac{1}{\ka}\int{d^{n}x\sqrt{-h}\bl{[}\frac{n-1}{l}+\frac{l\m{R}}{2(n-2)}\br{]}}-\int{d^{n}x\sqrt{-h}\Psi}~.
\end{equation}
Variando esta acci\'on, tenemos 
\begin{equation}
\label{BY1}\tau_{ab}=-\frac{1}{\kappa}\biggl{(}K_{ab}-h_{ab}K+\frac{n-1}%
{l}h_{ab}-\frac{l}{n-2} G_{ab}\biggr{)}-h_{ab}[\Psi]~.
\end{equation}
Sea la foliaci\'on time-like $r=R=constant$
\begin{equation}
h_{ab}dx^{a}dx^{b}=-N(R)dt^{2}+S(R)d\Sigma^{2}_{k}~,
\end{equation}
en donde se define el tensor de Einstein $G_{ab}$, que se calcula de las expresiones dadas en (\ref{Rh})
\begin{equation}
G_{ab}=\m{R}_{ab}-\frac{1}{2}\m{R}h_{ab}~, \qquad G_{tt}=\frac{(n-2)(n-1)}{2}\frac{kN}{S}~, \qquad G_{ij}=-\frac{(n-2)(n-3)}{2}k\nu_{ij}~.
\end{equation}
Entonces las componentes del stress tensor, en las coordenadas $(t,r,\Sigma_{k})$, las cuales se expanden al rededor del borde $r_{b}=\infty$, son\footnote{Estos t\'erminos son suficientes para dimensiones $D=3,4,5,6,7$.}
\begin{equation}
\tau_{tt}=-\frac{(n-1)}{\kappa}\bl{[}\frac{NS^{'}}{2S\sqrt{H}}-\frac{N}{l}\bl{(}1+\frac{l^{2}k}{2S}\br{)}\br{]}+N[\Psi]~,
\end{equation}
\begin{equation}
\tau_{ij}=\frac{\nu_{ij}}{\kappa}\bl{[}\frac{S}{2\sqrt{H}}\bl{(}\frac{N^{'}}{N}+\frac{S^{'}}{S}(n-2)\br{)}-\frac{(n-1)S}{l}-\frac{lk(n-3)}{2}\br{]}-\nu_{ij}S[\Psi]~,
\end{equation}
donde $\nu_{ij}$ son las componentes de la secci\'on transversal $d\Sigma_{k}^{2}$~.\\
Parca calcular la energ\'ia se considera que la m\'etrica del borde puede ser localmente escrita 
en la forma ADM:
\begin{equation}
h_{ab}dx^{a}dx^{b}=-L^{2}dt^{2}+\sigma_{ij}(dy^{i}+L^{i}dt)(dy^{j}+L^{j}dt)~,
\end{equation}
donde $L$ y $L^{i}$ son las funciones lapso y shift respectivamente, ${y^{i}}$ son las coordenadas intr\'insecas
en una hiper-superficie $\Sigma$. La geometr\'ia 
del borde tiene una isometr\'ia generada por el vector de 
Killing  $\xi^{a}=(\partial_{t})^{a}$ para el cual
la carga conservada es la energ\'ia gravitacional:
\begin{equation}
E=Q_{\frac{\partial}{\partial t}}=\oint_{\Sigma}d^{D-2}y\sqrt{\sigma}u^{a}%
\tau_{ab}\xi^{b}=\bl{(}\oint_{\Sigma} d^{2}y\sqrt{\nu}\br{)}\frac{S^{\frac{D-2}{2}}\tau_{tt}}{\sqrt{N}}=\frac{\sigma_{k,n-1}S^{\frac{D-2}{2}}}{\sqrt{N}}\tau_{tt}~,
\end{equation}
para la m\'etrica estacionaria, $L^{i}=0$, $L^{2}=N$, donde la integral esta asociada con la superficie $t=constante$,
cuya m\'etrica inducida es 
\begin{equation}
\sigma_{ij}dx^{i}dx^{j}=Sd\Sigma_{k}^{2}~.
\end{equation}
El vector normal  $u^{a}=(\partial_{t})^{a}/\sqrt{-g_{tt}}$.\\
\newpage
Es igualmente \'util tener las expresiones 
de las componentes del tensor hologr\'afico en
las coordenadas $(t,x,\Sigma_{k,n-1})$
\begin{equation}
\tau_{tt}=-\frac{(n-1)}{\kappa}\bl{[}\frac{f^{3/2}\Omega^{'}}{2\eta \sqrt{\Omega}}-\frac{\Omega f}{l}\bl{(}1+\frac{l^{2}k}{2\Omega}\br{)}\br{]}+\frac{\Omega f}{\kappa}[\Psi],
\end{equation}
\begin{equation}
\tau_{ij}=\frac{\nu_{ij}}{\kappa}\bl{[}\frac{(\Omega f)^{'}}{2\eta\sqrt{\Omega f}}+\frac{(n-2)}{2\eta}\frac{\Omega^{'}\sqrt{f}}{\sqrt{\Omega}}-\frac{(n-1)\Omega}{l}-\frac{lk(n-3)}{2}\br{]}-\frac{\nu_{ij}\Omega}{\kappa}[\Psi].
\end{equation}
Y la energ\'ia 
\begin{equation}
E=Q_{\frac{\partial}{\partial t}}=\oint_{\Sigma}d^{D-2}y\sqrt{\sigma}u^{a}%
\tau_{ab}\xi^{b}=\frac{\sigma_{k,n-1}\Omega^{\frac{D-2}{2}}}{\sqrt{\Omega f}}\tau_{tt}~,
\end{equation}
\chapter{C\'alculo del potencial on-shell}
\label{calculopotencial}
En esta parte mostrar\'e algunos detalles 
del c\'alculo no-trivial del potencial
para el agujero negro con pelo, con secci\'on
transversal plana $k=0$. Para 
el caso esf\'erico $k=1$ el procedimiento (pero m\'as tedioso) y resultado es el mismo. 
\newpage
\begin{align}
\et^{2}V(\phi)&=-\frac{f\Om^{''}}{\eta^{2}\Om^{2}}-\frac{f^{'}\Om^{'}}{\eta^{2}\Om^{2}} \\ \non
\et^{2}V(\phi)&=-\frac{1}{x^{2\nu-2}\nu^{4}}\bl{[}\frac{1}{l^{2}}+\frac{\alpha}{2}\bl{(}\frac{x^{2+\nu}-1}{2+\nu}+\frac{x^{2-\nu}-1}{2+\nu}-x^{2}+1\br{)}\br{]}\bl{[}\frac{x^{\nu-1}(\nu-1)^{2}\nu^{2}-x^{\nu-1}(\nu-1)\nu^{2}}{x^{2}(x^{\nu}-1)^{2}\eta^{2}} \\ \non
&-\frac{4x^{2\nu-1}(\nu-1)\nu^{3}+2x^{2\nu-1}\nu^{4}}{(x^{\nu}-1)^{4}\eta^{2}x^{2}}+\frac{2x^{2\nu-1}\nu^{3}-4x^{2\nu-1}(\nu-1) \nu^{3}}{x^{2}\eta^{2}(x^{\nu}-1)^{3}}\br{]}(x^{\nu}-1)^{4}\eta^{4} \\ \non
&-\frac{1}{2x^{2\nu-2}\nu^{4}}\bl{(}\frac{x^{\nu-1}(\nu-1)\nu^{2}}{x\eta^{2}(x^{\nu}-1)^{2}}-\frac{2x^{2\nu-1}\nu^{3}}{x\eta^{2}(x^{\nu}-1)^{3}}\br{)}\bl{(}x^{1+\nu}+x^{1-\nu}-2x\br{)} \\ \non
V(\phi)&=-\frac{f(x)\nu^{2}}{x^{2\nu-2}\nu^{4}}\bl{(}\frac{x^{3\nu-3}\nu^{2}+4x^{2\nu-3}\nu^{2}+x^{\nu-3}\nu^{2}+3x^{3\nu-3}\nu-3x^{\nu-3}\nu}{(x^{\nu}-1)^{4}\eta^{2}}\br{)}(x^{\nu}-1)^{4}\eta^{2} \\ \non
&-\frac{f(x)\nu^{2}}{x^{2\nu-2}\nu^{4}}\bl{(}\frac{2x^{3\nu-3}-4x^{2\nu-3}+2x^{\nu-3}}{(x^{\nu}-1)^{4}\eta^{2}}\br{)}(x^{\nu}-1)^{4}\eta^{2} \\ \non 
&-\frac{\al\eta^{4}(x^{\nu}-1)^{4}}{2x^{2\nu-2}\nu^{4}}\bl{(}-\frac{\nu^{2}}{\eta^{2}(x^{\nu}-1)^{3}}\br{)}\bl{(}x^{2\nu-2}(\nu+1)+x^{\nu-2}(\nu-1)\br{)}\bl{(}x^{1+\nu}+x^{1-\nu}-2x\br{)} \\ \non
V(\phi)&=-\frac{f(x)x^{-\nu}}{x\nu^{2}}\bl{(}x^{2\nu}(\nu+1)(\nu+2)+4x^{\nu}(\nu^{2}-1)+(\nu-1)(\nu-2)\br{)} \\ \non
&+\frac{\al(x^{\nu}-1)}{2\nu^{2}x^{\nu-1}}\frac{x^{2\nu-2}}{x^{\nu-1}}\bl{(}x^{1+\nu}+x^{1-\nu}-2x\br{)}\bl{(}x^{-\nu}(\nu-1)+(\nu+1)\br{)} \\ \non
V(\phi)&=-\frac{f(x)}{2x\nu^{2}}\bl{(}2x^{\nu}(\nu+1)(\nu+2)+8(\nu^{2}-1)+2x^{-\nu}(\nu-1)(\nu-2)\br{)} \\ \non
& +\frac{\al}{2\nu^{2}x^{\nu-1}}(x^{\nu}-1)\bl{(}1+\nu+x^{-\nu}(\nu-1)\br{)}(1+x^{2\nu}-2x^{\nu}) \\ \non
V(\phi)&=-\frac{f(x)}{2x\nu^{2}}\bl{(}2x^{\nu}(\nu+1)(\nu+2)+8(\nu^{2}-1)+2x^{-\nu}(\nu -1)(\nu -2)\br{)} \\ \non
&-\frac{\al}{2\nu^{2}x^{\nu -1}}(x^{\nu}-1)^{2}\bl{(}2-x^{\nu}(\nu +1)+x^{-\nu}(\nu -1)\br{)} \\ \non 
\end{align}
Donde $f(x)=f(e^{l_{\nu}\phi})$ es la funci\'on
m\'etrica, y es posible simplificar a\'un m\'as
\begin{align}
V(\phi)&=-\frac{e^{-l_{\nu}\phi}}{2\nu^{2}}\bl{[}\frac{1}{l^{2}}+\frac{\al}{2}\bl{(}\frac{e^{(2+\nu)l_{\nu}\phi}-1}{2+\nu}+\frac{e^{(2-\nu)l_{\nu}\phi}-1}{2-\nu}-x^{2}+1\br{)}\br{]}\bl{[}8(\nu^{2}-1)\\ \non
&+2(\nu+1)(\nu+2)e^{\nu l_{\nu}\phi}+ \\ \non
& 2(\nu-1)(\nu-2)e^{-\nu l_{\nu}\phi}\br{]}-\frac{\al e^{l_{\nu}\phi}}{2\nu^{2}}\bl{(}\exp{\frac{\nu l_{\nu}\phi}{2}}-\exp{-\frac{\nu l_{\nu}\phi}{2}}\br{)}^{2}\bl{[}2-e^{\nu l_{\nu}\phi}(\nu+1)+e^{-\nu l_{\nu}\phi}(\nu-1)\br{]} \\ \non
V(\phi)&=-\frac{(\nu^{2}-4)}{l^{2}\nu^{2}}\bl{[}\frac{\nu-1}{\nu+2}e^{-\phi l_{\nu}(\nu+1)}+\frac{\nu+1}{\nu-2}e^{\phi l_{\nu}(\nu-1)}+4\frac{\nu^{2}-1}{\nu^{2}-4}e^{-\phi l_{\nu}}\br{]} \\ \non
& -\frac{\al}{2\nu^{2}}\bl{[}\frac{e^{(\nu+2)l_{\nu}\phi}}{2+\nu}-\frac{e^{(2-\nu)l_{\nu}\phi}}{\nu-2}-e^{2l_{\nu}\phi}+\frac{\nu^{2}}{\nu^{2}-4}\br{]}\bl{[}4(\nu^{2}-1)e^{-l_{\nu}\phi}+(\nu+1)(\nu+2)e^{(\nu-1)l_{\nu}\phi}+ \\ \non
& (\nu-1)(\nu-2)e^{-(\nu+1)l_{\nu}\phi}\br{]}\\ \non
&-\frac{\al}{2\nu^{2}}(e^{\nu l_{\nu}\phi}-2+e^{-\nu l_{\nu}\phi})[2e^{l_{\nu}\phi}-e^{(\nu+1)l_{\nu}\phi}(\nu+1)+e^{(1-\nu)l_{\nu}\phi}(\nu-1)] \\ \non
V(\phi)&=V_{\Lambda}(\phi)-\frac{\al}{2(\nu^{2}-4)}\bl{[}\nu^{2}(e^{l_{\nu}\phi (\nu-1)}-e^{l_{\nu}\phi (\nu+1)}-e^{l_{\nu}\phi (\nu-1)}+e^{-l_{\nu}\phi (\nu+1)}+4e^{-l_{\nu}\phi}-4e^{l_{\nu}\phi})+ \\ \non
& 3\nu (e^{\phi l_{\nu}(\nu-1)}+e^{\phi l_{\nu}(\nu+1)}-e^{-\phi l_{\nu}(\nu-1)}-e^{-\phi l_{\nu}(\nu+1)}) \\ \non
& +2(e^{\phi l_{\nu}(\nu-1)}-e^{\phi l_{\nu}(\nu+1)}-e^{-\phi l_{\nu}(\nu-1)}+e^{-\phi l_{\nu}(\nu+1)})-4e^{-l_{\nu}\phi}+4e^{l_{\nu}\phi}\br{]} \\ \non
V(\phi)&=V_{\Lambda}(\phi)-\frac{\al}{2(\nu^{2}-4)}\bl{[}(2\nu^{2}+6\nu+4)\sinh{\phi l_{\nu}(\nu-1)}-(2\nu^{2}-6\nu+4)\sinh{\phi l_{\nu}(\nu+1)}\\ \non
&+8(1-\nu^{2})\sinh{\phi l_{\nu}}\br{]} \\ \non
V(\phi)&=\frac{\Lambda (\nu^{2}-4)}{3\nu^{2}}\bl{[}\frac{\nu-1}{\nu+2}e^{-\phi l_{\nu}(\nu+1)}+\frac{\nu+1}{\nu-2}e^{\phi l_{\nu}(\nu-1)}+4\frac{\nu^{2}-1}{\nu^{2}-4}e^{-\phi l_{\nu}}\br{]} \\ \non
& +\al\bl{[}\frac{\nu-1}{\nu+2}\sinh{\phi l_{\nu}(\nu+1)}-\frac{\nu+1}{\nu-2}\sinh{\phi l_{\nu}(\nu-1)}+4\frac{\nu^{2}-1}{\nu^{2}-4}\sinh{l_{\nu \phi}}\br{]} \\ \non    
\end{align}

Como puede verse, este potencial no depende
 de $\eta$ la constante de integraci\'on. Esta desaparece 
a medida que uno va simplificando. Entonces 
la teor\'ia queda definida por los par\'ametros 
$\Lambda, \alpha,\nu, G$. Donde $\Lambda$ es la
constante cosmol\'ogica, $\alpha$ es un nuevo par\'ametro de la teor\'ia, $\nu$ es el par\'ametro hairy y $G$ es la constante de Newton.
 
\chapter{Formas diferenciales}
\label{Formas11}
Una forma diferencial de orden $r$ 
o $r$-forma, es un tensor totalmente antisim\'etrico 
del tipo $(0,r)$ el cual se define en el espacio cotangente $T^{\star}\mathcal{M}$ de la variedad $\mathcal{M}$ de dimensi\'on $D$.\\
El producto cu\~na (que es una base para las $2$-formas) se define como,
\begin{equation}
dx^{\mu}\wedge dx^{\nu}=dx^{\mu}\otimes dx^{\nu}-dx^{\nu}\otimes dx^{\mu}.
\end{equation}
La base de un espacio vectorial de las $p$-formas
diferenciales, denotado por  $\Lambda^{p}(\mathcal{M})$.
\begin{equation}
dx^{\mu_{1}}\wedge\ldots\wedge dx^{\mu_{p}}=\sum_{\sigma}(-1)^{\vert\sigma\vert}dx^{\sigma(\mu_{1})}\otimes\ldots\otimes dx^{\sigma(\mu_{p})},
\end{equation}
donde $\sigma$ denota la permutaci\'on de \'indices. Entonces, sea la forma $H\in\Lambda^{p}(\mathcal{M})$
que puede ser escrita en la base
\begin{equation}
H=\frac{1}{p!}H_{\mu_{1}\ldots\mu_{p}}dx^{\mu_{1}}\wedge\ldots\wedge dx^{\mu_{p}}.
\end{equation}
Por ejemplo, el campo gauge del electromagnetismo
es una $1$-forma, 
\begin{equation}
A=A_{\mu}dx^{\mu}.
\end{equation}
El tensor de fuerza del campo electromagn\'etico es
es una $2$-forma,
\begin{equation}
F=\frac{1}{2}F_{\mu\nu}dx^{\mu}\wedge dx^{\nu}.
\end{equation}
La derivada exterior $d=dx^{\mu}\wedge\pa_{\mu}$
es un operador que mapea $d:\Lambda^{p}(\mathcal{M})\rightarrow\Lambda^{p+1}(\mathcal{M})$. Se puede mostrar para el campo electromagn\'etico que el tensor de fuerza $F$ es la derivada exterior del campo gauge, $F=dA$.\\
El dual de Hodge definida en la variedad $\mathcal{M}$ (de dimensi\'on $D$)
equipada con la m\'etrica $g_{\mu\nu}$ es el mapeo $\star:\Lambda(\mathcal{M})^{p}\rightarrow \Lambda(\mathcal{M})^{D-p}$. El dual de Hodge 
actuando sobre la base de las formas diferenciales es,
\begin{equation}
\star(dx^{\mu_{1}}\wedge\ldots\wedge dx^{\mu_{p}})=
\frac{\sqrt{-g}}{(D-p)!}\epsilon^{\mu_{1}\ldots\mu_{p}}_{~~~~~~~\nu_{p+1}\ldots\nu_{D}}dx^{\nu_{p+1}}\wedge\ldots\wedge dx^{\nu_{D}}.
\end{equation}
Donde $\epsilon_{\mu_{1}\ldots\mu_{D}}$ es la densidad
de Levi-Civita, tal que $\epsilon_{\mu_{1}\ldots\mu_{D}}=(+1,-1,0)$
dependiendo de las permutaciones. 
Finalmente, el elemento de volumen
se puede expresar como,
\begin{equation}
dV=d^{D}x\sqrt{-g}=\frac{\sqrt{-g}}{D!}\epsilon_{\mu_{1}\ldots\mu_{D}}dx^{\nu_{1}}\wedge\ldots\wedge dx^{\nu_{D}}.
\end{equation}